\documentclass[preprint, aps, prd, preprintnumbers, floatfix, nofootinbib, 11pt]{revtex4-1}

\usepackage[dvips]{graphics}
\usepackage{amsmath,amsfonts,amssymb,amsthm}
\usepackage{epsfig,bm}
\usepackage{feynmp}
\usepackage{graphicx,comment}
\usepackage{dcolumn}
\usepackage{mathtools}
\usepackage{fancyhdr}
\usepackage{cancel}
\usepackage{slashed}
\usepackage{verbatim}
\usepackage{listings}
\usepackage{latexsym}
\usepackage{rotating}
\usepackage{float}
\usepackage{enumerate}
\usepackage{array}
\usepackage{tabularx}
\usepackage{longtable}
\usepackage{booktabs}

\usepackage[usenames,dvipsnames]{color}
  \definecolor{hgreen}{rgb}{0,.5,0}
  \definecolor{hred}{rgb}{.5,0,0}
  \definecolor{hblue}{rgb}{0,0,.5}
\usepackage[colorlinks=true,linkcolor=hblue,citecolor=hgreen,filecolor=hblue,urlcolor=hred]{hyperref}

\usepackage[english]{babel}
\usepackage[utf8]{inputenc}

\allowdisplaybreaks

\def\beq{\begin{equation}}
\def\eeq{\end{equation}}
\def\bea{\begin{eqnarray}}
\def\eea{\end{eqnarray}}

\newcommand{\be}{\begin{equation}}
\newcommand{\ee}{\end{equation}}
\newcommand{\bear}{\begin{eqnarray}}
\newcommand{\eear}{\end{eqnarray}}
\newcommand{\ba}{\begin{array}}
\newcommand{\ea}{\end{array}}

\begin{document}


\title{Flavorful Two Higgs Doublet Models with a Twist}

\author{Wolfgang~Altmannshofer}
\email{wolfgang.altmannshofer@uc.edu}
\affiliation{Department of Physics, University of Cincinnati, Cincinnati OH, 45220, USA}

\author{Brian~Maddock}
\email{maddocbf@mail.uc.edu}
\affiliation{Department of Physics, University of Cincinnati, Cincinnati OH, 45220, USA}

\begin{abstract}
We explore Two Higgs Doublet Models with non-standard flavor structures.
In analogy to the four, well studied, models with natural flavor conservation (type~1, type~2, lepton-specific, flipped), we identify four models that preserve an approximate $U(2)^5$ flavor symmetry acting on the first two generations.
In all four models, the couplings of the 125~GeV Higgs are modified in characteristic flavor non-universal ways. The 
heavy neutral and charged Higgs bosons show an interesting non-standard phenomenology. We discuss their production and decay modes and identify the most sensitive search channels at the LHC.
We study the effects on low energy flavor violating processes finding relevant constraints from $B_d$ and $B_s$ meson oscillations and from the rare decay $B_s \to \mu^+ \mu^-$. We also find that lepton flavor violating $B$ meson decays like $B_s \to \tau \mu$ and $B \to K^{(*)} \tau \mu$ might have branching ratios at an observable level.  
\end{abstract}



\maketitle
\tableofcontents

\section{Introduction}

Measurements of Higgs rates at the LHC show that the Standard Model (SM) Higgs mechanism provides the bulk of the masses of the third generation fermions. The decay $h \to \tau^+\tau^-$ has been observed at a rate compatible with the SM prediction~\cite{Sirunyan:2017khh}. Similarly, evidence exists for a SM-like $h \to b\bar b$ decay~\cite{Sirunyan:2017elk, Aaboud:2017xsd}.
Recently, production of the Higgs in association with top quarks has been observed in agreement with the SM~\cite{Sirunyan:2018hoz}.

Much less is known about the origin of the first and second generation masses. With the exception of the muon, direct measurements of Higgs couplings to the light fermions are extremely challenging. It is therefore unknown if the light fermions obtain their mass from the Higgs boson.
A complementary approach to probe the origin of light fermion masses is to search for signatures of alternatives to the SM Higgs mechanism in which the light fermion masses originate from a new source of electroweak symmetry breaking.
The simplest realization of such a setup is the Two Higgs Doublet Model (2HDM).

In~\cite{Altmannshofer:2015esa} a 2HDM setup was proposed in which one Higgs doublet couples only to the third generation fermions, and a second Higgs doublet couples mainly to the first and second generation (see also~\cite{Ghosh:2015gpa,Botella:2016krk,Das:1995df,Blechman:2010cs}). A dynamical generation of such a coupling structure can be achieved using the flavor-locking mechanism~\cite{Knapen:2015hia,Altmannshofer:2017uvs}. The collider phenomenology of this ``flavorful'' 2HDM scenario was discussed in~\cite{Altmannshofer:2016zrn}. 

The proposed 2HDM goes beyond the principle of natural flavor conservation (NFC)~\cite{Glashow:1976nt} and introduces flavor changing neutral currents (FCNCs) at tree level. However, the Yukawa couplings of the first Higgs doublet to the third generation preserve a $U(2)^5$ flavor symmetry, which is only broken by the small couplings of the second Higgs doublet. The approximate $U(2)^5$ symmetry protects the most sensitive flavor violating transitions between the second and first generation. 
 
In this work we explore additional flavor structures for 2HDMs that approximately preserve a $U(2)^5$ flavor symmetry for the first two generations. Starting from the flavorful 2HDM scenario of~\cite{Altmannshofer:2015esa} we ``twist'' the Yukawa couplings of the down-type quarks and/or leptons by exchanging the Higgs doublets these fermions couple to.
In analogy to the four well studied 2HDMs with natural flavor conservation (type~1, type~2, lepton-specific, flipped) we obtain four flavorful 2HDMs in which the third and first two generations of each fermion type (up-type quarks, down-type quarks, leptons) obtain the bulk of their mass from a different source.
The non-standard flavor structures of these four 2HDMs lead to (i) distinct, flavor non-universal modifications of all Higgs couplings with respect to the models with NFC, (ii) potentially sizable flavor violating Higgs couplings involving the third generation fermions. This implies an interesting characteristic collider and flavor phenomenology. 
(For recent work on 2HDMs with other non-standard flavor structures see~\cite{Crivellin:2013wna, Bauer:2015fxa, Botella:2015hoa, Bauer:2015kzy, Buschmann:2016uzg, Sher:2016rhh, Primulando:2016eod, Alves:2017xmk, Gori:2017tvg, Kohda:2017fkn, Dery:2017axi, Crivellin:2017upt, Das:2018qjb}.)

The paper is structured as follows:
In section~\ref{F2HDM} we introduce the four flavorful 2HDMs, discuss the Yukawa textures and the couplings of the fermions to the various Higgs boson mass eigenstates. In section~\ref{LHP} we consider the phenomenology of the 125~GeV Higgs boson, comparing the predicted production and decay rates in our models to measurements at the LHC. In sections~\ref{HHP} and \ref{CHP} we evaluate the production cross sections and decay branching ratios of the heavy neutral and charged Higgs bosons. We then compare the model predictions to the limits from current searches for extra Higgs bosons that are being performed at the LHC and identify the most sensitive collider probes of the models. In section~\ref{MesonPheno} we investigate the characteristic effects of the new sources of flavor violation on low energy flavor violating processes such as meson mixing and rare $B$ meson decays. We conclude in section~\ref{sec:conclusions}.

\section{Flavorful Two Higgs Doublet Models}
\label{F2HDM}

One of the simplest realizations of a viable alternative framework of mass generation are 2HDMs with one doublet coupling only to the third generation, and a second doublet coupling mainly to the first and second generation. Such a setup was proposed in~\cite{Altmannshofer:2015esa} (see also~\cite{Ghosh:2015gpa,Botella:2016krk,Das:1995df,Blechman:2010cs}). The masses of the SM fermions arise from two sources, the vacuum expectation values of two Higgs doublets $\phi$ and $\phi^\prime$. The relevant part of the 2HDM Lagrangian is
\begin{eqnarray}
 -{\cal L}_\text{2HDM} &\supset& \sum_{i,j} \left( \lambda^u_{ij} (\bar q_i u_j) \tilde\phi + \lambda^d_{ij} (\bar q_i d_j) \phi + \lambda^e_{ij} (\bar\ell_i e_j) \phi \right) ~ + \text{h.c.} \nonumber \\
 && + \sum_{i,j} \left( \lambda^{\prime u}_{ij} (\bar q_i u_j) \tilde \phi' + \lambda^{\prime d}_{ij} (\bar q_i d_j) \phi' + \lambda^{\prime e}_{ij} (\bar\ell_i e_j) \phi' \right)~ + \text{h.c.} ~, \label{eq:Lagrangian}
\end{eqnarray}
where $\tilde \phi^{(\prime)} = i \sigma_2 (\phi^{(\prime)})^*$.
The three generations of quark and lepton doublets are denoted by $q_i$, $\ell_i$, and $u_i$, $d_i$, $e_i$ are the up quark, down quark, and charged lepton singlets. The $\lambda$ and $\lambda^\prime$ matrices are the Yukawa couplings.\footnote{We do not consider neutrino masses and mixing in this work. Neutrino masses could for example originate from a standard see-saw mechanism (with heavy right-handed neutrinos far above the TeV scale). In such a case none of the observables considered in our study will be affected in any significant way.}

The above setup for the Higgs couplings violates the principle of natural flavor conservation. Both of the Higgs doublets couple to the leptons, the up-type quarks, and the down-type quarks, leading to FCNCs at tree level.
 
\subsection{Yukawa textures}

We are interested in Yukawa couplings beyond NFC that do not introduce an unacceptably large amount of flavor violation. 
This can be achieved by demanding that one set of the Yukawa couplings preserves a $U(2)^5$ flavor symmetry, acting on the first two generations. In this case, flavor transitions between the first and second generation are protected. Such transitions are absent at first order in flavor symmetry breaking and arise only at second order as an effective $(2 \to 3) \times (3 \to 1)$ transition. As we will discuss in section~\ref{MesonPheno}, effects in neutral Kaon and $D$ meson oscillations are indeed typically well below present constraints.

We consider the following set of Yukawa matrices in the flavor basis\footnote{In this work we discuss the phenomenological implications of this specific set of Yukawa matrices. There are certainly other Yukawa textures that preserve an approximate $U(2)^5$ flavor symmetry and that can reproduce the observed fermion masses and mixings. Such textures might lead to a different phenomenology.}
\begin{subequations}
\begin{eqnarray} &&
\lambda_{u_{1,2}} \sim \frac{\sqrt{2}}{v_{u_{1,2}}}
\begin{pmatrix}
~m_u~& ~m_u ~& ~m_u~ \\
~m_u ~& ~m_c ~& ~m_c~ \\
~m_u~&~ m_c~ & ~m_c~
\end{pmatrix}
\label{YukawaTextures}~,~~~~~~~
 \lambda_{u_{3}} \sim \frac{\sqrt{2}}{v_{u_3}}
\begin{pmatrix}
~0~& ~0 ~& ~0~ \\
~0 ~& ~0 ~& ~0~ \\
~0~&~ 0~ & ~m_t~  
\end{pmatrix} ~, \\ &&
 \lambda_{d_{1,2}} \sim \frac{\sqrt{2}}{v_{d_{1,2}}}
\begin{pmatrix}
~m_d~& ~\lambda m_s ~& ~\lambda^3 m_b~ \\
~m_d ~& ~m_s ~& ~\lambda^2 m_b~ \\
~m_d~&~ m_s~ & ~m_s~
\end{pmatrix}\label{DownYukawaTextures} ~,~~~
\lambda_{d_3} \sim \frac{\sqrt{2}}{v_{d_3}}
\begin{pmatrix}
~0~& ~0 ~& ~0~ \\
~0 ~& ~0 ~& ~0~ \\
~0~&~ 0~ & ~m_b~
\end{pmatrix} ~, \\ &&
\label{EndYukawaTextures}
 \lambda_{\ell_{1,2}} \sim \frac{\sqrt{2}}{v_{\ell_{1,2}}}
\begin{pmatrix}
~m_e~& ~m_e ~& ~m_e~ \\
~m_e ~& ~m_\mu ~& ~m_\mu~ \\
~m_e~&~ m_\mu~ & ~m_\mu~
\end{pmatrix} ~,~~~~~~~~
 \lambda_{\ell_3} \sim \frac{\sqrt{2}}{v_{\ell_3}}
\begin{pmatrix}
~0~& ~0 ~& ~0~ \\
~0 ~& ~0 ~& ~0~ \\
~0~&~ 0~ & ~m_\tau~
\end{pmatrix} ~. 
\end{eqnarray}
\end{subequations}
Due to the rank 1 nature of the $\lambda_{u_3}$, $\lambda_{d_3}$, $\lambda_{e_3}$ Yukawa couplings, the $U(2)^5$ flavor symmetry acting on the first two generations is only broken by the small $\lambda_{u_{1,2}}$, $\lambda_{d_{1,2}}$, $\lambda_{e_{1,2}}$ Yukawa couplings.
Such a pattern of textures can be obtained using for example the flavor locking mechanism~\cite{Knapen:2015hia,Altmannshofer:2017uvs}.
Note that the above Yukawa couplings contain additional structure that is not dictated by the approximate $U(2)^5$ flavor symmetry. Our choice is motivated on the one hand by simplicity (the Yukawa matrices do not contain any unnecessary hierarchies) and on the other hand by robustness: The entries in the Yukawa couplings of the first and second generations are chosen such that the mass eigenvalues reproduce the observed values without any tuning. The structure in the down sector leads naturally to the observed pattern in the Cabibbo-Kobayashi-Maskawa (CKM) matrix elements. Alternatively, the CKM matrix could also be generated in the up sector, but we will not consider this option here as it requires additional hierarchies in the up Yukawa coupling.
The entries in the above matrices are given up to $O(1)$ factors that, in all generality, can be complex.

\begin{table}
\begin{center}
\begin{tabular}{l cccccc}
\hline\hline
Model & ~~~~~$u_{1,2}$~~~~~ & ~~~~~$u_{3}$~~~~~ & ~~~~~$d_{1,2}$~~~~~ & ~~~~~$d_3$~~~~~ & ~~~~~$e_{1,2}$~~~~~ &  ~~~~~$e^{3}_R$~~~~~ \\ 
\hline
Type 1A & $\Phi$& $\Phi$&$\Phi$& $\Phi$& $\Phi$&$\Phi$ \\
Type 1B & $\Phi'$& $\Phi$&$\Phi'$& $\Phi$& $\Phi'$&$\Phi$ \\[8pt]
Type 2A& $\Phi$& $\Phi$& $\Phi'$& $\Phi'$& $\Phi'$& $\Phi'$\\
Type 2B& $\Phi'$& $\Phi$& $\Phi$& $\Phi'$& $\Phi$& $\Phi'$\\[8pt]
Flipped A& $\Phi$& $\Phi$&$\Phi'$& $\Phi'$& $\Phi$&$\Phi$ \\
Flipped B& $\Phi'$& $\Phi$&$\Phi$& $\Phi'$& $\Phi'$&$\Phi$ \\[8pt]
Lepton-Specific A &$ \Phi$& $\Phi$& $\Phi$ &$ \Phi$& $\Phi'$& $\Phi'$\\
Lepton-Specific B &$ \Phi'$& $\Phi$& $\Phi'$ &$ \Phi$& $\Phi$& $\Phi'$ \\
\hline\hline
\end{tabular}
\end{center}
\caption{Summary of the way in which the SM quarks and leptons couple to the two Higgs doublets $\phi$ and $\phi^\prime$  in each of the considered models. In the models with natural flavor conservation (A), all generations of each type of fermion couple to the same Higgs doublet. In the flavorful models (B), the third generation and the first two generation couple to different Higgs doublets.}
\label{PhiCouplings}
\end{table}

The vacuum expectation values $v_i$ in Eqs.~(\ref{YukawaTextures})~-~(\ref{EndYukawaTextures}) correspond to either $v$ or $v^\prime$, depending on the model under consideration.
The Yukawa couplings for the third or first two generations are identified with the $\lambda$ and $\lambda^\prime$ couplings introduced in Eq.~(\ref{eq:Lagrangian}), accordingly.
Without loss of generality, we denote the Higgs doublet that couples to the top quark with $\phi$~\cite{Davidson:2005cw}, i.e. $v_{u_3} = v$, $v_{u_{1,2}} = v^\prime$ and $\lambda_{u_3} = \lambda^u$, $\lambda_{u_{1,2}} = \lambda^{\prime\,u}$.
This leaves us with four distinct ``flavorful'' possibilities to assign the two Higgs doublets to the down-quarks and leptons.
In analogy to the four well known 2HDMs with natural flavor conservation (that we refer to as type~1A, type~2A, lepton-specific~A, and flipped~A, in the following) we denote our four flavorful models as {\it type~1B}, {\it type~2B}, {\it lepton-specific~B}, and {\it flipped~B}. The type~1B model was studied in some detail in~\cite{Altmannshofer:2015esa,Altmannshofer:2016zrn,Altmannshofer:2017uvs}. The coupling structure of all four flavorful models is summarized in table~\ref{PhiCouplings}.

Rotating the fermions into mass eigenstates, we define the following mass parameters
\begin{equation}
 m^u_{q q^\prime} = \frac{v}{\sqrt{2}} \langle q_L | \lambda^u | q^\prime_R \rangle ~,~~ m^{\prime \,u}_{q q^\prime} = \frac{v^\prime}{\sqrt{2}}  \langle q_L | \lambda^{\prime \,u} | q^\prime_R \rangle ~,
\end{equation}
with quark mass eigenstates $q,q^\prime = u,c,t$. These mass parameters obey $m^u_{q q^\prime} + m^{\prime \,u}_{q q^\prime} = m_q \delta_{q q^\prime}$, where $m_q$ are the observed up-type quark masses. Analogous definitions and identities hold for the down-type quarks and the charged leptons. We derive expressions for the $m^\prime$ mass parameters in the mass eigenstate basis that automatically reproduce the observed fermion masses and CKM matrix elements.
We find the following values for the up mass parameters in all four types of flavorful models
\begin{subequations}
\label{upMassParametersFull}
\begin{eqnarray}
\label{upMassParameters}
 m_{uu}^\prime &=& m_u + O(1)\times\frac{m_u^2}{m_t} ~,~~~~~~~~~~~~~~~ m_{cc}^\prime = m_c + O(1)\times\frac{m_c^2}{m_t} ~,~~ m_{tt}^\prime = O(1) \times m_c ~, \\ \label{upMassParametersb}
 m_{uc}^\prime &=& \frac{m_{ut}^\prime m_{tc}^\prime}{m_t} \left(1 + O(1) \times \frac{m_c}{m_t} \right) ~,~~ m_{ut}^\prime = O(1) \times m_u ~,~~~~~~~~~~ m_{ct}^\prime = O(1) \times m_c ~,  \\ \label{upMassParametersc}
 m_{cu}^\prime &=& \frac{m_{ct}^\prime m_{tu}^\prime}{m_t} \left(1 + O(1) \times \frac{m_c}{m_t} \right) ~,~~ m_{tu}^\prime = O(1) \times m_u ~,~~~~~~~~~~ m_{tc}^\prime = O(1) \times m_c ~.
\end{eqnarray}
\end{subequations}
For leptons we find analogous expressions for the off-diagonal mass parameters in all four types
\begin{subequations}
\label{leptonMassParametersFull}
\begin{eqnarray}
\label{leptonMassParameters1}
 m_{e\mu}^\prime &=& \frac{m_{e\tau}^\prime m_{\tau\mu}^\prime}{m_\tau} \left(1 + O(1) \times \frac{m_\mu}{m_\tau} \right) ~,~~ m_{e\tau}^\prime = O(1) \times m_e ~,~~ m_{\mu\tau}^\prime = O(1) \times m_\mu ~,  \\
 m_{\mu e}^\prime &=& \frac{m_{\mu\tau}^\prime m_{\tau e}^\prime}{m_\tau} \left(1 + O(1) \times \frac{m_\mu}{m_\tau} \right) ~,~~ m_{\tau e}^\prime = O(1) \times m_e ~,~~ m_{\tau\mu}^\prime = O(1) \times m_\mu ~.
\end{eqnarray}
However, the diagonal mass terms depend on the type of flavorful model
\begin{eqnarray}
\label{leptonMassParameters2}
 m_{ee}^\prime &=& \begin{cases} m_e + O(1)\times\frac{m_e^2}{m_\tau} ~~~~~~~~\text{type~1B~,~ flipped~B}\\ ~~~~~~~~ O(1)\times\frac{m_e^2}{m_\tau} ~~~~~~~~\text{type~2B~,~ lepton-specific~B} \end{cases}~, \\
 m_{\mu\mu}^\prime &=& \begin{cases} m_\mu + O(1)\times\frac{m_\mu^2}{m_\tau} ~~~~~~~~\text{type~1B~,~ flipped~B}\\ ~~~~~~~~ O(1)\times\frac{m_\mu^2}{m_\tau} ~~~~~~~~\text{type~2B~,~ lepton-specific~B} \end{cases}~, \\
 m_{\tau\tau}^\prime &=& \begin{cases} ~~~~~~~~ O(1)\times m_\mu ~~~~~~~~\text{type~1B~,~ flipped~B}\\ m_\tau + O(1)\times m_\mu ~~~~~~~~\text{type~2B~,~ lepton-specific~B} \end{cases} ~.
\end{eqnarray}
\end{subequations}
Finally, for the down quarks we find for all four types
\begin{subequations}
\label{downMassParametersFull}
\begin{eqnarray}
\label{downMassParameters1}
 m_{bs}^\prime &=& O(1)\times m_s  ~,~~ m_{ds}^\prime = m_{bs}^\prime V_{td}^* \left( 1 + O(1)\times \frac{m_s}{m_b}\right) ~, \\ \label{downMassParameters1b}
 m_{bd}^\prime &=& O(1)\times m_d ~,~~m_{sd}^\prime = m_{bd}^\prime V_{ts}^* \left( 1 + O(1)\times \frac{m_s}{m_b}\right) ~.
\end{eqnarray}
The diagonal entries and the remaining off-diagonal entries depend on the type of model
\begin{eqnarray}
\label{downMassParameters2}
 m_{dd}^\prime &=& \begin{cases} m_d - m_{bd}^\prime V_{td}^* \left( 1 + O(1)\times\frac{m_s}{m_b} \right) ~~~~~~~~\text{type~1B~,~ lepton-specific~B}\\ ~~~~~ - m_{bd}^\prime V_{td}^* \left( 1 + O(1)\times\frac{m_s}{m_b} \right) ~~~~~~~~\text{type~2B~,~ flipped~B} \end{cases}~, \\
 m_{ss}^\prime &=& \begin{cases} m_s - m_{bs}^\prime V_{ts}^* \left( 1 + O(1)\times\frac{m_s}{m_b} \right) ~~~~~~~~\text{type~1B~,~ lepton-specific~B}\\ ~~~~~ - m_{bs}^\prime V_{ts}^* \left( 1 + O(1)\times\frac{m_s}{m_b} \right) ~~~~~~~~\text{type~2B~,~ flipped~B} \end{cases}~, \\
 m_{bb}^\prime &=& \begin{cases} ~~~~~~~~ O(1)\times m_s ~~~~~~~~~~~~~~~~~~~~~~~~~~~~ \text{type~1B~,~ lepton-specific~B}\\ m_b + O(1)\times m_s ~~~~~~~~~~~~~~~~~~~~~~~~~~~~ \text{type~2B~,~ flipped~B} \end{cases} ~,\\
 m_{sb}^\prime &=& \begin{cases} - V_{ts}^* m_b \left( 1 + O(1)\times\frac{m_s}{m_b} \right) ~~~~~~~~~~~~~~~~\text{type~1B~,~ lepton-specific~B}\\ + V_{ts}^* m_b \left( 1 + O(1)\times\frac{m_s}{m_b} \right) ~~~~~~~~~~~~~~~~\text{type~2B~,~ flipped~B} \end{cases} ~,\label{downMassParameters2f} \\
 m_{db}^\prime &=& \begin{cases} - V_{td}^* m_b \left( 1 + O(1)\times\frac{m_s}{m_b} \right) ~~~~~~~~~~~~~~~~\text{type~1B~,~ lepton-specific~B}\\ + V_{td}^* m_b \left( 1 + O(1)\times\frac{m_s}{m_b} \right) ~~~~~~~~~~~~~~~~\text{type~2B~,~ flipped~B} \end{cases} ~. \label{downMassParameters2g}
 \end{eqnarray}
\end{subequations}
As we assume that the CKM matrix is generated in the down sector, the CKM elements $V_{ts}$ and $V_{td}$ appear in several of the down-type mass parameters. 

The $O(1)$ terms in the above expressions are free parameters that in general can be complex.
It is worth noting that due to those $O(1)$ terms, the off-diagonal mass parameters $m_{ff'}$ and $m_{f'f}$ need not be the same for any type of fermion.
It is also important to note that in all cases the mass parameters that are responsible for flavor mixing between the first and second generation are suppressed by small mass ratios and not independent from the mass entries that parameterize mixing with the third generation. All $(2\to 1)$ mixing is given by an effective $(2\to3)\times(3\to1)$ mixing. This is a consequence of the breaking of the $U(2)^5$ symmetry by only one set of Yukawa couplings.

\subsection{Couplings of the Higgs bosons}

\begin{table}
\begin{center}
\begin{tabular}{l ccccccc}
\hline\hline
Model & ~~~~$\kappa_V^h$~~~~ & ~~~~$\kappa_{u_{3}}^h$~~~~ & ~~~~$\kappa_{u_{1,2}}^h$~~~~ & ~~~~$\kappa_{d_{3}}^h$~~~~ &  ~~~~$\kappa_{d_{1,2}}^h$~~~~ & ~~~~$\kappa_{\ell_{3}}^h$~~~~ & ~~~~$\kappa_{\ell_{1,2}}^h$~~~~ \\ 
\hline
Type 1A& $s_{\beta-\alpha}$ & $c_\alpha/s_\beta$ & $c_\alpha/s_\beta$ & $c_\alpha/s_\beta$ & $c_\alpha/s_\beta$ & $c_\alpha/s_\beta$ & $c_\alpha/s_\beta$   \\
Type 1B& $s_{\beta-\alpha}$ & $c_\alpha/s_\beta$&$-s_\alpha/c_\beta$ &$c_\alpha/s_\beta$&$-s_\alpha/c_\beta$&$c_\alpha/s_\beta$&$-s_\alpha/c_\beta$   \\[8pt]
Type 2A& $s_{\beta-\alpha}$  &$c_\alpha/s_\beta$&$c_\alpha/s_\beta$ &$-s_\alpha/c_\beta$ &$-s_\alpha/c_\beta$ & $-s_\alpha/c_\beta$& $-s_\alpha/c_\beta$\\
Type 2B& $s_{\beta-\alpha}$  & $c_\alpha/s_\beta$& $-s_\alpha/c_\beta$& $-s_\alpha/c_\beta$& $c_\alpha/s_\beta$& $-s_\alpha/c_\beta$& $c_\alpha/s_\beta$\\[8pt]
Flipped A & $s_{\beta-\alpha}$  &$c_\alpha/s_\beta$  &$c_\alpha/s_\beta$ &$-s_\alpha/c_\beta$ &$-s_\alpha/c_\beta$ & $c_\alpha/s_\beta$ & $c_\alpha/s_\beta$ \\
Flipped B & $s_{\beta-\alpha}$  & $c_\alpha/s_\beta$& $-s_\alpha/c_\beta$& $-s_\alpha/c_\beta$& $c_\alpha/s_\beta$& $c_\alpha/s_\beta$& $-s_\alpha/c_\beta$  \\ [8pt]
Lepton-Specific A& $s_{\beta-\alpha}$  & $c_\alpha/s_\beta$ &$c_\alpha/s_\beta$ &$c_\alpha/s_\beta$  &$c_\alpha/s_\beta$  & $-s_\alpha/c_\beta$& $-s_\alpha/c_\beta$ \\
Lepton-Specific B& $s_{\beta-\alpha}$  &$c_\alpha/s_\beta$ &$-s_\alpha/c_\beta$ &$c_\alpha/s_\beta$ &$-s_\alpha/c_\beta$ &$-s_\alpha/c_\beta$ & $c_\alpha/s_\beta$\\
\hline\hline
\end{tabular}
\end{center}
\caption{The leading order flavor diagonal coupling modifiers of the $125$~GeV Higgs $h$. }
\label{kappaExpressionsSM}
\end{table}
\begin{table}
\begin{center}
\begin{tabular}{l ccccccc}
\hline\hline
Model & ~~~~$\kappa_V^H$~~~~ & ~~~~$\kappa_{u_{3}}^H$~~~~ & ~~~~$\kappa_{u_{1,2}}^H$~~~~ & ~~~~$\kappa_{d_{3}}^H$~~~~ &  ~~~~$\kappa_{d_{1,2}}^H$~~~~ & ~~~~$\kappa_{\ell_{3}}^H$~~~~ & ~~~~$\kappa_{\ell_{1,2}}^H$~~~~ \\ 
\hline
Type 1A & $c_{\beta-\alpha}$ &$s_\alpha/s_\beta$ & $s_\alpha/s_\beta$ &$s_\alpha/s_\beta$  & $s_\alpha/s_\beta$ & $s_\alpha/s_\beta$ & $s_\alpha/s_\beta$  \\
Type 1B & $c_{\beta-\alpha}$ & $s_\alpha/s_\beta$&$c_\alpha/c_\beta$ &$s_\alpha/s_\beta$&$c_\alpha/c_\beta$&$s_\alpha/s_\beta$&$c_\alpha/c_\beta$   \\[8pt]
Type 2A & $c_{\beta-\alpha}$ &$s_\alpha/s_\beta$&$s_\alpha/s_\beta$ &$c_\alpha/c_\beta$ &$c_\alpha/c_\beta$ & $c_\alpha/c_\beta$& $c_\alpha/c_\beta$\\
Type 2B & $c_{\beta-\alpha}$ & $s_\alpha/s_\beta$& $c_\alpha/c_\beta$& $c_\alpha/c_\beta$& $s_\alpha/s_\beta$& $c_\alpha/c_\beta$& $s_\alpha/s_\beta$\\[8pt]
Flipped A & $c_{\beta-\alpha}$ &$s_\alpha/s_\beta$  &$s_\alpha/s_\beta$ &$c_\alpha/c_\beta$ &$c_\alpha/c_\beta$ & $s_\alpha/s_\beta$ & $s_\alpha/s_\beta$ \\
Flipped B & $c_{\beta-\alpha}$ & $s_\alpha/s_\beta$& $c_\alpha/c_\beta$& $c_\alpha/c_\beta$& $s_\alpha/s_\beta$& $s_\alpha/s_\beta$& $c_\alpha/c_\beta$ \\[8pt]
Lepton-Specific A & $c_{\beta-\alpha}$ & $s_\alpha/s_\beta$ &$s_\alpha/s_\beta$ &$s_\alpha/s_\beta$  &$s_\alpha/s_\beta$  & $c_\alpha/c_\beta$& $c_\alpha/c_\beta$ \\
Lepton-Specific B & $c_{\beta-\alpha}$ &$s_\alpha/s_\beta$ &$c_\alpha/c_\beta$ &$s_\alpha/s_\beta$ &$c_\alpha/c_\beta$ &$c_\alpha/c_\beta$ & $s_\alpha/s_\beta$\\
\hline\hline
\end{tabular}
\end{center}
\caption{The leading order flavor diagonal coupling modifiers of the heavy scalar Higgs $H$.}
\label{kappaExpressionsHeavy}
\end{table}
\begin{table}
\begin{center}
\begin{tabular}{l cccccc}
\hline\hline
Model & ~$\kappa_{u_{3}}^A$, $\kappa_{d_{i}u_{3}}^\pm$~ & ~$\kappa_{u_{1,2}}^A$, $\kappa_{d_i u_{1,2} }^\pm$~ & ~$\kappa_{d_{3}}^A$, $\kappa_{u_{i} d_{3}}^\pm$~ & ~$\kappa_{d_{1,2}}^A$, $\kappa_{u_i d_{1,2}}^\pm$~ & ~$\kappa_{\ell_{3}}^A$, $\kappa_{\nu_{3}\ell_{3}}^\pm$~ & ~$\kappa_{\ell_{1,2}}^A$, $\kappa_{\nu_{1,2}\ell_{1,2}}^\pm$~ \\ 
\hline
Type 1A&$-1/t_\beta$ & $-1/t_\beta$ &$-1/t_\beta$  & $-1/t_\beta$ & $-1/t_{\beta}$ & $-1/t_{\beta}$  \\
Type 1B& $-1/t_\beta$&$t_{\beta}$ &$-1/t_\beta$&$-1/t_\beta$&$-1/t_{\beta}$&$t_{\beta}$   \\[8pt]
Type 2A&$-1/t_\beta$&$-1/t_\beta$ &$t_{\beta}$ &$t_{\beta}$ & $t_{\beta}$& $t_{\beta}$\\
Type 2B& $-1/t_\beta$& $t_{\beta}$& $t_{\beta}$& $-1/t_\beta$& $t_{\beta}$& $-1/t_{\beta}$\\[8pt]
Flipped A&$-1/t_\beta$  &$-1/t_\beta$ &$t_{\beta}$ &$t_{\beta}$ & $-1/t_{\beta}$ & $-1/t_{\beta}$ \\
Flipped B& $-1/t_\beta$& $t_{\beta}$& $t_{\beta}$& $-1/t_\beta$& $-1/t_{\beta}$& $t_{\beta}$ \\[8pt]
Lepton-Specific A& $-1/t_\beta$ &$-1/t_\beta$ &$-1/t_\beta$  &$-1/t_\beta$  & $t_{\beta}$& $t_{\beta}$ \\
Lepton-Specific B&$-1/t_\beta$ &$t_{\beta}$ &$-1/t_\beta$ &$t_\beta$ &$t_{\beta}$ & $-1/t_{\beta}$\\
\hline\hline
\end{tabular}
\end{center}
\caption{The leading order flavor diagonal coupling modifiers of the psuedoscalar Higgs $A$ and charged Higgs $H^\pm$.}
\label{kappaExpressionsCharged}
\end{table}

Next, we discuss the couplings of the physical Higgs bosons in the four different models.
We largely follow the notation and conventions in~\cite{Altmannshofer:2016zrn} and state only the relevant results.

The part of the Lagrangian that parametrizes the couplings to the three neutral scalars, $h$, $H$, and $A$ (we identify $h$ with the 125~GeV Higgs), as well as the charged Higgs $H^\pm$ to mass eigenstate fermions is written as
\begin{eqnarray}
  \mathcal{L} \subset 
  &-& \sum_{f=d,\ell} \sum_{i,j} (\bar{f}_i P_R f_j) \Big(h(Y_h^f)_{ij}+H(Y^f_H)_{ij}-iA(Y_A^f)_{ij} \Big) + \text{h.c.}  \nonumber \\
  &-& \sum_{i,j} (\bar{u}_i P_R u_j) \Big(h(Y_h^u)_{ij}+H(Y^u_H)_{ij}+iA(Y_A^u)_{ij} \Big) + \text{h.c.} \\
  &-& \sqrt{2} \sum_{i,j} \Big( (\bar{d_i}P_R u_j)H^-(Y_{\pm}^u)_{ij}-(\bar{u_i}P_R d_j)H^+(Y^d_{\pm})_{ij}-(\bar{\nu_i}P_R \ell_j)H^+(Y^\ell_{\pm})_{ij} \Big) + \text{h.c.} ~. \nonumber
\end{eqnarray}
For the flavor diagonal and off-diagonal couplings of the neutral Higgs bosons to leptons one finds 
\begin{eqnarray}  \label{DiagonalCouplings}
 \kappa_{\ell_i \ell_j}^h \frac{m_{\ell_j}}{v_W} \equiv (Y_h^\ell)_{ij} = \frac{m_{\ell_j}}{v_W}\Bigg ( \frac{c_\alpha}{s_\beta} \delta_{ij} - \frac{m^\prime_{{\ell_i}{\ell_j}}}{m_{\ell_j}}\frac{c_{\beta-\alpha}}{s_\beta c_\beta}\Bigg) ~,  \\ \label{DiagonalCouplingsH}
 \kappa_{\ell_i \ell_j}^H \frac{m_{\ell_j}}{v_W} \equiv (Y_H^\ell)_{ij} = \frac{m_{\ell_j}}{v_W}\Bigg ( \frac{s_\alpha}{s_\beta} \delta_{ij} +\frac{m^\prime_{{\ell_i}{\ell_j}}}{m_{\ell_j}}\frac{s_{\beta-\alpha}}{s_\beta c_\beta}\Bigg) ~, \\ \label{DiagonalCouplingsA}
 \kappa_{\ell_i \ell_j}^A \frac{m_{\ell_j}}{v_W} \equiv (Y_A^\ell)_{ij} = \frac{m_{\ell_j}}{v_W}\Bigg ( -\frac{1}{t_\beta} \delta_{ij} + \frac{m^\prime_{{\ell_i}{\ell_j}}}{m_{\ell_j}}\frac{1}{s_\beta c_\beta}\Bigg) ~,
\end{eqnarray}
where we introduced the coupling modifiers $\kappa$ with respect to the SM Higgs couplings.
We use the notation $c_\phi = \cos\phi$, $s_\phi = \sin\phi$, and $t_\phi = \tan\phi$. The angle $\alpha$ parametrizes the mixing between the neutral CP-even components of the two Higgs doublets and $\tan\beta = v / v^\prime$ is the ratio of Higgs vacuum expectation values. 
Completely analogous expressions hold for the neutral Higgs couplings to the up-type and down-type quarks.

Ignoring neutrino mixing (which is of no relevance for our study) one finds for the charged Higgs couplings to leptons
\begin{eqnarray}
 \kappa_{\nu_i \ell_j}^\pm \frac{m_{\ell_j}}{v_W} \equiv (Y_\pm^\ell)_{ij} =  \frac{m_{\ell_j}}{v_W} \Bigg(-\frac{1}{t_\beta} \delta_{ij} + \frac{m^\prime_{\ell_i \ell_j}}{m_{\ell_j}}\frac{1}{s_\beta c_\beta}\Bigg) ~. 
\end{eqnarray}
In the expressions for the charged Higgs couplings to quarks, the CKM matrix $V$ enters. We find
\begin{eqnarray} \label{eq:Hpm_couplings}
 \kappa_{d_i u_j}^\pm \frac{m_{u_j}}{v_W} V_{u_j d_i}^* \equiv (Y_\pm^u)_{ij} = \frac{m_{u_j}}{v_W} V_{u_j d_i}^* \Bigg(-\frac{1}{t_\beta} + \sum_k \frac{m^\prime_{u_k u_j}}{m_{u_j}} \frac{V_{u_k d_i}^*}{V_{u_j d_i}^*} \frac{1}{s_\beta c_\beta}  \Bigg) ~, \\
 \kappa_{u_i d_j}^\pm \frac{m_{d_j}}{v_W} V_{u_i d_j} \equiv (Y_\pm^d)_{ij} = \frac{m_{d_j}}{v_W} V_{u_i d_j} \Bigg(-\frac{1}{t_\beta} + \sum_k \frac{m^\prime_{d_k d_j}}{m_{d_j}} \frac{V_{u_i d_k}}{V_{u_i d_j}} \frac{1}{s_\beta c_\beta}  \Bigg) ~.
 \end{eqnarray}
All of these expressions for the couplings are completely generic and can be applied to any of our flavorful models. The only terms that change in the different models are the $m^\prime$ mass parameters, as given in Eqs.~(\ref{upMassParametersFull}), (\ref{leptonMassParametersFull}), and (\ref{downMassParametersFull}). 

In tables~\ref{kappaExpressionsSM},~\ref{kappaExpressionsHeavy}, and~\ref{kappaExpressionsCharged}, we show the leading order coupling modifiers for the flavor diagonal couplings of the Higgs bosons $\kappa_i \equiv \kappa_{ii}$ as an expansion in $1/m_3$, where $m_3 = m_t, m_b, m_\tau$. We compare the coupling modifiers of all four flavorful 2HDM types to those of the four 2HDM types with natural flavor conservation.
As is well known, the coupling modifiers are flavor universal in the models with natural flavor conservation. In the flavorful models the modifiers are flavor dependent and differentiate between the third generation and the first two generations.

\section{Light Higgs phenomenology}
\label{LHP}

\subsection{Constraints from Higgs signal strength measurements}
\label{SignalStrengthSection}

The introduction of a second doublet alters the couplings to the 125~GeV Higgs boson $h$ as shown in table~\ref{kappaExpressionsSM} as well as Eq.~(\ref{DiagonalCouplings}). 
We can compare the Higgs production and decay rates predicted by our models to those measured by ATLAS and CMS in order to constrain the new physics parameter space.

To determine the constraints from the measured Higgs signals we construct a $\chi^2$ function
\begin{eqnarray} \label{eq:chi2}
 \chi^2 = \sum_{i,j} \left(\frac{(\sigma\times \text{BR})_i^\text{exp}}{(\sigma\times \text{BR})_i^\text{SM}} - \frac{(\sigma\times \text{BR})_i^\text{BSM}}{(\sigma\times \text{BR})_i^\text{SM}} \right) \left(\frac{(\sigma\times \text{BR})_j^\text{exp}}{(\sigma\times \text{BR})_j^\text{SM}} - \frac{(\sigma\times \text{BR})_j^\text{BSM}}{(\sigma\times \text{BR})_j^\text{SM}} \right) \big( \text{cov} \big)^{-1}_{ij} ~,
\end{eqnarray}
where $(\sigma\times \text{BR})_i^\text{exp}$, $(\sigma\times \text{BR})_i^\text{SM}$, and $(\sigma\times \text{BR})_i^\text{BSM}$ are the experimental measurements, the Standard Model predictions, and flavorful 2HDM predictions for the production cross sections times branching ratio of the various measured channels.

The ratios of experimental measurements and SM predictions that enter Eq.~(\ref{eq:chi2}) are given by the signal strength modifiers that are reported by ATLAS and CMS. 
The SM predictions for the production cross sections and branching ratios are taken from~\cite{deFlorian:2016spz}.
The ratios of BSM and SM predictions for individual channels can be obtained in a straight-forward way as functions of the coupling modifiers. For the gluon-gluon fusion production (ggf), vector boson fusion production (VBF), production in association with $W$ and $Z$ bosons (Wh, Zh), and production in association with top quarks (tth), we have
\begin{eqnarray}
 && \frac{\sigma_\text{ggf}^\text{BSM}}{\sigma_\text{ggf}^\text{SM}} \simeq 1.065 (\kappa_t^h)^2 + 0.002 (\kappa_b^h)^2 - 0.067 (\kappa_b^h)(\kappa_t^h) ~, \\
 && \frac{\sigma_\text{VBF}^\text{BSM}}{\sigma_\text{VBF}^\text{SM}} = \frac{\sigma_\text{Wh}^\text{BSM}}{\sigma_\text{Wh}^\text{SM}} = \frac{\sigma_\text{Zh}^\text{BSM}}{\sigma_\text{Zh}^\text{SM}} = (\kappa^h_V)^2 ~, \qquad
 \frac{\sigma_\text{tth}^\text{BSM}}{\sigma_\text{tth}^\text{SM}} = (\kappa_t^h)^2 ~,
\end{eqnarray}
where for the loop induced gluon-gluon fusion we take into account top and bottom contributions at 1-loop.
For tree level decays, the partial widths simply scale with the appropriate coupling modifiers. 
In the case of the loop induced $h \to gg$ decay width we take into account top and bottom contributions at 1-loop (We explicitly checked that loops with lighter quarks do not lead to any appreciable effects). For $h \to \gamma \gamma$ we consider $W$, top, and bottom loops. We neglect charged Higgs loops, that are typically tiny~\cite{Altmannshofer:2012ar}
\begin{eqnarray}
 && \frac{\Gamma_{WW^*}^\text{BSM}}{\Gamma_{WW^*}^\text{SM}} = \frac{\Gamma_{ZZ^*}^\text{BSM}}{\Gamma_{ZZ^*}^\text{SM}} = (\kappa_V^h)^2 ~, \qquad \frac{\Gamma_{f \bar f}^\text{BSM}}{\Gamma_{f \bar f}^\text{SM}} = (\kappa_f^h)^2 ~,~~ \text{for}~ f = b, \tau, c, s, \mu ~,  \\
 && \frac{\Gamma_{gg}^\text{BSM}}{\Gamma_{gg}^\text{SM}} \simeq 1.065 (\kappa_t^h)^2 + 0.002 (\kappa_b^h)^2 - 0.067 (\kappa_b^h)(\kappa_t^h) ~, \\
 && \frac{\Gamma_{\gamma\gamma}^\text{BSM}}{\Gamma_{\gamma\gamma}^\text{SM}} \simeq 1.640 (\kappa_V^h)^2 + 0.080 (\kappa_t^h)^2 - 0.725 (\kappa_V^h)(\kappa_t^h) + 0.006 (\kappa_V^h)(\kappa_b^h) - 0.001 (\kappa_b^h)(\kappa_t^h) ~.
\end{eqnarray}

The covariance matrix in Eq.~(\ref{eq:chi2}) contains the experimental uncertainties and (where available) the correlations among the uncertainties. We assume that theory uncertainties in the ratio of BSM and SM predictions are negligible compared to current experimental uncertainties. We take into account the Higgs signal strengths from the LHC run~1 combination~\cite{Khachatryan:2016vau}, as well as several individual run~2 results, in particular measurements of $h \to ZZ^*$~\cite{Sirunyan:2017exp, Aaboud:2017vzb}, $h \to WW^*$~\cite{CMS:2017pzi, ATLAS:2018gcr}, $h \to \gamma\gamma$~\cite{Aaboud:2018xdt, Sirunyan:2018ouh}, $h\to \tau^+\tau^-$~\cite{Sirunyan:2017khh}, $h \to b \bar b$~\cite{Sirunyan:2017elk, Aaboud:2017xsd}, and $h \to \mu^+\mu^-$~\cite{CMS:2017qgo, Aaboud:2017ojs}.
We also include results on Higgs production is association with top quarks~\cite{CMS:2017lgc, Aaboud:2017rss, Sirunyan:2018mvw}. (See~\cite{Chowdhury:2017aav,Haller:2018nnx} for recent Higgs signal strength studies of 2HDMs with natural flavor conservation.)

\begin{figure}[tb]
\begin{center} 
\includegraphics[width=0.47\textwidth]{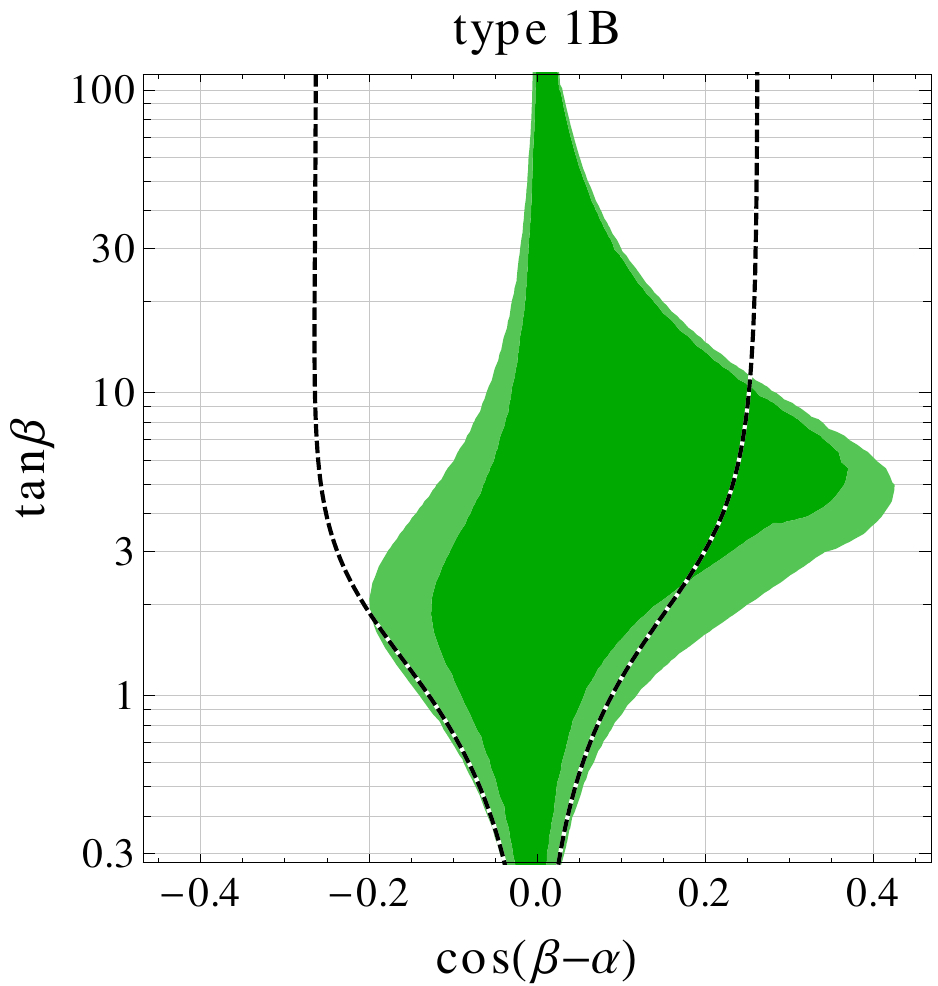} ~~~~~
\includegraphics[width=0.47\textwidth]{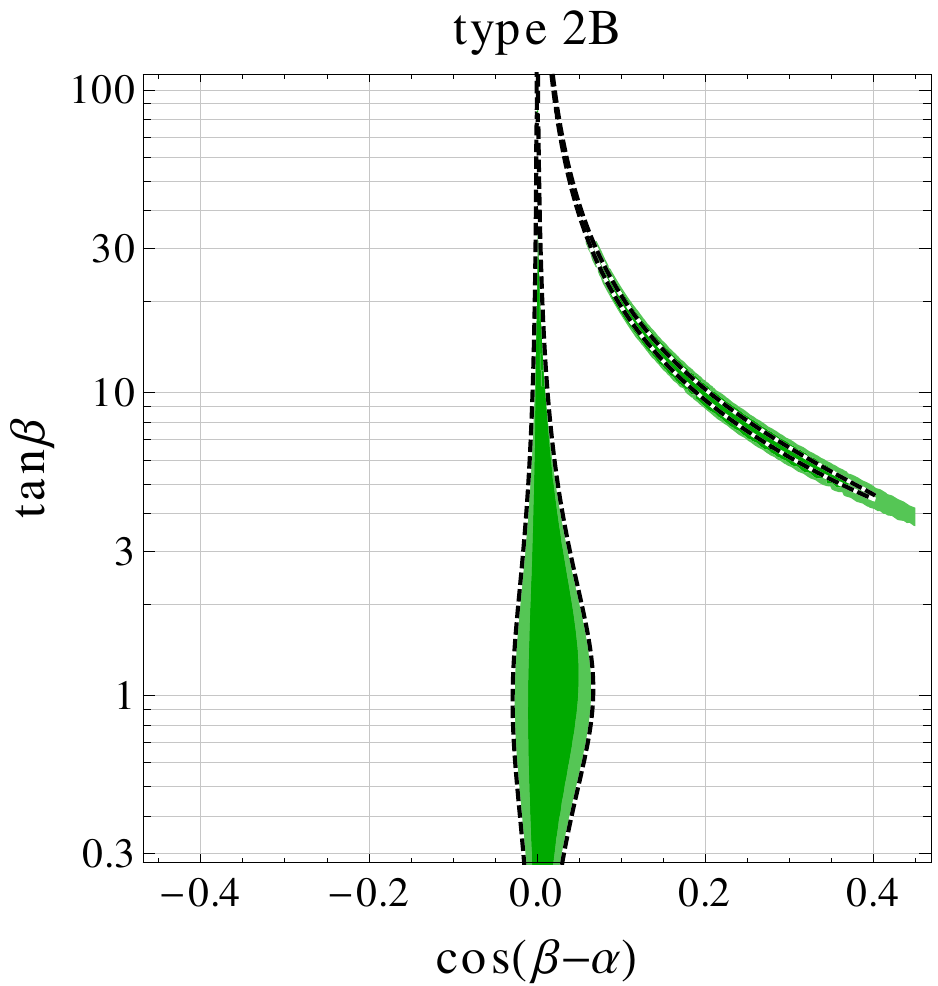} \\[16pt]
\includegraphics[width=0.47\textwidth]{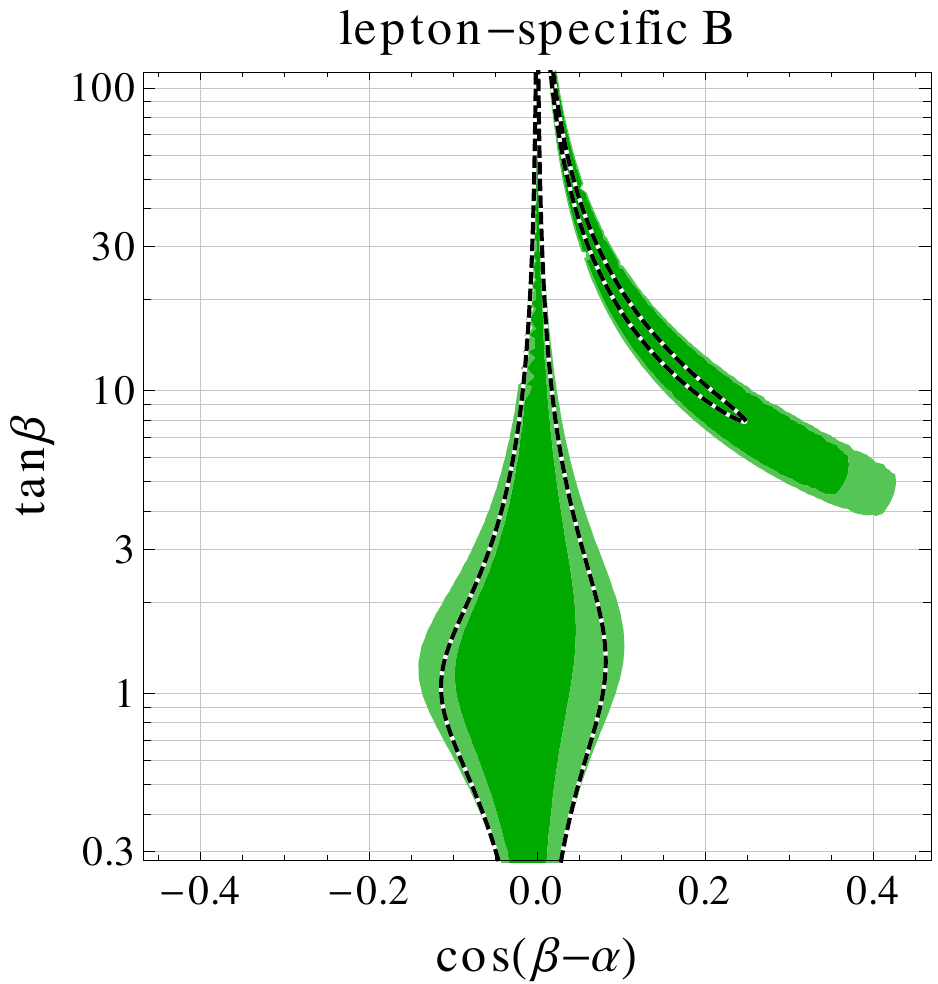} ~~~~~
\includegraphics[width=0.47\textwidth]{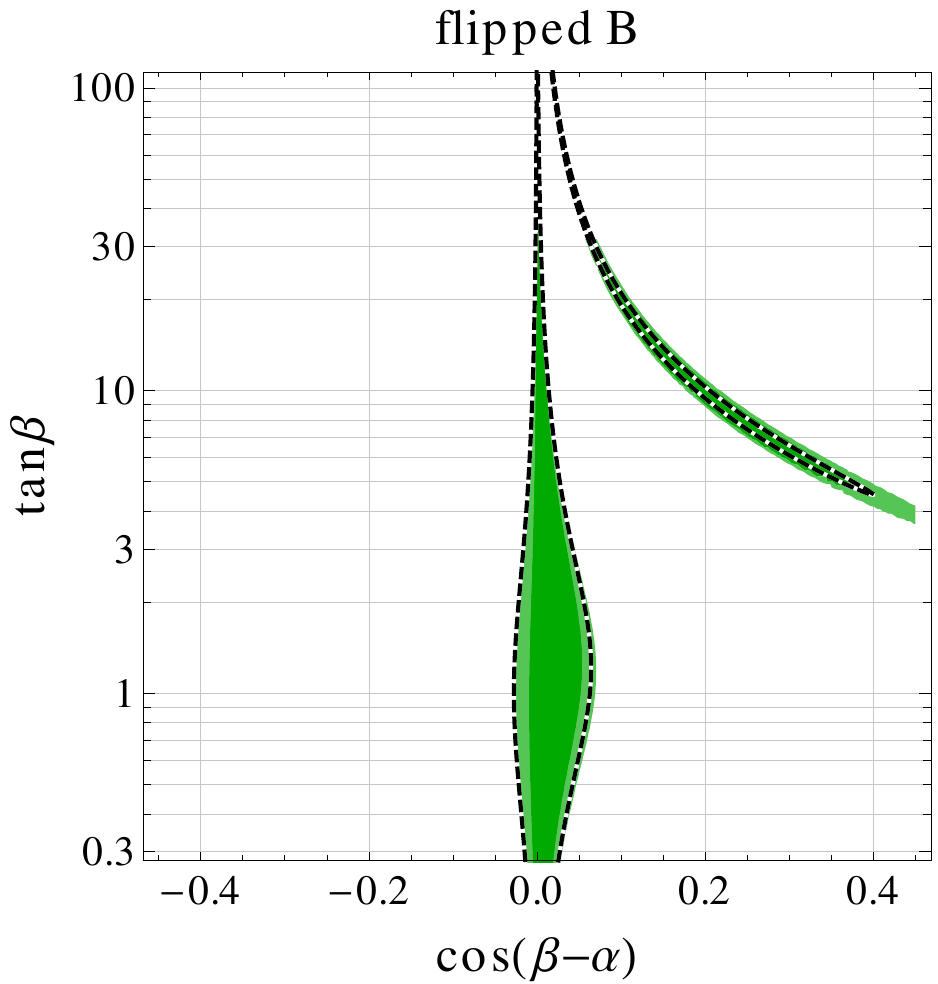}
\caption{Constraints on $\cos(\beta-\alpha)$ vs. $\tan\beta$ based on the results from the LHC measurements of the 125~GeV Higgs signal strengths. We show both 1$\sigma$ and 2$\sigma$ regions for the four flavorful models in green. We allow the mass parameters to vary up to a factor of 3 times their expected values. For comparison, the 2$\sigma$ regions in the corresponding models with natural flavor conservation are shown by the dashed contours.}
\label{SMHiggsAlphaBetaConstraints}
\end{center}
\end{figure}

The couplings of $h$ are largely determined by the parameters $\alpha$ and $\beta$. Subleading corrections enter through the $m^\prime$ mass parameters, see Eq.~(\ref{DiagonalCouplings}). We use the $\chi^2$ function to put constraints on  the $\alpha$ and $\beta$ parameters allowing the $O(1)$ coefficients in the subleading corrections to vary in the range $(-3,3)$. The allowed regions in the $\cos(\beta-\alpha)$ vs. $\tan\beta$ plane that we obtain in this way are shown in Fig.~\ref{SMHiggsAlphaBetaConstraints}. The dark (light) green regions correspond to the $1\sigma$ and $2\sigma$ allowed regions (that we define as $\Delta \chi^2 = \chi^2 - \chi^2_\text{SM} < 1,4$) in the four flavorful models. We also compare these regions to the 2$\sigma$ constraint in the corresponding models with natural flavor conservation (dashed contours). The plot for the type~1B model updates the corresponding result in~\cite{Altmannshofer:2016zrn}.

The couplings of the 125~GeV Higgs to the third generation fermions are already constrained by current data to be SM-like at the level of 10\% - 20\%. By coupling the $\tau$ and/or $b$ to the second doublet (as in the type~2B, lepton-specific~B, and flipped~B models), we therefore find the parameter space to be more strongly constrained than in the type~1B model.
Note that in those models there are two distinct regions of parameter space: one region close to the {\it alignment limit} $\cos(\beta-\alpha) \simeq 0$, where the mixing between the 125~GeV Higgs and heavy Higgs is tiny and all $h$ couplings become SM-like, and a second narrow strip where the bottom and/or the tau coupling have opposite sign with respect to the SM prediction. 
The constraints for the type~2B and flipped~B models are very similar, implying that the bottom coupling (which largely determines the total width of $h$) is the most important factor in determining the parameter space of these models.
Generally, as $\tan\beta$ gets very large or very small the $\kappa$ values can deviate substantially from 1, resulting in strong constraints. Moderate values for $\tan\beta$ are the least constrained.

Currently, the only decay of the Higgs into a non-third generation fermion which has been constrained in a relevant way at the LHC is the decay to $\mu^+\mu^-$~\cite{Aaboud:2017ojs,CMS:2017qgo}. However, the current sensitivities to the $h \to \mu^+\mu^-$ decay are not sufficient to impose strong constraints on our parameter space, yet. 
Future precise measurements of $h \to \mu^+\mu^-$ can potentially constrain large parts of the open parameter space of the type~1B model. The type~2B and flipped~B models will be mainly constrained by improved measurements of $h \to b \bar b$. For the lepton-specific~B model, future precision measurements of $h \to \tau^+\tau^-$ will give the most relevant constraints.  

\subsection{Flavor violating decays}

Along with altering the flavor diagonal couplings of the light Higgs, the introduction of the second doublet also introduces flavor violating couplings of $h$ to the fermions.
We expect in our models a number of FCNC decays that are extremely suppressed in the SM, most notably rare top decays $t \to ch$ and $t \to uh$ as well as lepton flavor violating Higgs decays $h\to \tau \mu$, $h \to \tau e$, and $h\to \mu e$.

In all four flavorful models the branching ratio of $t \to c h$ is given by
\begin{eqnarray} \label{eq:raretop}
 \text{BR}(t \to h c) &=& \frac{m_c^2 }{2m_t^2} \frac{c_{\beta - \alpha}^2}{s_\beta^2 c_\beta^2} \left( \frac{|m^\prime_{tc}|^2}{m_c^2} + \frac{|m^\prime_{ct}|^2}{m_c^2} \right) \frac{\left( 1-{m_h^2 / m_t^2} \right)^2}{ \left(1- {m_W^2/m_t^2}\right)^2 \left(1+{2 m_W^2 / m_t^2}\right) } \nonumber \\
 &\simeq& 1.7 \times 10^{-6} \times \frac{c_{\beta - \alpha}^2}{s_\beta^2 c_\beta^2} \left( \frac{|m^\prime_{tc}|^2}{m_c^2} + \frac{|m^\prime_{ct}|^2}{m_c^2} \right) ~.
\end{eqnarray}
From our study of Higgs signal strength measurements described in section~\ref{SignalStrengthSection} we find in all four flavorful models the constraint $\frac{c_{\beta-\alpha}}{s_\beta c_\beta} \lesssim 2.5$. Combined with the generic expectation $m^\prime_{tc} \sim m^\prime_{ct} \sim m_c$, this implies that BR$(t \to h c)$ is typically not larger than $\sim 10^{-5}$. While this is much larger than the SM prediction of $O(10^{-15})$~\cite{AguilarSaavedra:2004wm}, it is below the current and expected sensitivities at the LHC~\cite{Aaboud:2017mfd,Sirunyan:2017uae}. The decay $t \to h u$ is further suppressed by the up quark mass and generically not larger than $10^{-10}$, i.e. far below any foreseeable experimental sensitivity. Rare top decays could have much larger branching ratios if the CKM matrix is generated in the up-sector. 

The branching ratio for the rare Higgs decay $h \to \tau\mu = h \to \tau^+\mu^- + \tau^-\mu^+$ is given by
\begin{eqnarray}
 \text{BR}(h \to \tau\mu) &=& \frac{m_h}{8\pi \Gamma_h} \frac{m_\mu^2}{v_W^2} \frac{c_{\beta-\alpha}^2}{s_\beta^2 c_\beta^2} \left( \frac{|m_{\tau\mu}^\prime|^2}{m_\mu^2} + \frac{|m_{\mu\tau}^\prime|^2}{m_\mu^2} \right) \nonumber \\ &\simeq& 2.3 \times 10^{-4} \times \frac{c_{\beta-\alpha}^2}{s_\beta^2 c_\beta^2} \left( \frac{|m_{\tau\mu}^\prime|^2}{m_\mu^2} + \frac{|m_{\mu\tau}^\prime|^2}{m_\mu^2} \right) ~,
\end{eqnarray}
where $\Gamma_h \simeq 4$~MeV is the total Higgs width.
This expression holds in all four flavorful models, and we generically expect branching ratios up to $\sim 10^{-3}$.
This has to be compared to the current bounds on this branching ratio from CMS~\cite{Sirunyan:2017xzt} and ATLAS~\cite{Aad:2016blu}
\begin{equation}
 \text{BR}(h \to \tau\mu)_\text{CMS} < 2.5\times 10^{-3} ~,\qquad \text{BR}(h \to \tau\mu)_\text{ATLAS} < 1.43\times 10^{-2} ~.
\end{equation}
Future searches for $h \to \tau\mu$ will start to probe interesting new physics parameter space. 

In all our models, the branching ratio of $h \to \tau e$ is suppressed by a factor of $m_e^2/m_\mu^2 \sim 10^{-5}$ compared to $h \to \tau\mu$ and therefore outside the reach of foreseeable experiments. The branching ratio of $h \to \mu e$ is further suppressed and generically not larger than $10^{-10}$.

\section{Heavy neutral Higgs production and decays}
\label{HHP}

We expect a distinct collider phenomenology for the heavy Higgs bosons in each of our models.
In contrast to models with natural flavor conservation, flavor alignment, or minimal flavor violation~\cite{Glashow:1976nt,DAmbrosio:2002vsn,Pich:2009sp,Buras:2010mh,Altmannshofer:2012ar,Gori:2017qwg}, the coupling modifiers of the heavy Higgs bosons to fermions are not flavor universal. The difference is particularly striking for moderate and large $\tan\beta$.
As shown in table~\ref{kappaExpressionsHeavy}, for $\cos(\beta-\alpha) \simeq 0$ and $\tan\beta \gg 1$,  whenever the coupling to a third generation fermion is suppressed by a factor $\frac{\sin\alpha}{\sin\beta} \simeq \frac{1}{\tan\beta}$,
the couplings to the corresponding first and second generation fermions are enhanced by a factor $\frac{\cos\alpha}{\cos\beta} \simeq \tan\beta$, and vice versa.  
Depending on the type of flavorful model, a specific set of fermions can dominate the decay of the heavy Higgs bosons
and cause different types of production modes to be more or less relevant.
In the following we will focus on the type~2B, lepton-specific~B, and flipped~B models.
The collider phenomenology of the type~1B model has been discussed previously in~\cite{Altmannshofer:2016zrn}.

For the numerical results that will be presented in this section as well as in the subsequent charged Higgs section we will consider a fixed set of $m^\prime$ mass parameters.
To choose $m^\prime$ parameters in the up and lepton sectors, we start with the Yukawa textures from Eqs.~(\ref{YukawaTextures}) and~(\ref{EndYukawaTextures}) setting all free $O(1)$ parameters to $+1$. The precise values for $m_{u,c,t}$ and $m_{e,\mu,\tau}$ in Eqs.~(\ref{YukawaTextures}) and~(\ref{EndYukawaTextures}) are then fully determined by demanding that the mass eigenvalues reproduce the known fermion masses (we use $\overline{\text{MS}}$ masses at a scale of 500~GeV). In the down sector, the entries in Eq.~(\ref{DownYukawaTextures}) of $O(\lambda m_s)$, $O(\lambda^2 m_b)$, and $O(\lambda^3 m_b)$ are chosen to reproduce the CKM matrix. The $m_{d,s,b}$ parameters in Eq.~(\ref{DownYukawaTextures}) are determined by the known down quark masses, setting the remaining free $O(1)$ parameters to $+1$.
Generically, choosing different $O(1)$ parameters does not lead to a qualitative change of the heavy neutral and charged Higgs phenomenology. We will discuss the quantitative impact of varying the $O(1)$ parameters where appropriate.

\subsection{Production cross sections}

As we have seen in section~\ref{SignalStrengthSection}, the type~2B, lepton-specific~B, and flipped~B models are strongly constrained by Higgs signal strength measurements. In order to have maximal freedom in choosing a value for $\tan\beta$, we will limit our discussion to the decoupling limit and thus set $\cos(\beta -\alpha) = 0$\footnote{As shown in Fig.~\ref{SMHiggsAlphaBetaConstraints}, there are also tuned narrow strips of parameter space beyond the decoupling limit of the type~2B, lepton-specific~B, and flipped~B models that are allowed by Higgs signal strength data. In those regions of parameter space the bottom and/or the tau couplings of $h$ have opposite sign with respect to the SM prediction. A detailed study of the heavy Higgs phenomenology in these ``flipped sign scenarios'' is beyond the scope of this work. (See e.g.~\cite{Ferreira:2014naa,Modak:2016cdm,Ferreira:2017bnx,Coyle:2018ydo} for corresponding studies in other 2HDM scenarios.)}. In this limit the couplings of the heavy scalar and 
pseudoscalar Higgs to 
fermions are identical and, furthermore, their couplings to gauge bosons vanish. The main production modes of the heavy neutral Higgs bosons are therefore: gluon-gluon fusion, production in association with tops or bottoms, and direct production from a $q \bar q^\prime$ initial state. Vector boson fusion and production in association with gauge bosons is absent.

We compute the cross section of the $q\bar q^\prime \to H$ processes by convoluting the leading order parton level cross section with the appropriate MMHT~2014 quark parton distribution functions (PDF)~\cite{Harland-Lang:2014zoa}
\begin{equation}
\sigma(q_i \bar q_j \to H) =  \frac{\pi}{12 s}\Big(|(Y_H^q)_{ij}|^2 + |(Y_H^q)_{ji}|^2 \Big) \int_{\frac{m_{H}^2}{s}}^1 \frac{dx}{x} f_q(x) f_{\bar q^\prime}\left(\frac{m_{H}^2}{x s}\right).  
\label{HeavyHiggsCS}
\end{equation}
where $s$ is the center of mass energy of the protons. 
We take into account $c\bar c$, $b \bar b$, $b \bar s$, and $s \bar b$ initial states. Given the small couplings to the lighter quark generations, we find that the remaining possible quark combinations are always sub-dominant (despite the larger PDF's).
We do not include higher order corrections where one or two $b$ quarks appear in the final state, keeping in mind that such processes might modify our $b\bar b \to H$ and $bs \to H$ results by an $O(1)$ amount~\cite{Harlander:2003ai}. 
In the type~2B and flipped~B models, we expect that the $b\bar b \to H$ production is the most relevant for moderate and large $\tan\beta$, thanks to the enhanced couplings to bottom quarks. Production from initial state charm benefits from slightly larger PDF's but is suppressed by the significantly smaller charm mass.
In the lepton-specific~B model instead, we expect $c \bar c \to H$ to dominate for moderate and large $\tan\beta$, as the couplings to bottom are suppressed. 

We estimate the $gg \to H$ production cross section by scaling the corresponding cross section of a heavy Higgs with SM-like couplings from~\cite{deFlorian:2016spz} by the ratio of the leading order $H \to gg$ partial width in our model and a heavy Higgs with SM-like couplings. We take the expression for the partial width from~\cite{Giardino:2013bma}. 

 \begin{figure}[tb]
 \begin{center}
 \includegraphics[width=0.48\textwidth]{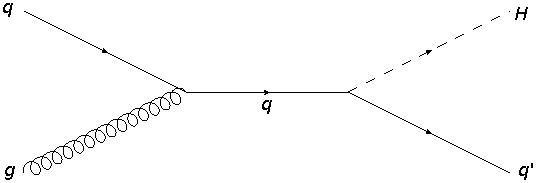} ~~~~~~
 \includegraphics[width=0.32\textwidth]{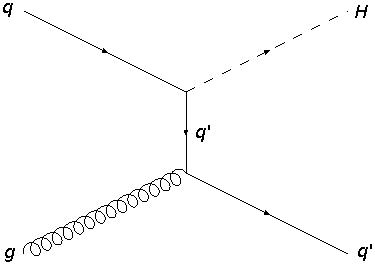} \\[16pt]
 \caption{Feynman diagrams for the production of a Higgs boson in association with a quark.}
 \label{tH1Feyn}
 \end{center}
 \end{figure}

Top associated production arises from diagrams like those shown in Fig.~\ref{tH1Feyn}. The corresponding cross section is identical for all four flavorful models. We use the cross section from~\cite{Altmannshofer:2016zrn}, which was obtained by summing over the initial state quarks $u$ and $c$ and convoluting the parton cross section with the appropriate PDF's.

 \begin{figure}[tb]
 \begin{center}
 \includegraphics[width=0.36\textwidth]{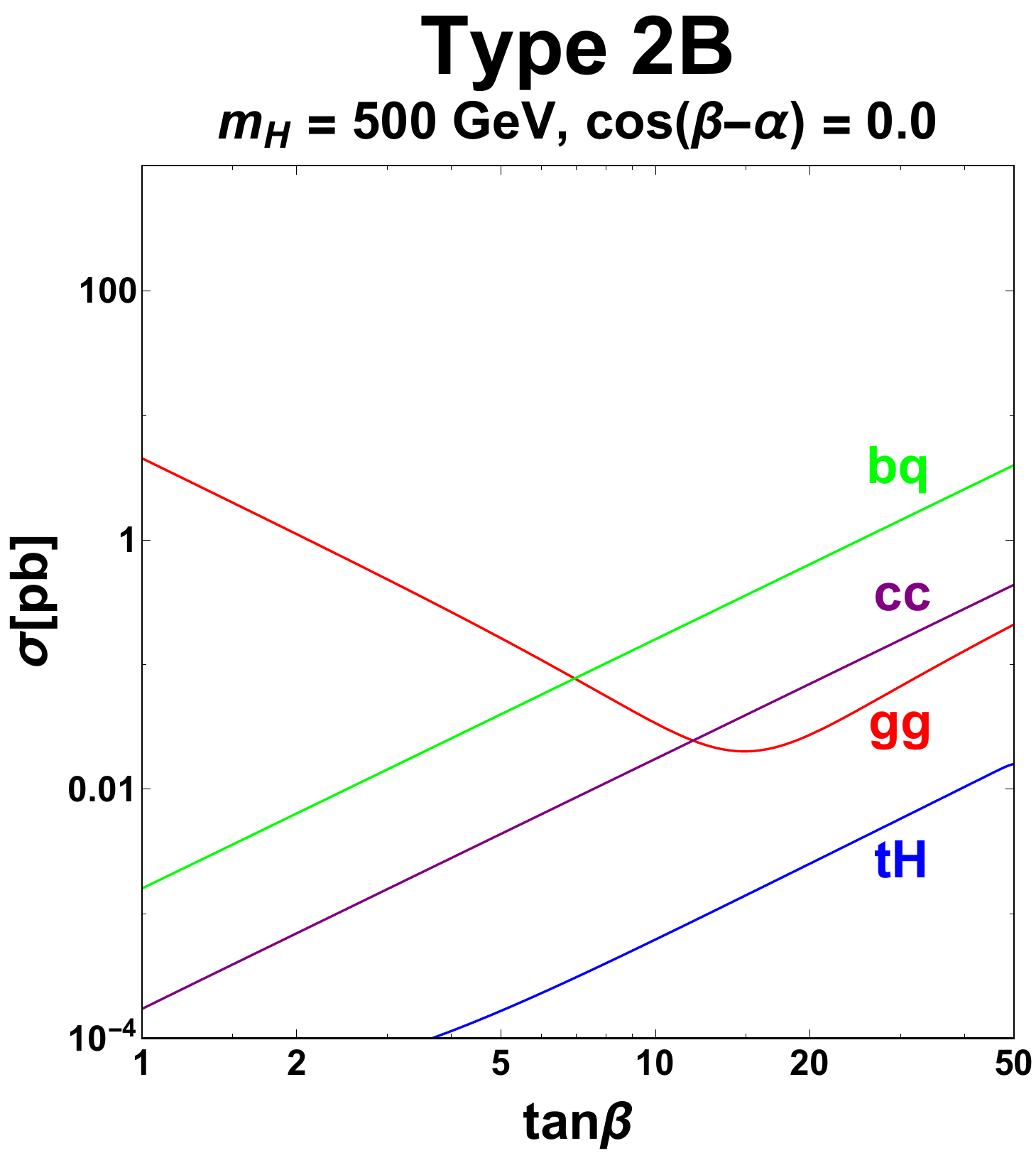} ~~~~~
 \includegraphics[width=0.37\textwidth]{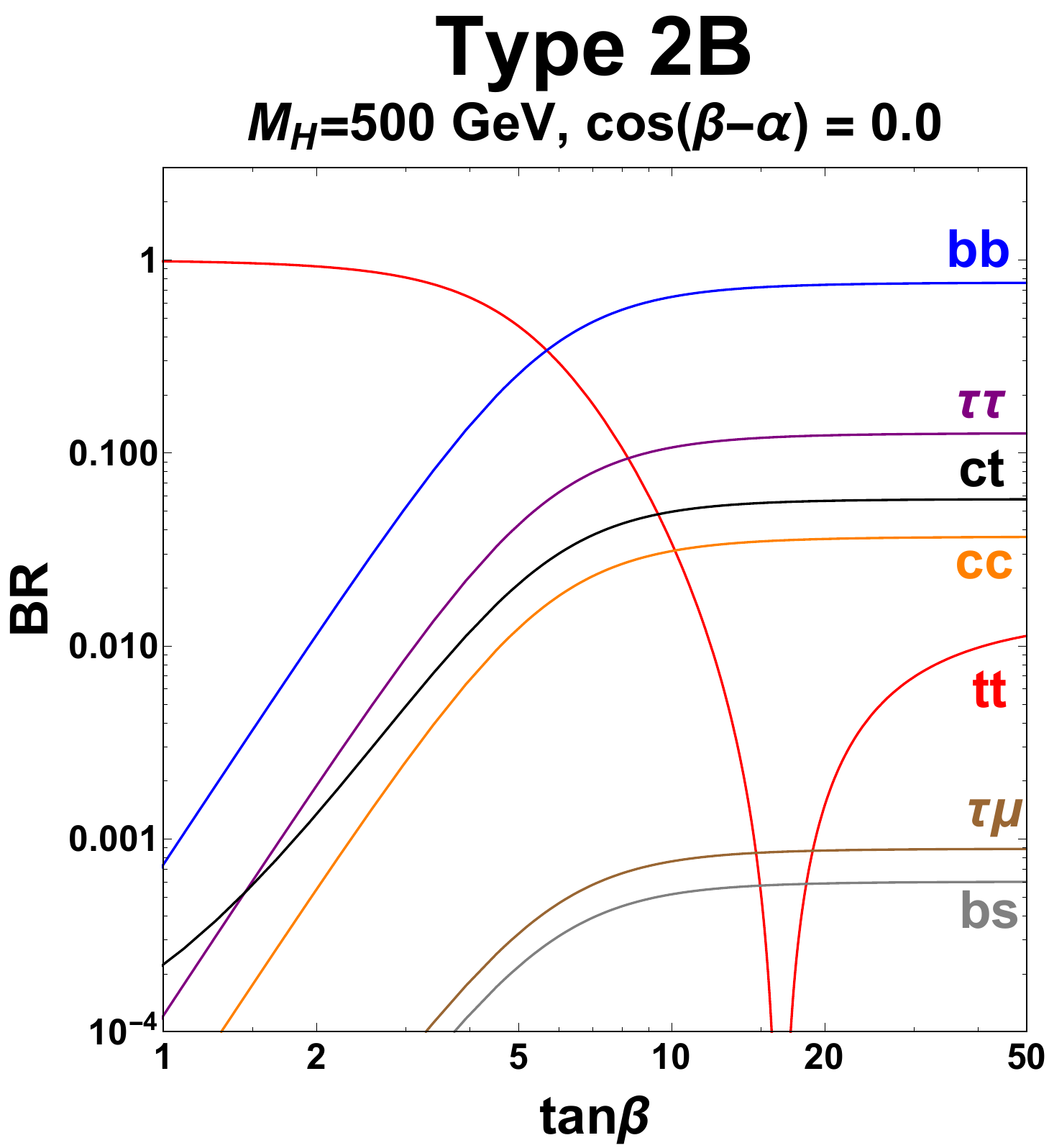} \\[12pt]
 \includegraphics[width=0.36\textwidth]{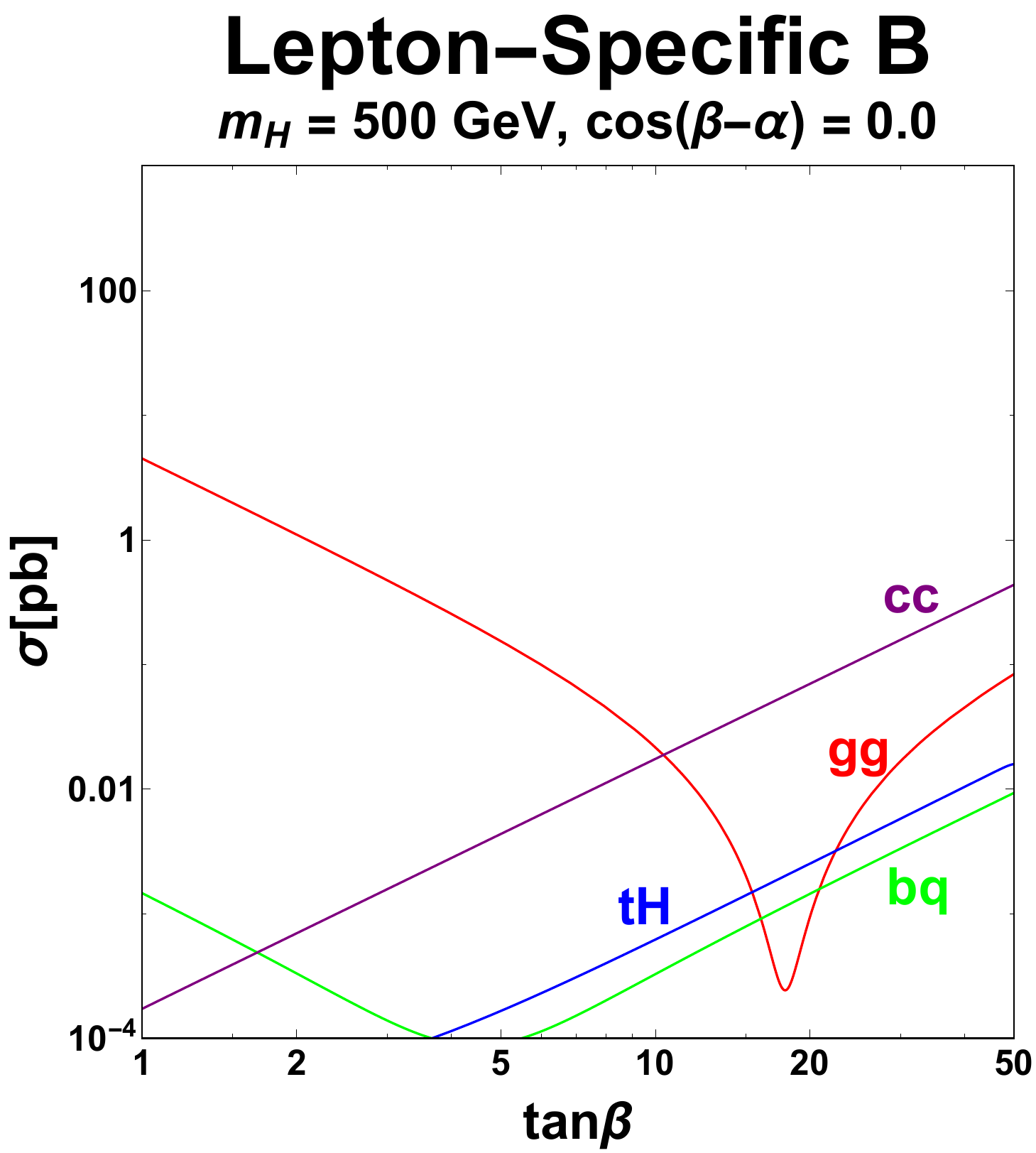} ~~~~~
 \includegraphics[width=0.37\textwidth]{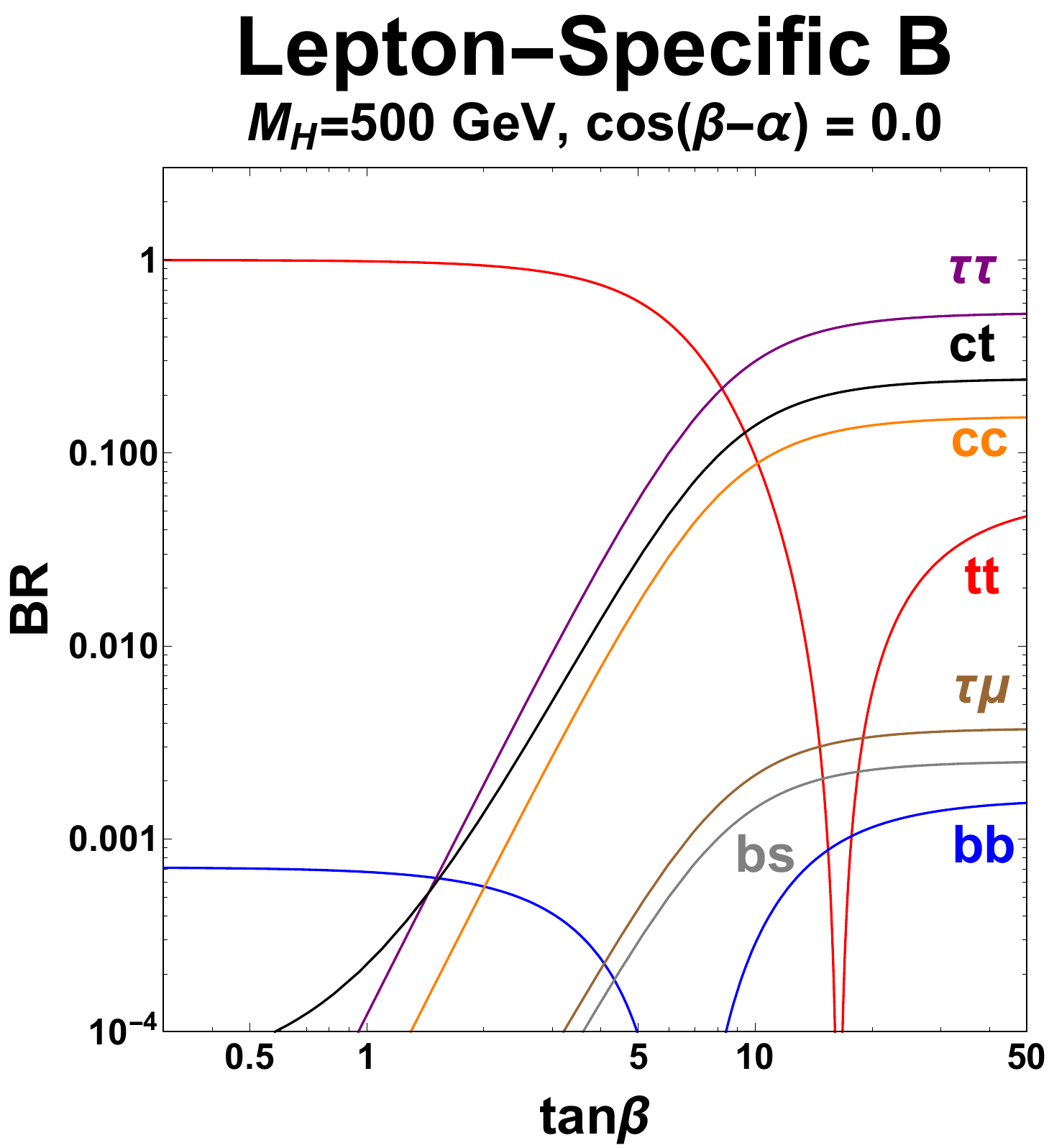} \\[12pt]
 \includegraphics[width=0.36\textwidth]{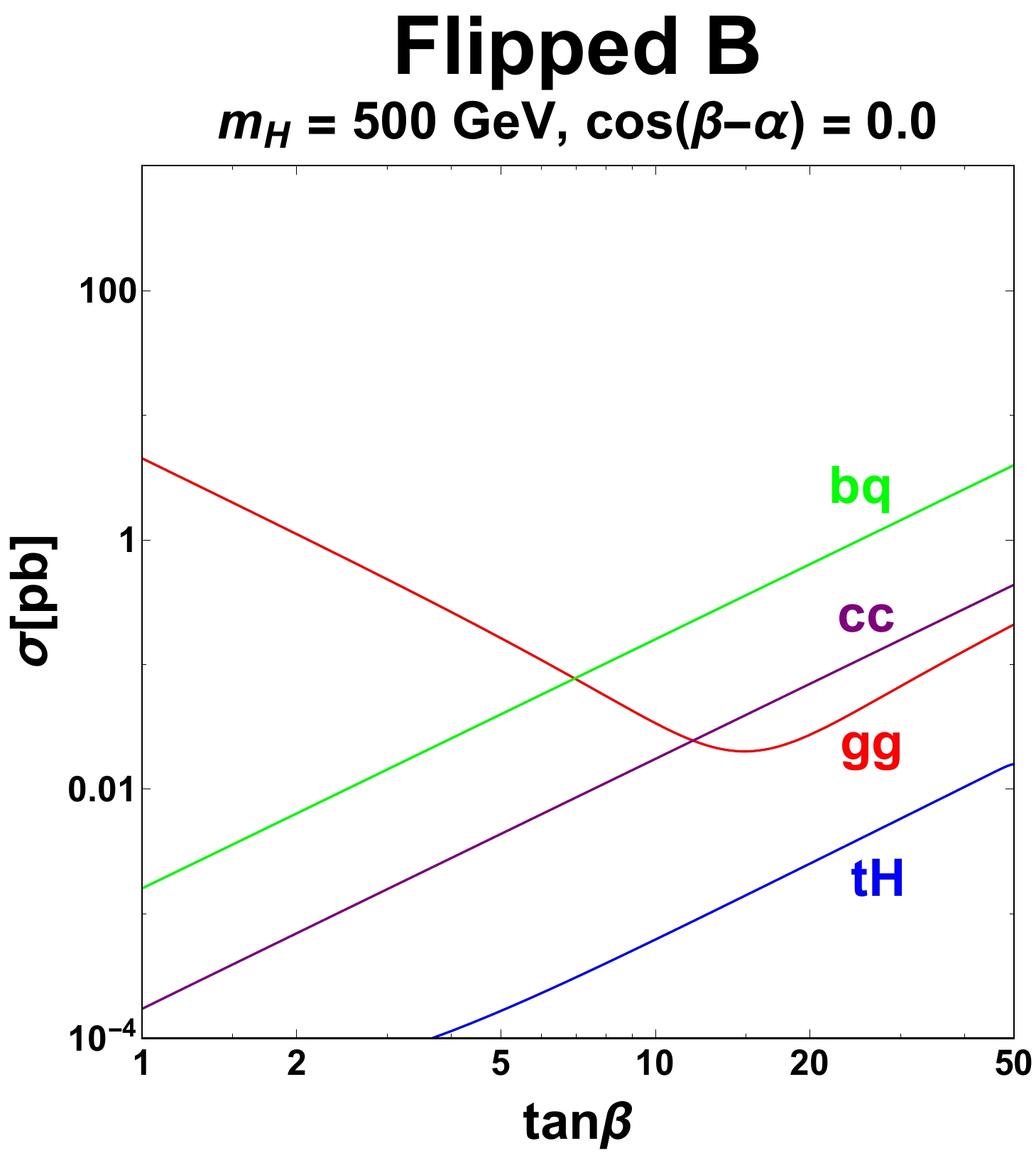} ~~~~~
 \includegraphics[width=0.37\textwidth]{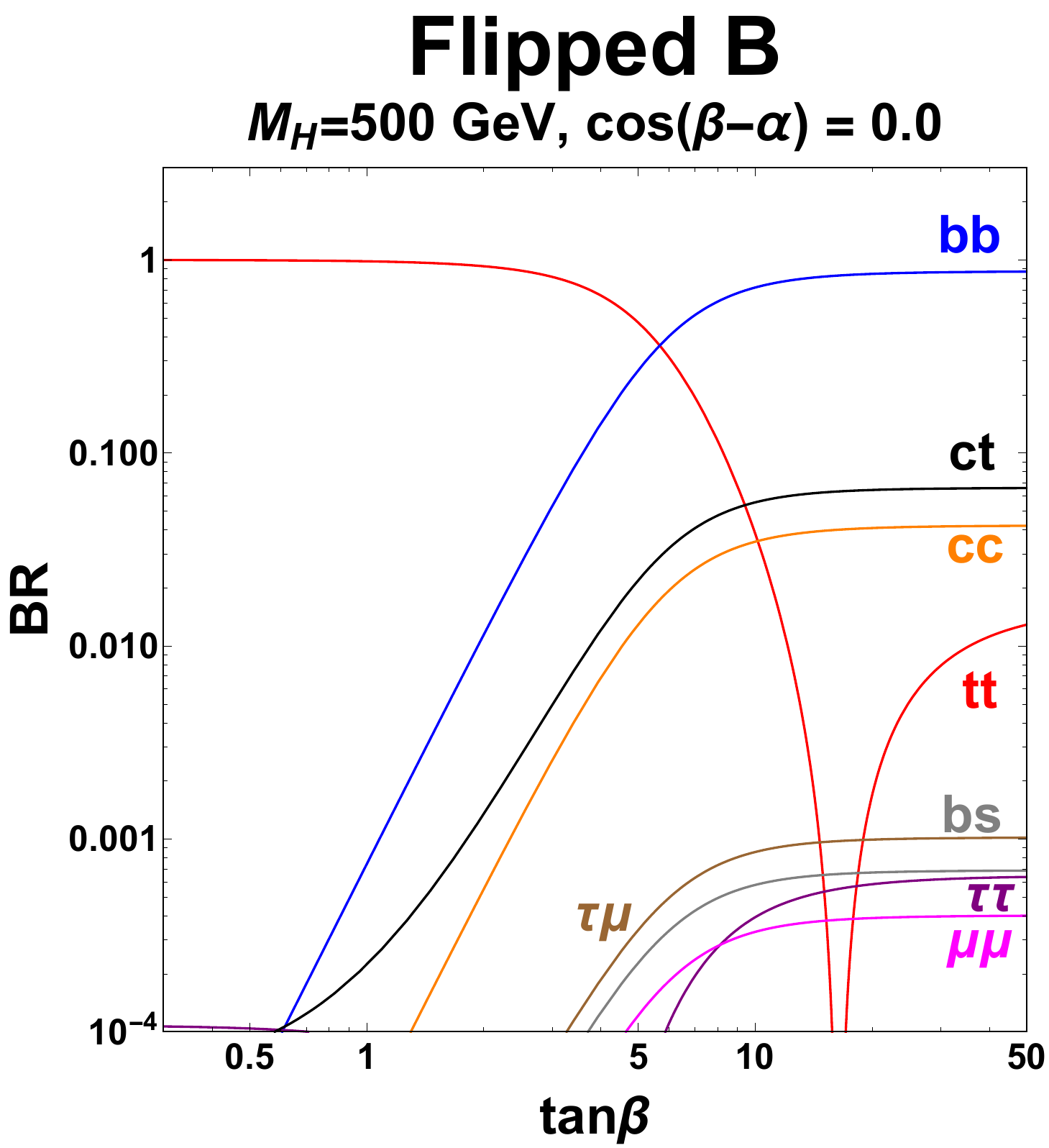} 
 \caption{Production cross sections at 13 TeV proton-proton collisions (left) and branching ratios (right) of the heavy scalar Higgs with mass $m_H = 500$~GeV in the type~2B model (top), lepton-specific~B model (center), and flipped~B model (bottom) as function of $\tan\beta$. In all plots we set $\cos(\beta-\alpha) = 0$.}
 \label{HiggstanB}
 \end{center}
 \end{figure}

The plots on the left hand side of Fig.~\ref{HiggstanB} show the various production cross sections for the three considered types of models as function of $\tan\beta$, for fixed Higgs mass of $m_H = 500$~GeV, and $\cos(\beta-\alpha) = 0$. 
In the type~2B and flipped~B models, production involving bottom quarks is typically most relevant, while in the lepton-specific~B model either production from $c\bar c$ or gluon-gluon fusion dominates. 
For large $\tan\beta$, the gluon-gluon fusion production is sub-dominant in all cases due to the suppressed Higgs coupling to tops. Gluon-gluon fusion is minimal for intermediate values of $\tan\beta \sim 15$, where the heavy Higgs coupling to tops accidentally vanishes. The precise location of the minimum depends on the choice of $m^\prime$ parameters and can shift by an $O(1)$ factor. 
For large and small $\tan\beta$ the shown production cross sections are robust with respect to $O(1)$ changes in the $m^\prime$ parameters.

Overall, the total production cross section of a heavy Higgs of mass 500~GeV ranges from several hundred $fb$ to several $pb$ in the type~2B and flipped~B models, and from tens of $fb$ to several $pb$ in the lepton-specific~B model. 
The results for the type~2B and flipped~B models are very similar to the corresponding models with natural flavor conservation. The reason is that the dominant production modes are governed by the top and bottom couplings that behave very similarly in those type~A and~B models.
The results for the lepton-specific~B model, however, differ markedly from the corresponding results of the lepton-specific~A model. In the type~A model, all couplings to quarks are universally suppressed by $1/\tan\beta$ leading to tiny production cross sections. In the type~B model the couplings to charm are enhanced, leading to an appreciable amount of heavy Higgs production.

\subsection{Branching ratios}
 
The heavy Higgs bosons can in principle decay to SM fermions, to the SM gauge bosons, and to other Higgs bosons.
In the decoupling limit $\cos(\beta-\alpha) = 0$, the decays of $H$ and $A$ to final states with massive vector bosons vanish. Decays into photons and gluons are loop suppressed and typically tiny. We assume that the heavy Higgs bosons are sufficiently degenerate, such that decays into each other are kinematically forbidden. The decay into two light Higgs bosons is in principle possible. The corresponding trilinear couplings depend on the couplings in the Higgs potential and can be made arbitrarily small.
In the following, we will only consider decays into fermions.
Generically, the decay widths of the heavy scalar $H$ to two fermions are
\begin{equation}
 \Gamma(H \to f_i f_j) = \Gamma(H \to f_i \bar f_j + f_j \bar f_i) = \frac{N_c m_H}{8 \pi} \Big(|(Y_H^f)_{ij}|^2 + |(Y_H^f)_{ji}|^2 \Big) ~, 
 \label{HeavyHiggsBR}
\end{equation}
where we assumed that the mass of the fermions is negligible $m_{f_i,\bar f_j} \ll m_H$. 
The color factor is $N_c=1$ for leptons and $N_c=3$ for quarks. 
This expression is sufficiently generic to describe both flavor conserving and flavor violating decays. 
In the case where one or both of the fermions is a top quark, top mass effects have to be included
\begin{eqnarray}
 \Gamma(H \to t u_i) &=& \frac{N_c m_H}{8 \pi} \left( 1 - \frac{m_t^2}{m_H^2} \right)^2 \Big(|(Y_H^u)_{3i}|^2 + |(Y_H^u)_{i3}|^2 \Big) ~, \\
 \Gamma(H \to t \bar t) &=& \frac{N_c m_H}{8 \pi} \left[ \left(1 - \frac{4m_t^2}{m_H^2} \right)^{\frac{3}{2}} \text{Re}((Y_H^u)_{33})^2 + \left(1 - \frac{4m_t^2}{m_H^2} \right)^{\frac{1}{2}} \text{Im}((Y_H^u)_{33})^2 \right] ~.
\end{eqnarray}
We show the branching ratios of the heavy Higgs as function of $\tan\beta$ in the plots on the right hand side of Fig.~\ref{HiggstanB}. The heavy Higgs mass is set to $m_H = 500$~GeV and $\cos(\beta-\alpha) = 0$. 
The main decay modes of the heavy Higgs to the fermions are easily understood from table~\ref{PhiCouplings}, that shows to which fermions the $\phi^\prime$ doublet couples. 
In the type~2B and flipped~B models we expect the $b\bar{b}$ decay to dominate at large $\tan\beta$. For the lepton-specific setup we expect the $\tau^+\tau^-$ decay to be the primary branching ratio. In the flipped~B model, the $\tau^+\tau^-$ decay is instead strongly suppressed.
For low $\tan\beta$, decays into $t\bar t$ dominate (if kinematically allowed). 
These are the same patterns as in the models with natural flavor conservation.

In contrast to the models with natural flavor conservation, decays involving charm quarks ($c\bar c$ and $ct$) can have branching ratios of $O(10\%)$ in all three flavorful models.
Also the decay into $t\bar t$ has branching ratios of several $\%$ for large $\tan\beta$, due to terms in the coupling of the heavy Higgs to tops that are proprotional to $m_c \tan\beta$. For $\tan\beta \simeq 15$ there can be a cancellation between the leading $1/\tan\beta$ suppressed term and the $m_c$ correction, leading to an accidental vanishing of the $t \bar t$ branching ratio.

Also lepton flavor violating decays can arise. 
In the lepton-specific~B model, we find the decay $\tau \mu$ can have branching ratios of up to $\sim 1\%$. In the type~2B and flipped~B model, the branching ratio of this decay mode is smaller by a factor of few, as it has to compete with the dominant decay into $b \bar b$.

The branching ratios of flavor diagonal decay modes like $b \bar b$, $\tau^+\tau^-$, and $c \bar c$ are fairly robust against changes in the $m^\prime$ mass parameters. The branching ratios of flavor violating decays can change by a factor of few if the relevant $m^\prime$ parameters are modified by an $O(1)$ amount.

In the decoupling limit, the scalar and pseudoscalar Higgs couplings are identical. Consequently, the production cross sections and branching ratios of the pseudoscalar Higgs are very similar to the scalar Higgs and we do not show the plots for the pseudoscalar. 

\subsection{Constraints from direct searches}
\label{BiggsConst}

 \begin{figure}[tb]
 \begin{center} 
 \includegraphics[width=0.44\textwidth]{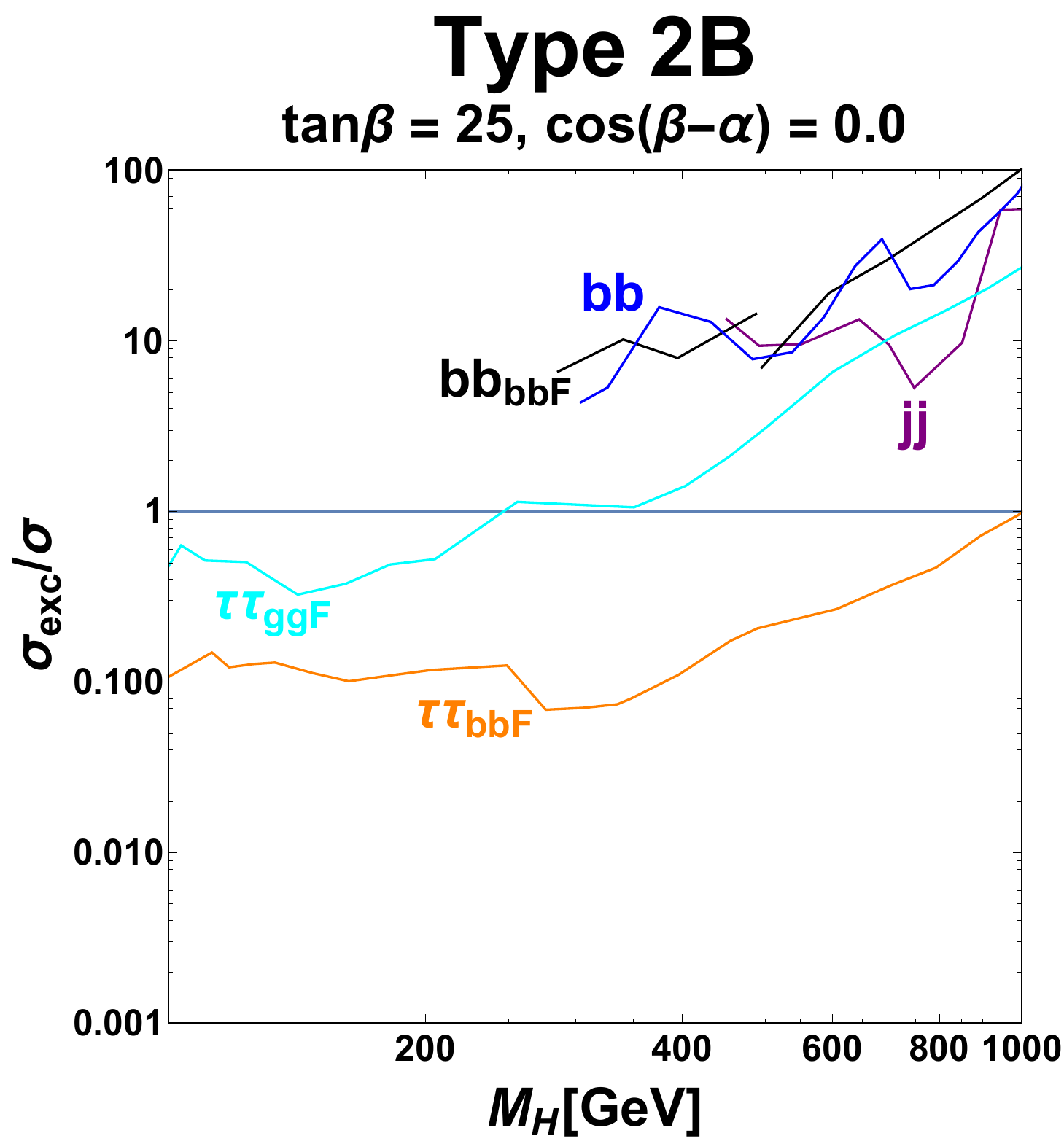} ~~
 \includegraphics[width=0.44\textwidth]{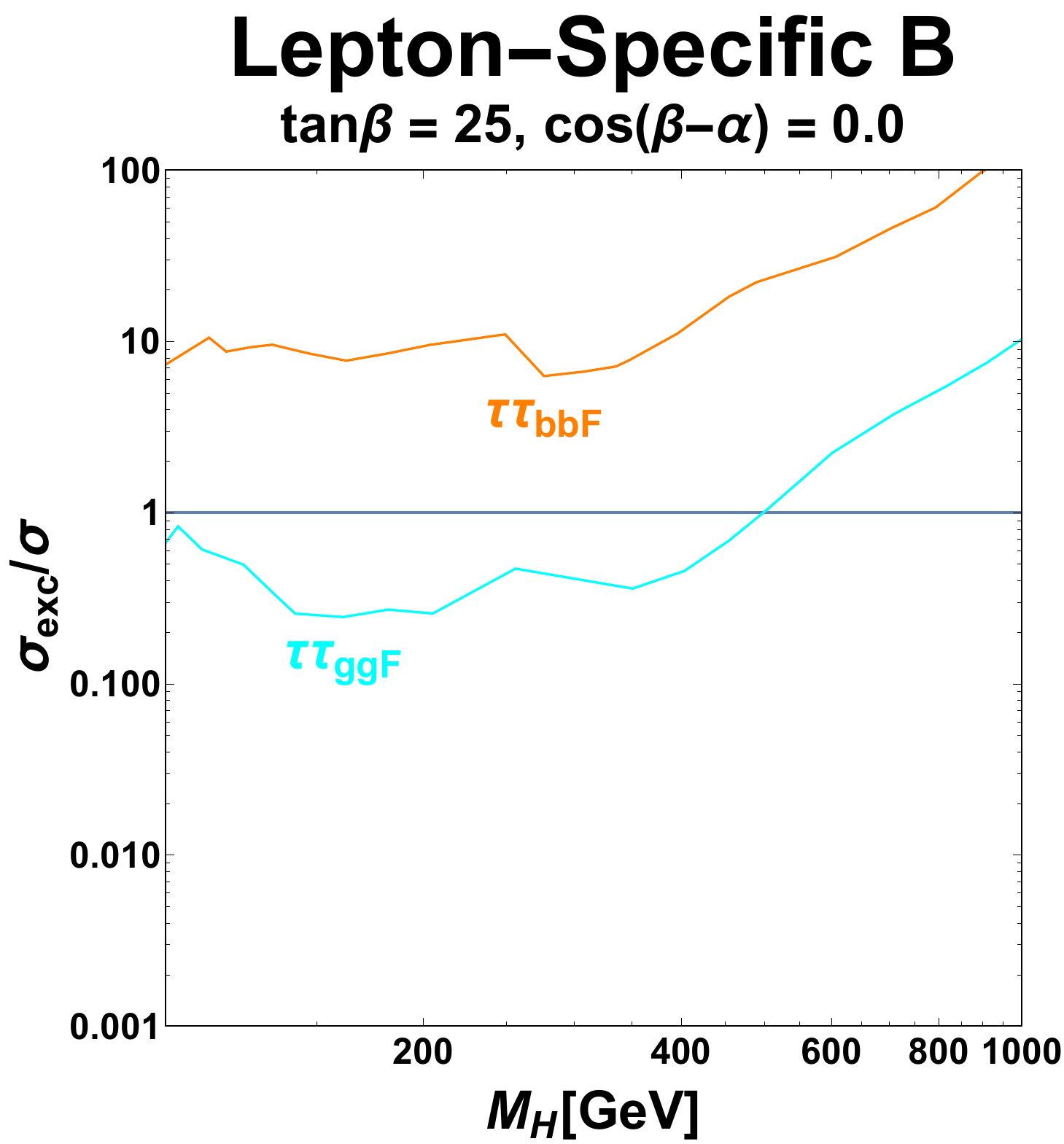} \\
 \includegraphics[width=0.44\textwidth]{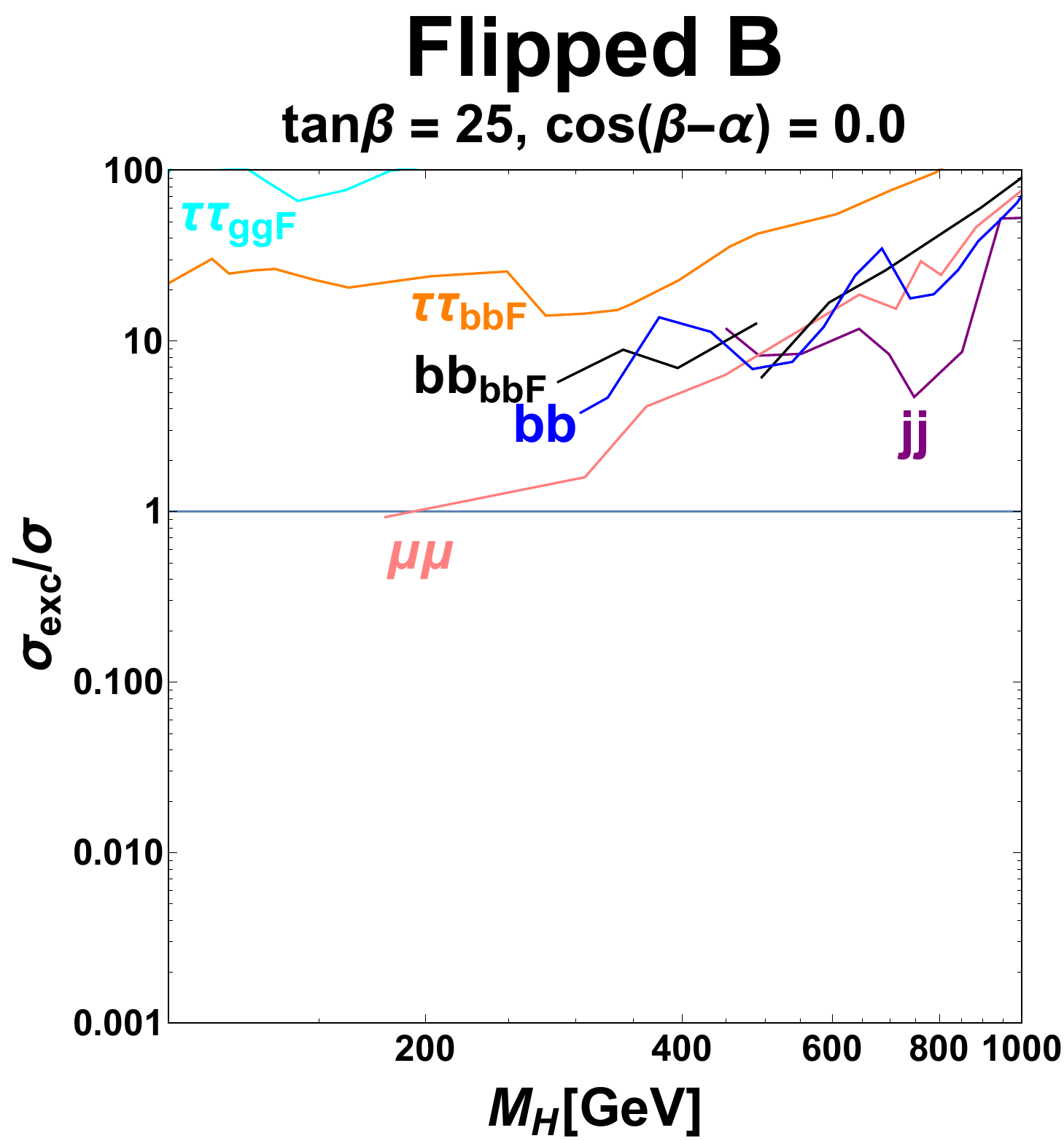}
 \caption{Exclusions for the heavy Higgs as a function of its mass $m_H$ for $\tan{\beta} = 25$ and $\cos(\beta-\alpha) = 0$. Cross section ratios smaller than 1 are experimentally excluded.}
 \label{ExclusionBiggs}
 \end{center}
 \end{figure}

Having examined the main production and decay modes of the heavy neutral Higgs bosons of the flavorful models we now compare results from current heavy Higgs searches at the LHC with the model predictions.  
We find the most relevant constraints come from 
\begin{itemize}
\item searches for $H \to \tau^+ \tau^-$ with the Higgs produced either in gluon-gluon fusion, or in association with $b$ quarks (ATLAS 13\,TeV with 36.1\,fb$^{-1}$~\cite{Aaboud:2017sjh} and CMS 13\,TeV with 2.2\,fb$^{-1}$~\cite{Khachatryan:2016qkc});
\item searches for low mass di-jet resonances (ATLAS 13\,TeV with 3.6 and 29.3\,fb$^{-1}$~\cite{Aaboud:2018fzt});
\item searches for $b \bar b$ resonances (CMS 13\,TeV with 35.7\,fb$^{-1}$~\cite{CMS-PAS-HIG-16-018} and CMS 8\,TeV with 19.7\,fb$^{-1}$~\cite{Sirunyan:2018pas});
\item searches for di-muon resonances (ATLAS 13\,TeV with 36\,fb$^{-1}$~\cite{Aaboud:2017buh} and CMS 13\,TeV with 36\,fb$^{-1}$~\cite{Sirunyan:2018exx}).
\end{itemize}

In Fig.~\ref{ExclusionBiggs} we show the ratio of the experimentally excluded rate $(\sigma \times \text{BR})_\text{exp}$ to the rate predicted in our flavorful 2HDMs $(\sigma \times \text{BR})_{\text{BSM}}$ as function of the heavy Higgs mass for a benchmark scenario with $\tan\beta = 25$ and $\cos(\beta-\alpha) = 0$. 
If this ratio is below 1, the model is excluded for the given set of parameters.

Concerning, the experimental searches that target Higgs production in association with bottom quarks, we estimate the theoretical production cross section from $b \bar b \to H$, keeping in mind that higher order corrections might change the result by an $O(1)$ amount. The corresponding constraints in the plots of Fig.~\ref{ExclusionBiggs} are labeled with the subscript ``bbF''. If experimental constraints assume gluon-gluon fusion production, we take into account both gluon-gluon fusion and also production from $c \bar c$, which should lead to the same experimental signature. The corresponding constraints are labeled ``ggF''.
If no particular production mode is singled out by the experimental search, we add up all the production mechanisms.
For each individual channel we show the strongest constraint among the considered experimental analyses.

We observe that for $\tan\beta = 25$ the type~2B and the lepton-specific~B models are strongly constrained by searches for heavy Higgs decaying to a $\tau^+\tau^-$ final state. Heavy Higgs masses up to $\sim 1$~TeV (type~2B) and up to $\sim 500$~GeV (lepton-specific~B) are already excluded in this case. The constraints are much weaker in the flipped~B model. Searches for di-muon, $b \bar{b}$ and di-jet resonance searches have sensitivities that start to approach the model predictions, but currently do not exclude parameter space with $\tan\beta = 25$. 

Note that the excluded mass ranges are extremely sensitive to the values of $\tan\beta$. For large $\tan\beta$ the production cross sections in all models are approximately proportional to $\tan^2\beta$. So, the cross section ratios quickly go below the exclusion line. However, as $\tan{\beta}$ becomes small the constraints generically get weaker and the constraints in the type~2B and lepton-specific~B case can be easily avoided. 

\section{Charged Higgs production and decays}
\label{CHP}

The collider phenomenology of the charged Higgs in the type~1B model has been discussed previously in~\cite{Altmannshofer:2016zrn}. Here we discuss the phenomenology of the charged Higgs in the type~2B, lepton-specific~B, and flipped~B models.
 
\subsection{Production cross sections}

 \begin{figure}[tb]
 \begin{center}
 \includegraphics[width=0.38\textwidth]{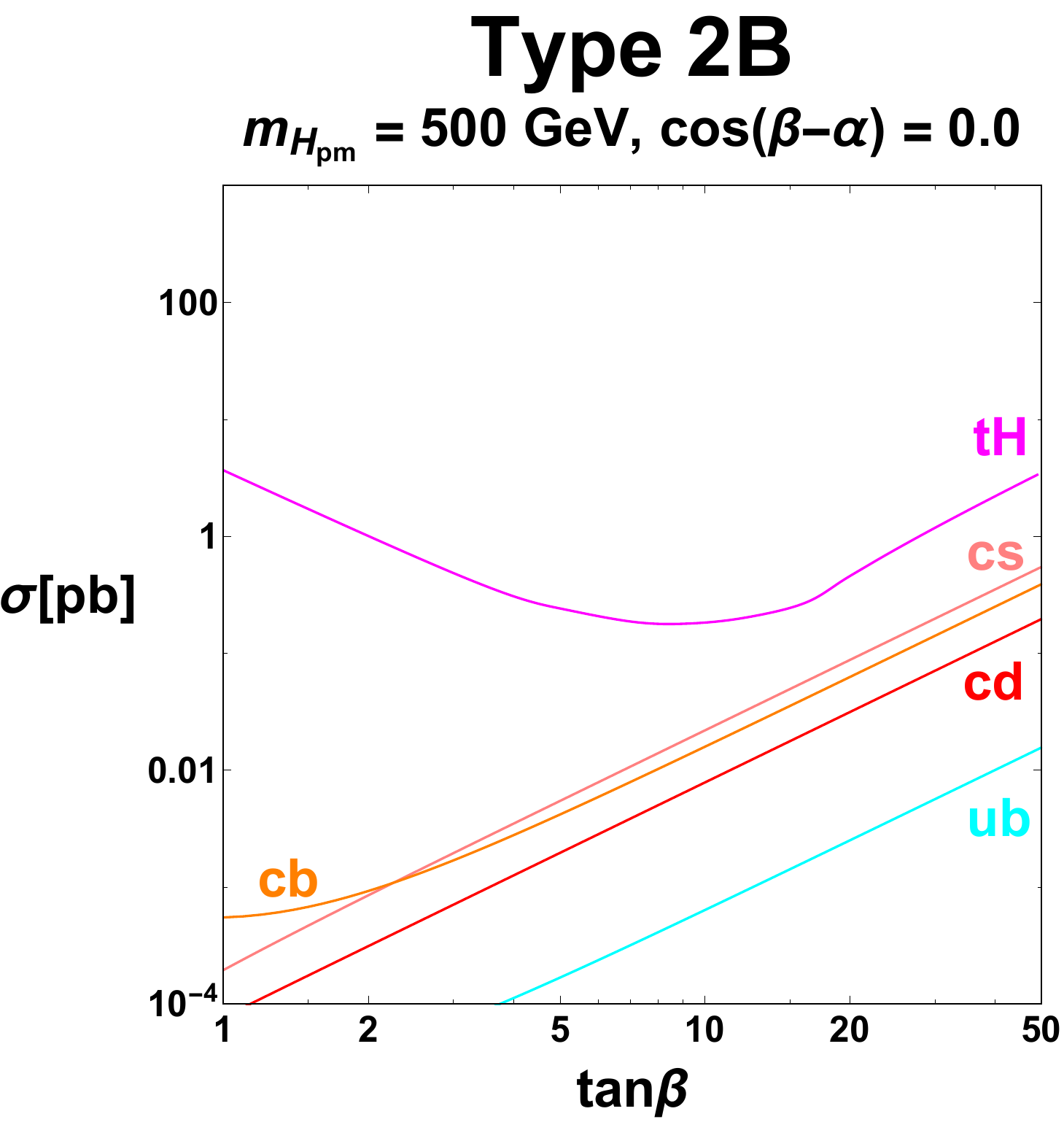} ~~~~~
 \includegraphics[width=0.36\textwidth]{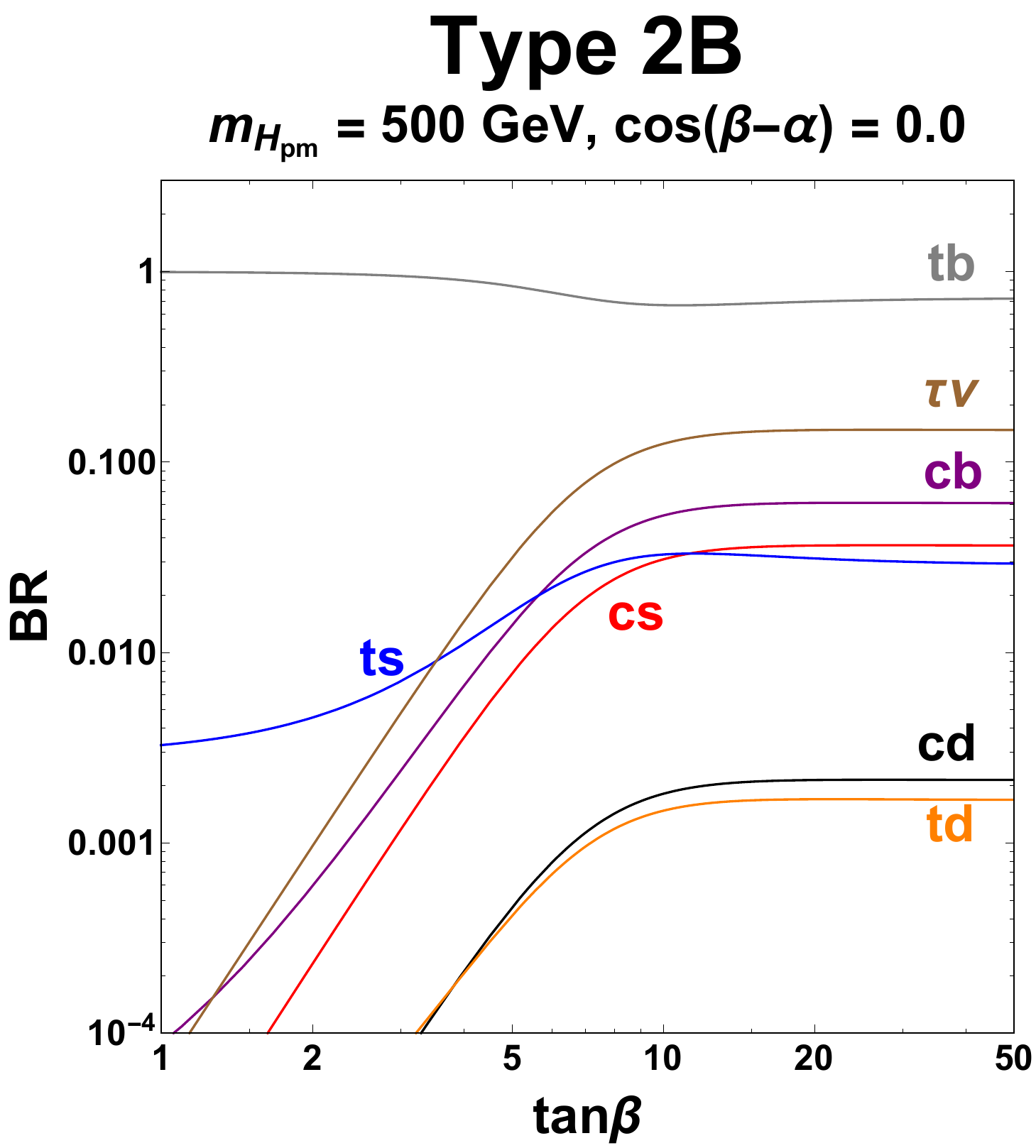} \\[12pt]
 \includegraphics[width=0.38\textwidth]{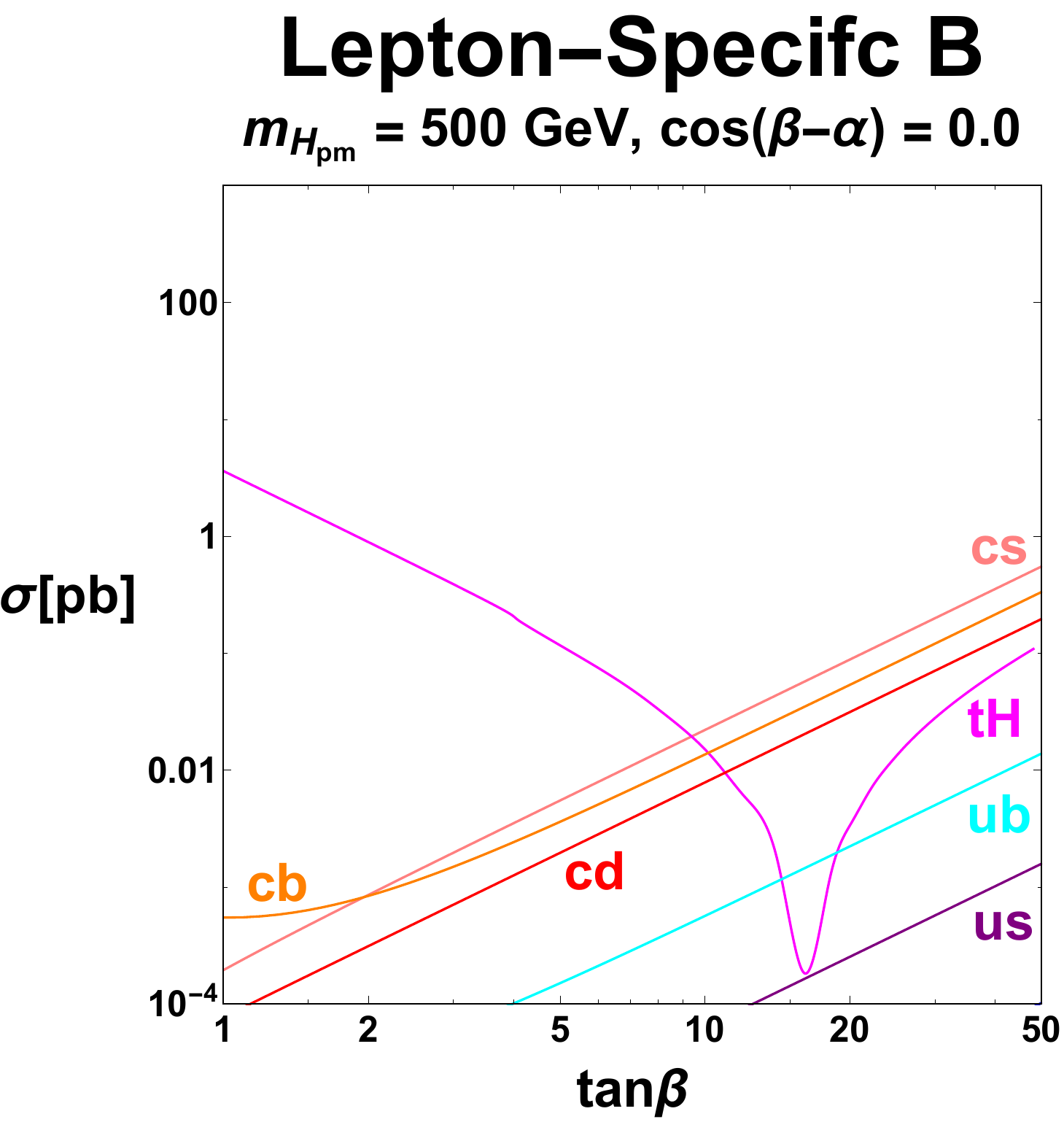} ~~~~~
 \includegraphics[width=0.36\textwidth]{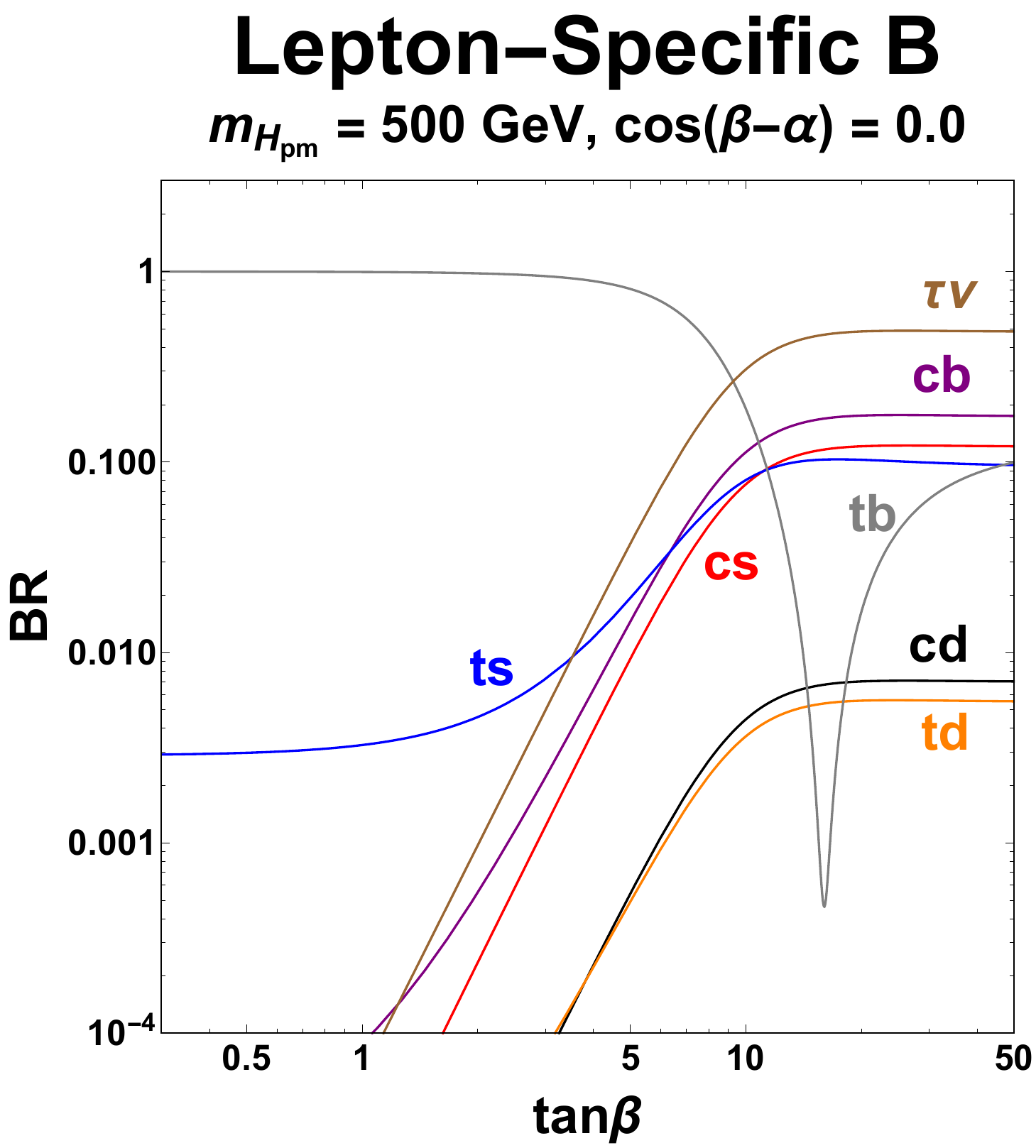} \\[12pt]
 \includegraphics[width=0.38\textwidth]{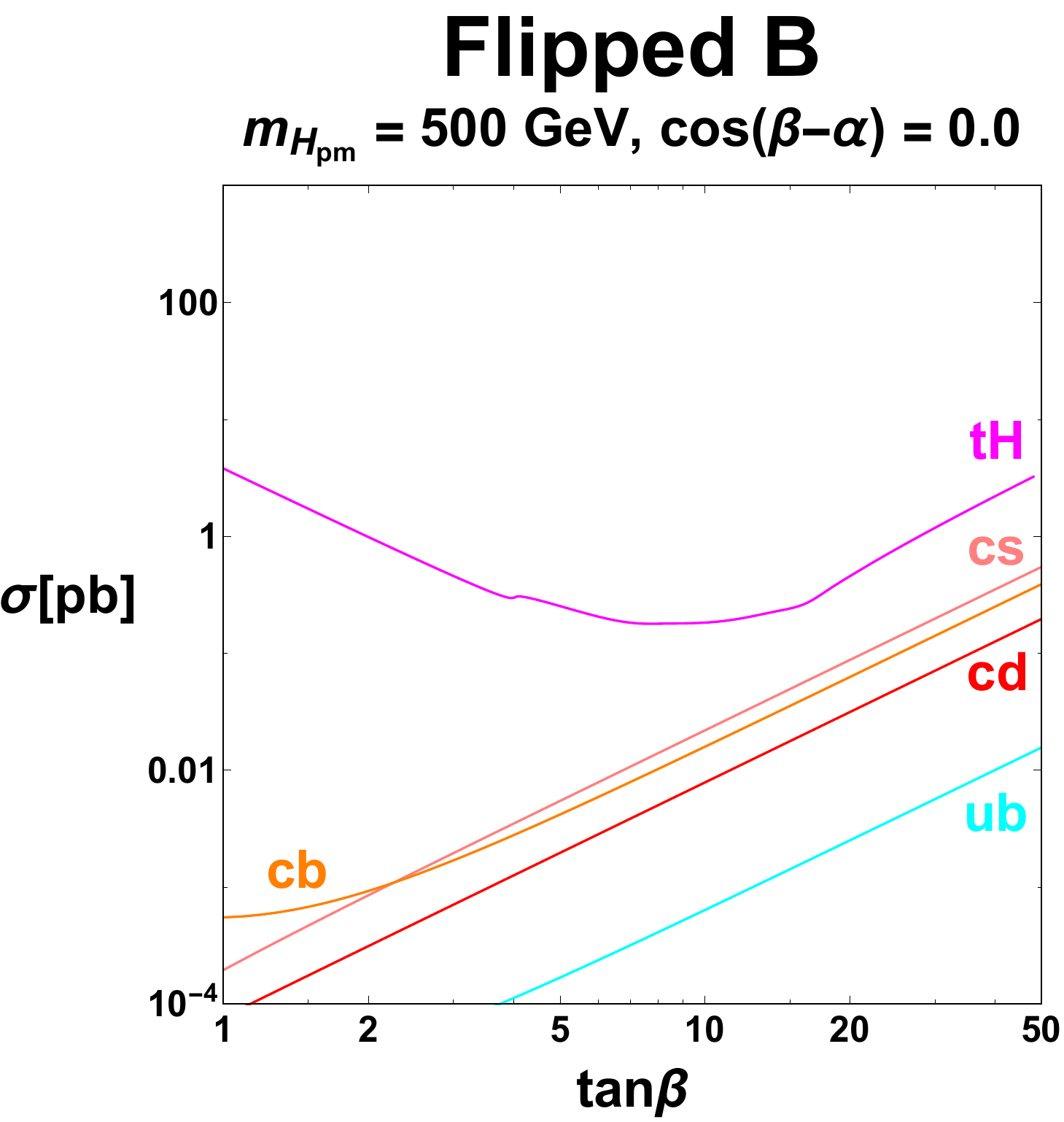} ~~~~~
 \includegraphics[width=0.36\textwidth]{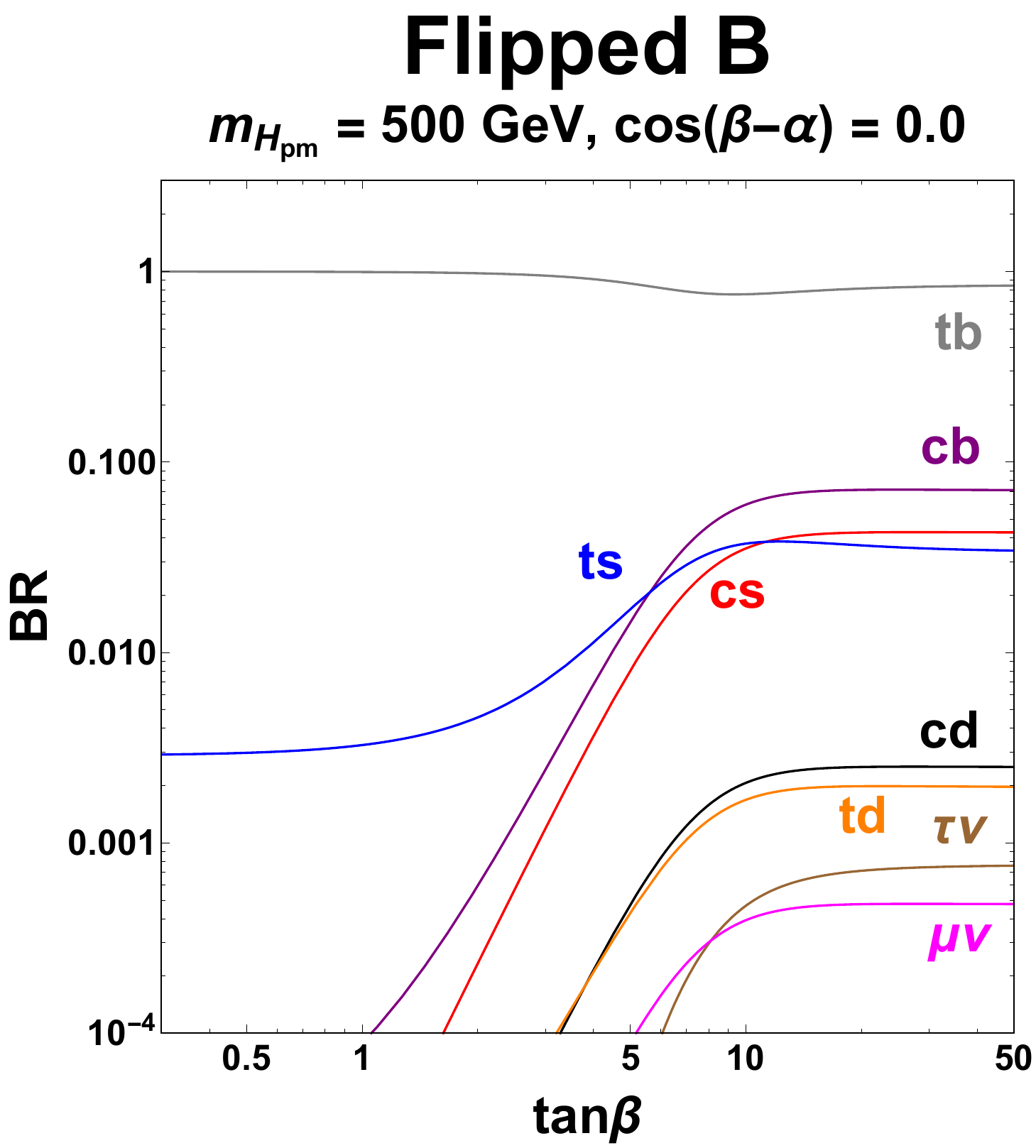}
 \caption{Production cross sections at 13TeV proton-proton collisions (left) and branching ratios (right) of the charged Higgs with mas $m_{H^\pm} = 500$~GeV in the type~2B model (top), lepton-specific~B model (center), and flipped~B model (bottom) as a function of $\tan\beta$. In all plots we set $\cos(\beta-\alpha)=0$. }
 \label{CHiggstanB}
 \end{center}
 \end{figure}

As for the neutral Higgses, the main production mode is again primarily from $q \bar q^\prime$ fusion. We estimate these cross sections using an expression analogous to Eq.~(\ref{HeavyHiggsCS}) along with the MMHT 2014 PDFs~\cite{Harland-Lang:2014zoa}. Also production in association with a top quark (see diagrams in Fig.~\ref{tH1Feyn}) can become important. 
The corresponding production cross section is taken from~\cite{Djouadi:2005gj}.

We show the production cross sections as function of $\tan\beta$ in Fig.~\ref{CHiggstanB}. As an example, we use the charged Higgs mass $m_{H^\pm} = 500$~GeV and set $\cos(\beta-\alpha) = 0$.

At low $\tan\beta$, the production in association with a top quark dominates in all three flavorful models.
In the type~2B and flipped~B models production in association with a top quark remains dominant also for large $\tan\beta$ due to the enhanced couplings to bottom in this region of parameter space.
In the lepton-specific~B model, however, large $\tan\beta$ implies suppression of both top and bottom couplings and the top associated charged Higgs production is suppressed.

We find that the charged Higgs production from $q \bar q^\prime$ fusion is dominated by initial states containing charm quarks. All three combinations $cb$, $cs$, and $cd$ have production cross sections of the same order of magnitude. While the coupling to $cd$ is suppressed by a factor of $\sim V_{cd}$ compared to the $cb$ and $cs$ couplings, this suppression is partially compensated by the larger down PDF.  
Furthermore, the $q \bar q^\prime$  production cross sections are mainly determined by couplings of the charged Higgs involving right handed charm quarks. Those couplings have the same scaling with $\tan\beta$ for all three flavorful models and we indeed observe that also the corresponding cross sections are approximately equal in the three models.

This is particularly interesting for the lepton-specific~B case. In the lepton-specific~A model, all couplings to quarks are suppressed at large $\tan\beta$, and charged Higgs production is tiny. In the ``B-type'' of the model, however, the enhanced couplings to charm open up the possibility to directly probe this region of parameter space at the LHC.

\subsection{Branching Ratios}

In the considered scenario with $\cos(\beta - \alpha) = 0$, the charged Higgs decays either to quarks or leptons. The decay to $W^\pm h$ is absent. The decay rate to fermions is given analogous to the neutral Higgs, Eq.~(\ref{HeavyHiggsBR}). 

In the type~2B and flipped~B models we expect the dominant branching ratio to be $tb$ both for small $\tan\beta$ (where the coupling to top is large) and at large $\tan\beta$ (where the coupling to bottom is enhanced).
This can be clearly seen in the plots of Fig.~\ref{CHiggstanB} that show the most relevant branching ratios as function of $\tan\beta$ for $m_{H^\pm} = 500$~GeV and $\cos(\beta - \alpha) = 0$.

In the type~2B model, the $\tau\nu$ decay mode has the second largest branching ratio at large $\tan\beta$. This is very similar to the type~2A model with natural flavor conservation. In contrast to the type~2A, decay modes including charm quarks, like $cb$ and $cs$, can have branching ratios of several~$\%$ in the flavorful type~2B model. 
Also in the flipped~B model, $cb$ and $cs$ can have branching ratios of several~$\%$. The decay to $\tau\nu$ on the other hand is strongly suppressed. The rather clean $\mu\nu$ final state can reach branching ratios of $O(10^{-3})$, which is orders of magnitude larger than in the flipped~A model.

In the lepton-specific~B model, the branching ratio to $\tau\nu$ dominates at large $\tan\beta$ and is typically around $50\%$. Decay modes involving charm ($cs$ and $cb$) as well as top ($ts$ and $tb$) have typical branching ratios of $O(10\%)$. 

For $\tan\beta$ above $\sim 10$ most branching ratios stay approximately constant. One exception is the $tb$ branching ratio in the lepton-specific~B model which changes considerably with $\tan\beta$. For $\tan\beta \sim 15$ the relevant coupling of the charged Higgs to $tb$ vanishes, due to an accidental cancellation between the $1/\tan\beta$ term and the term of $O(m_c)$ in Eq.~(\ref{eq:Hpm_couplings}). The same cancellation is also responsible for the dip in the top associated production in the lepton-specific~B model shown on the left-hand side of Fig.~\ref{CHiggstanB}. 
The precise value of $\tan\beta$ where this cancellation happens depends on the sign and exact size of the free $O(1)$ parameters in the $m^\prime$ mass parameters, see Eq.~(\ref{upMassParameters}). 
In general, variation of the $m^\prime$ mass parameters can change the branching ratios of flavor violating decays by a factor of few.

\subsection{Constraints from direct searches}

\begin{figure}[tb]
 \begin{center}
 \includegraphics[width=0.35\textwidth]{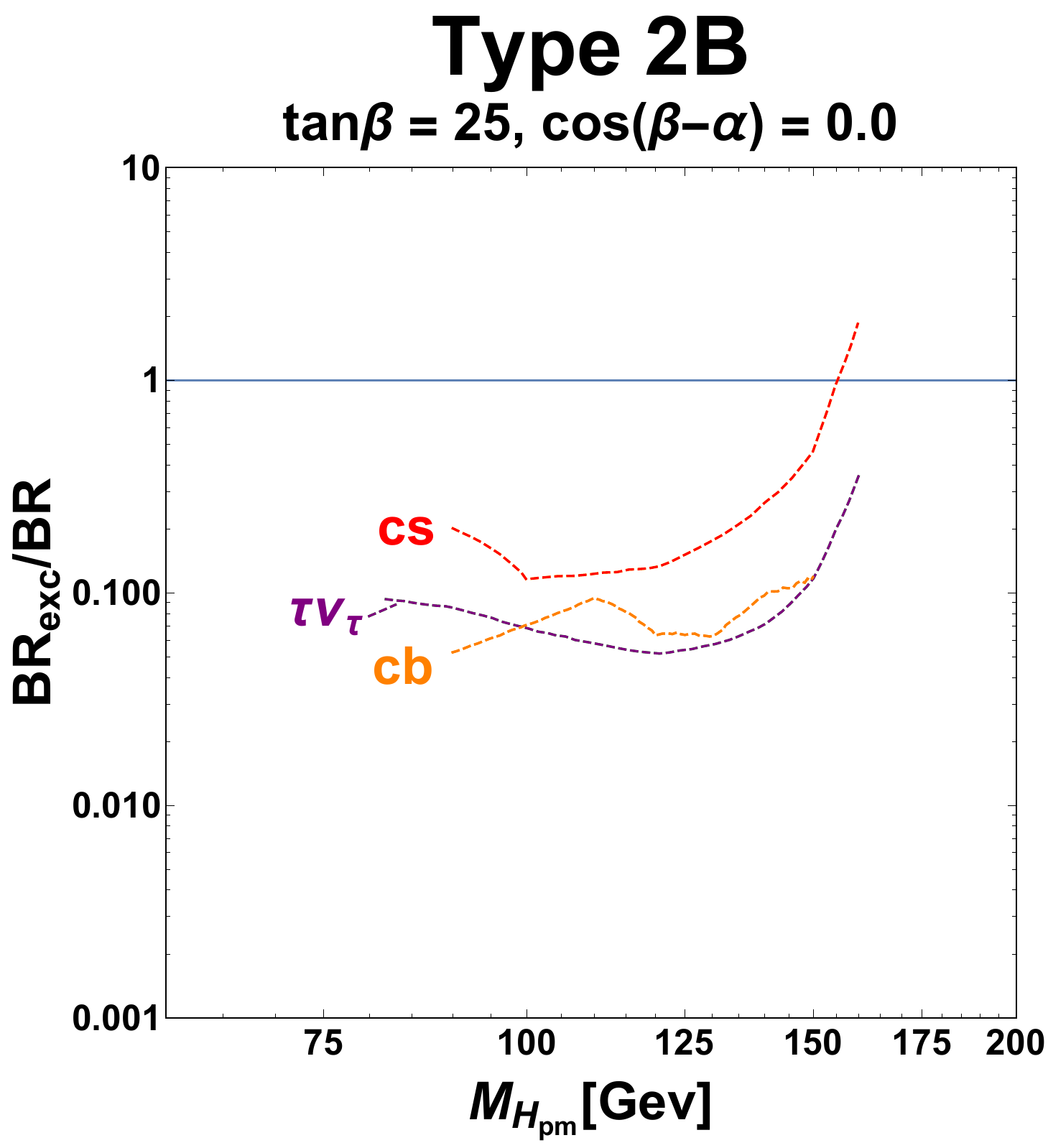} ~~~~~
 \includegraphics[width=0.35\textwidth]{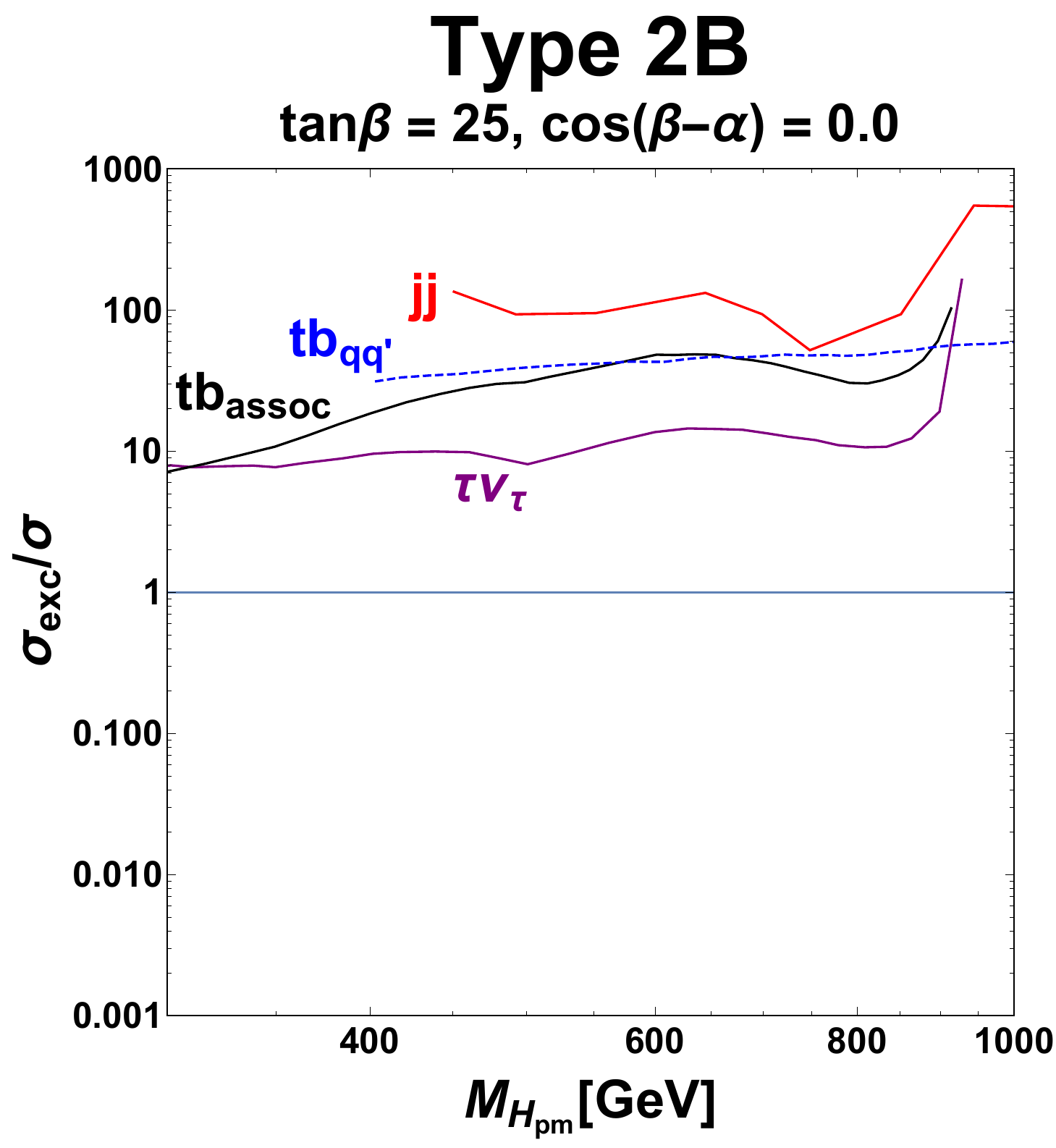} \\[12pt]
 \includegraphics[width=0.35\textwidth]{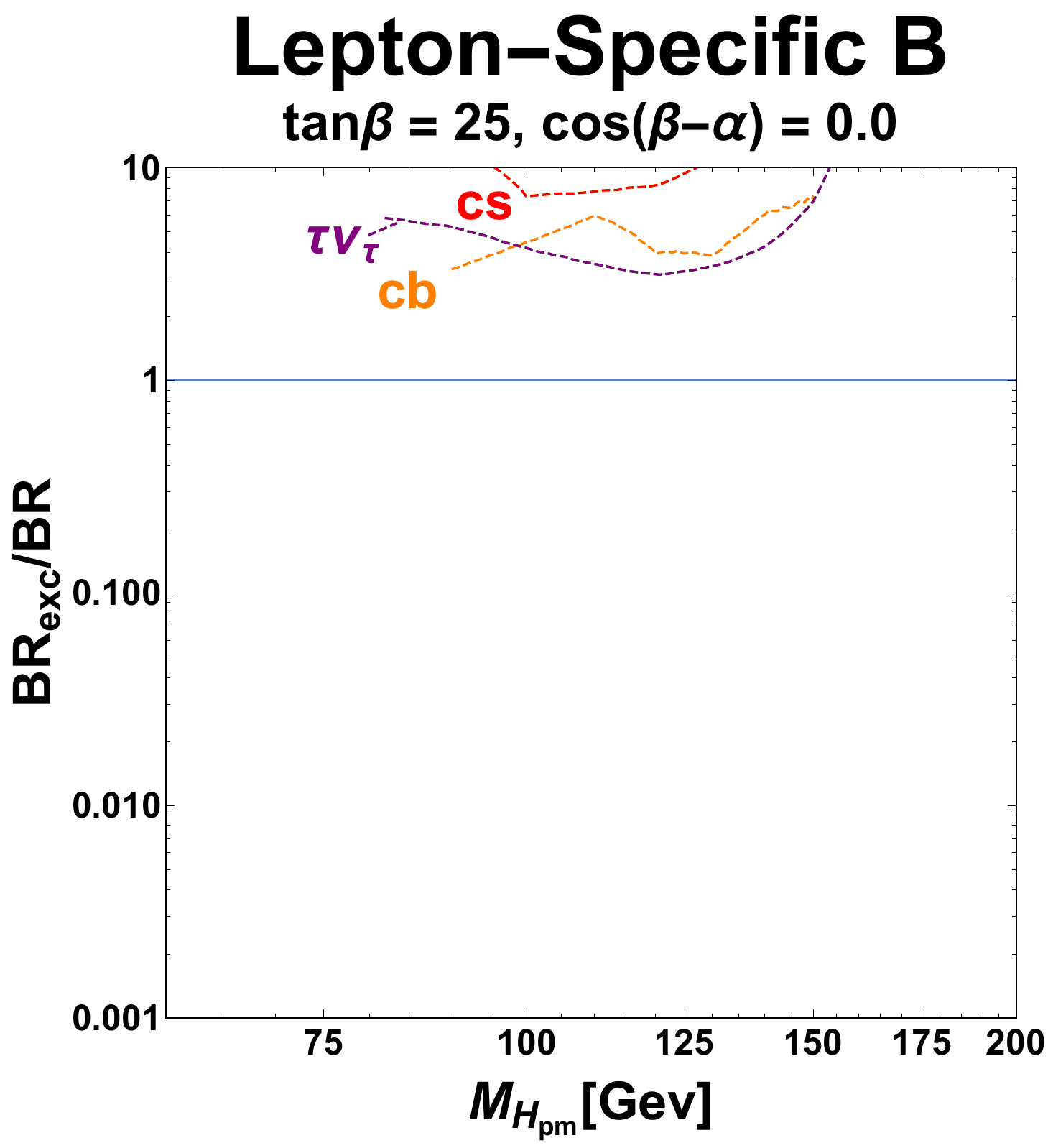} ~~~~~
 \includegraphics[width=0.35\textwidth]{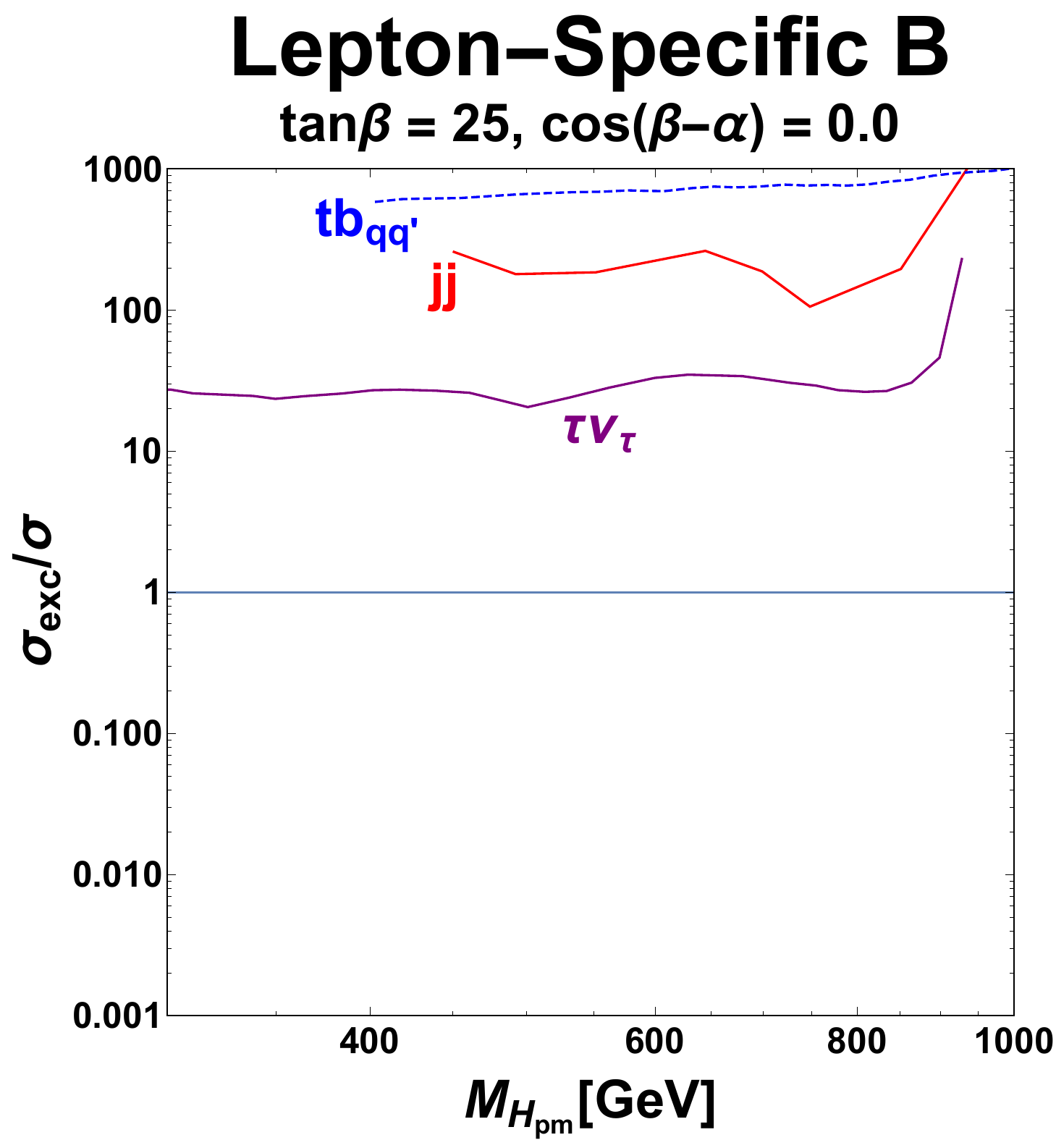} \\[12pt]
 \includegraphics[width=0.35\textwidth]{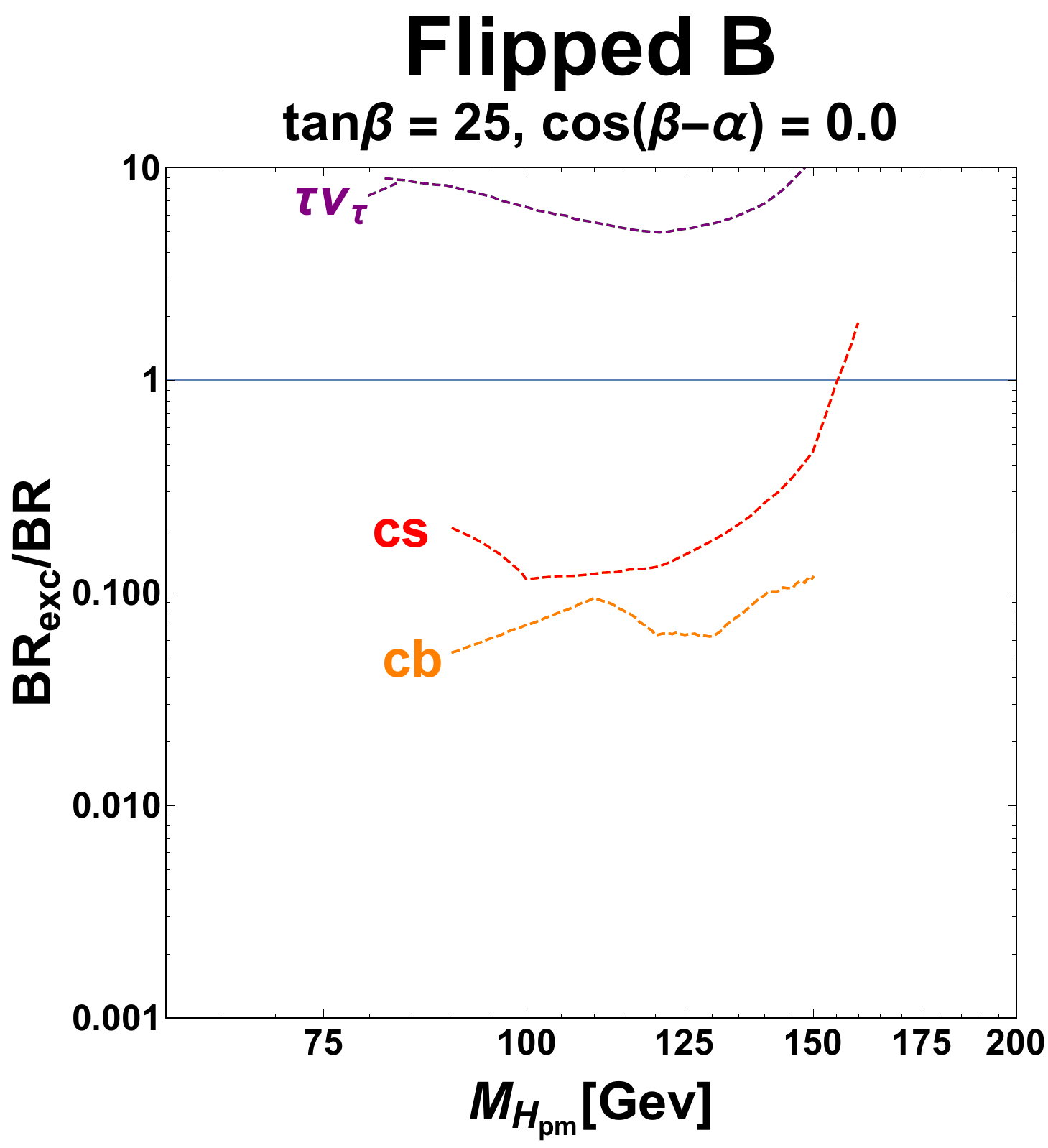} ~~~~~
 \includegraphics[width=0.35\textwidth]{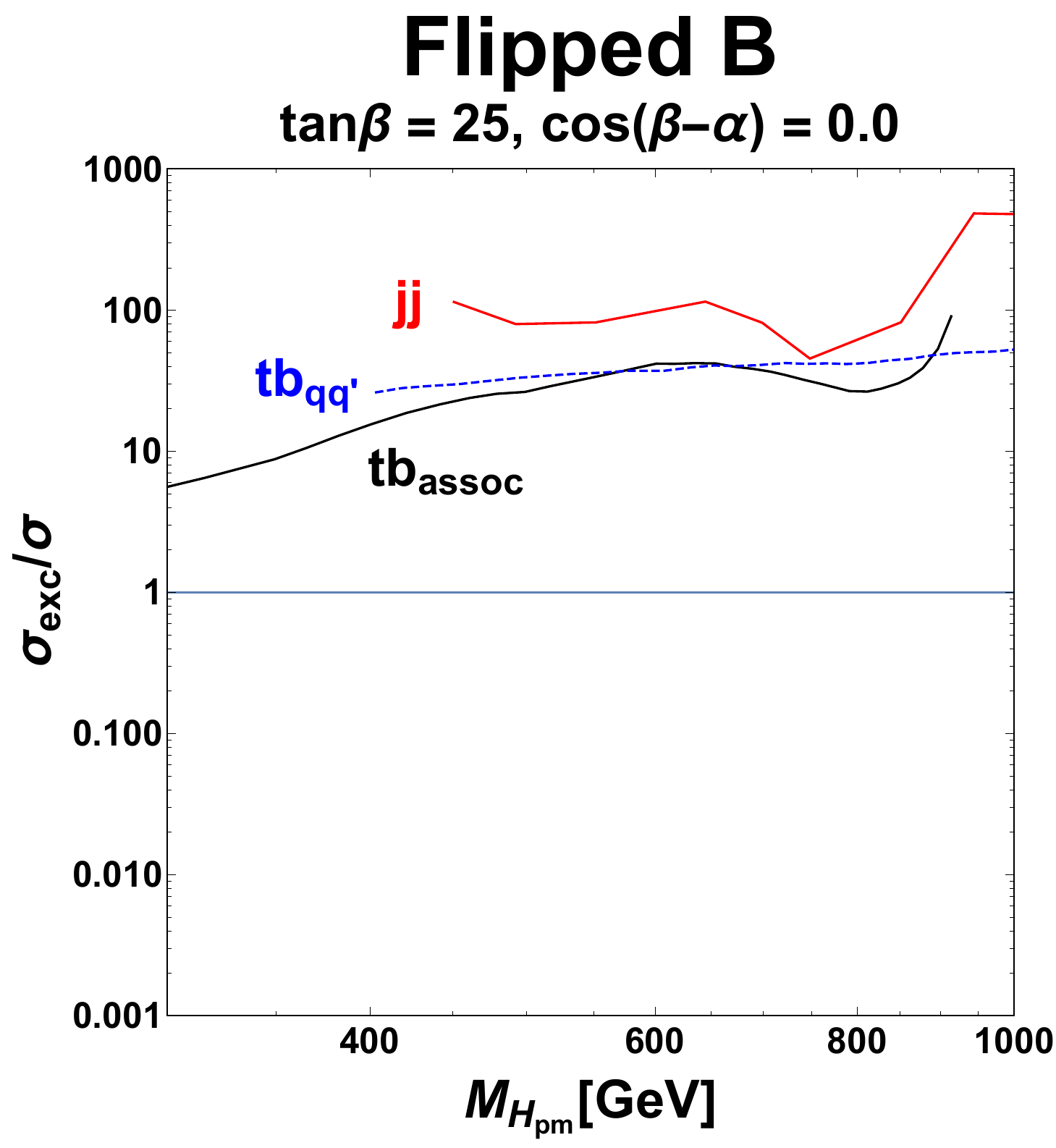}
 \caption{Exclusions of the charged Higgs for the low mass region (left) based on top decays and high mass regions (right) based on direct charged Higgs production as a function of the charged Higgs mass $m_{H^\pm}$ for $\tan\beta = 25$ and $\cos(\beta-\alpha) = 0$. Cross section ratios smaller than 1 are experimentally excluded.}
 \label{ExlusionChiggs}
 \end{center}
 \end{figure}

The constraints in this section are implemented with the same process we used in section~\ref{BiggsConst}. The strongest constraints come from 
\begin{itemize}
\item searches for light charged Higgs bosons that are produced from top decays and that decay into $cs$ (CMS 8\,TeV with 19.7\,fb$^{-1}$~\cite{Khachatryan:2015uua}), into $cb$ (CMS 8\,TeV with 19.7\,fb$^{-1}$~\cite{CMS:2016qoa}), or into $\tau\nu$ (CMS 8\,TeV with 19.7\,fb$^{-1}$~\cite{Khachatryan:2015qxa} and ATLAS 8\,TeV with 19.5\,fb$^{-1}$~\cite{Aad:2014kga});
\item searches for charged Higgs bosons produced in association with a top quark and decaying into $\tau\nu$ (CMS 8\,TeV with 19.7\,fb$^{-1}$~\cite{Khachatryan:2015qxa}, ATLAS 8\,TeV with 19.5\,fb$^{-1}$~\cite{Aad:2014kga}, and ATLAS 13\,TeV, 3.2\,fb$^{-1}$~\cite{Aaboud:2016dig});
\item searches for charged Higgs bosons produced in association with a top quark and decaying into $tb$ (ATLAS 8\,TeV with 20.3\,fb$^{-1}$~\cite{Aad:2015typ} and ATLAS 13\,TeV with 13.2\,fb$^{-1}$~\cite{ATLAS:2016qiq});
\item generic searches for low mass di-jet resonances (ATLAS 13\,TeV with 3.6 and 29.3\,fb$^{-1}$~\cite{Aaboud:2018fzt}).
\end{itemize}

For low mass charged Higgs at $\tan\beta = 25$, the type~2B and flipped~B models are ruled out due to $cq$ decays.
However, in the lepton-specific~B case the parameter space for charged Higgs bosons lighter than the top quark is still open, motivating continued search for charged Higgs bosons in top decays $t \to H^\pm b$. 
For $\tan\beta=25$ the high mass region is still largely unconstrained. For the flipped~B and type~2B models, searches for $H^\pm \to tb$ need to improve by approximately an order of magnitude to begin to probe the high mass region. The type~2B and lepton-specific~B models can also be probed by $H^\pm \to \tau \nu$ searches if their sensitivities improve one order or magnitude in the future.  

\section{Effects on flavor violating processes}\label{MesonPheno}

The flavor violating couplings of the neutral Higgs bosons also affect low energy flavor observables like meson mixing and rare meson decays. In the following we consider neutral $B$ meson, Kaon, and $D$ meson mixing as well as the branching ratios of several rare meson decays $B_s \to \mu^+\mu^-$, $B_s \to \tau \mu$, $B \to K \tau \mu$, $B \to K^\ast \tau \mu$, and $B_s \to \phi \tau \mu$. 

\subsection{Meson oscillations}

The SM Higgs, as well as the heavy scalar and pseudoscalar Higgs add contributions to neutral $B$~meson mixing at tree level. For the new physics contribution to the $B_s$ mixing amplitude normalized to the SM amplitude we have~\cite{Altmannshofer:2017uvs}
\begin{eqnarray} \label{eq:Bmix}
 \frac{M_{12}^{\text{NP}}}{M_{12}^{\text{SM}}}&=&\frac{m_{B_s}^2}{s_\beta^2 c_\beta^2} \Bigg( \frac{16\pi^2}{g_2^2}\Bigg)\frac{1}{S_0} \Bigg[ 2 X_4\Bigg(  \frac{c_{\beta-\alpha}^2}{m_h^2}+\frac{s_{\beta-\alpha}^2}{m_H^2}+\frac{1}{m_A^2}
 \Bigg)\frac{m_{bs}'^\ast m_{sb}'}{m_b^2 (V_{tb} V^\ast_{ts})^2}\nonumber \\
 &+&(X_2+X_3)\Bigg(  \frac{c_{\beta-\alpha}^2}{m_h^2}+\frac{s_{\beta-\alpha}^2}{m_H^2}-\frac{1}{m_A^2}\Bigg( \frac{(m_{bs}'^\ast)^2+ (m_{sb}')^2}{m_b^2(V_{tb}^\ast V_{ts})^2}  \Bigg)\Bigg] ~,
\end{eqnarray}
where $S_0 \simeq 2.3$ is a SM loop function.
The corresponding expression for the $B_d$ mixing amplitude is analogous. 
Note that this expression holds for all four flavorful 2HDMs.
The $X_i$ factors in Eq.~(\ref{eq:Bmix}) contain leading order QCD running corrections and ratios of hadronic matrix elements $X_2 = -0.47(-0.47)$, $X_3=-0.005(-0.005)$, $X_4 = 0.99(1.03)$, see~\cite{Altmannshofer:2017uvs}. 
The first value listed corresponds to $B_s$ and the second to $B_d$. From the above new physics contribution we can find values for the meson oscillation frequencies as well as the mixing phases
\begin{equation}
 \Delta M_q = \Delta M_q^{\text{SM}} \times \Bigg| 1+ \frac{M_{12}^{\text{NP}}}{M_{12}^{\text{SM}}} \Bigg| ~,~~~~ \phi_q = \phi_q^\text{SM} + \text{Arg}\Bigg( 1+ \frac{M_{12}^{\text{NP}}}{M_{12}^{\text{SM}}} \Bigg) ~. 
\end{equation}
We confront our models with experimental constraints by constructing a $\chi^2$ function that includes the mass differences and mixing phases in $B_s$ and $B_d$ mixing. The SM predictions and experimental results are taken from~\cite{Altmannshofer:2017uvs} (see also~\cite{DiLuzio:2017fdq} for a recent discussion of $B_s$ mixing constraints).
Note that in our models the $m_{sb}'$ and $m_{db}'$ mass parameters are largely fixed by the CKM matrix, see Eqs.~(\ref{downMassParameters2f}) and~(\ref{downMassParameters2g}). Thus we use the $B$ mixing observables to constrain the free $m_{bs}^\prime$ and $m_{bd}^\prime$ mass parameters, setting $m_{sb}' = \pm V_{ts}^* m_b$ and $m_{db}' = \pm V_{td}^* m_b$ (with the signs depending on the type of flavorful model).

\begin{figure}
\begin{center} 
  \includegraphics[width=0.47\textwidth]{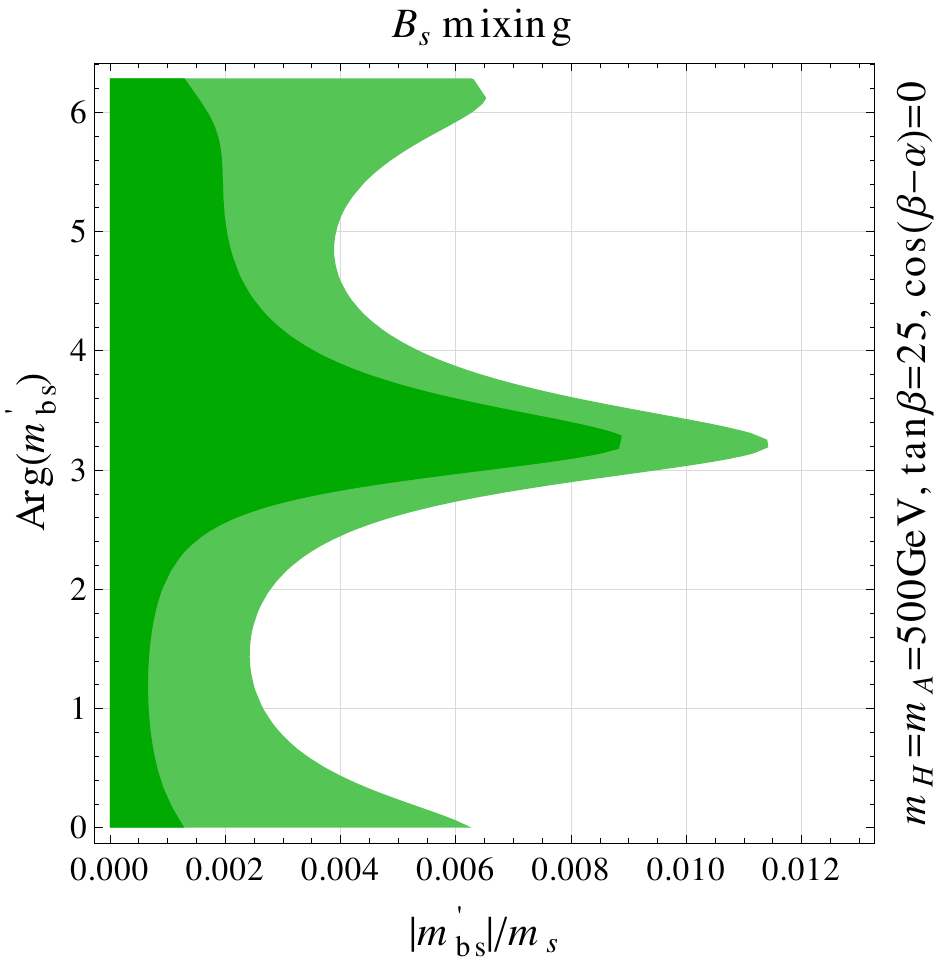} ~~~~~
  \includegraphics[width=0.47\textwidth]{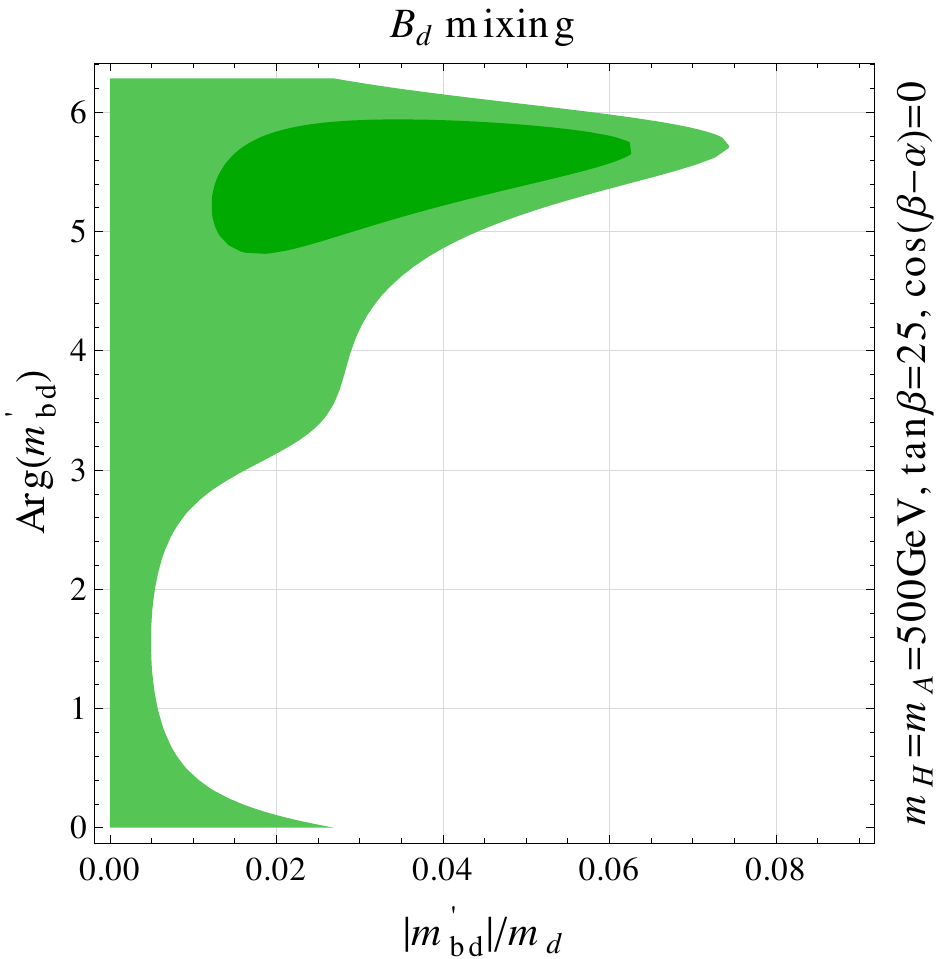}
  \caption{Meson mixing constraints on the mass parameters $m_{bs}'$ (left) and $m_{bd}'$ (right). The 1$\sigma$ and 2$\sigma$ allowed regions are shaded in green. We set $\cos(\beta - \alpha) = 0$, $\tan\beta = 25$, and $m_H = m_A = 500$~GeV. The shown regions correspond to the type~1B and lepton-specific~B models. In the type~2B and flipped~B models the allowed regions are shifted in phase by Arg$(m_{bq}^\prime) \to$ Arg$(m_{bq}^\prime) + \pi$.}
\label{fig:Bmix}
\end{center}
\end{figure}

In Fig.~\ref{fig:Bmix} we show constraints on the absolute values and phases of $m_{bs}^\prime$ (left) and $m_{bd}^\prime$ (right) for a benchmark scenario with $\cos(\beta-\alpha) = 0$ (as favored by the Higgs signal strengths measurements, see section~\ref{LHP}), $\tan\beta = 25$, and  $m_H = m_A = 500$~GeV.
The constraints on the $m'$ parameter scale approximately as $m_A^2/\tan\beta^2$, i.e. they become weaker for larger Higgs masses and stronger for larger $\tan\beta$. 
The shown constraints hold in the type~1B and lepton-specific~B models. In the type~2B and flipped~B models, the $m_{sb}'$ and $m_{db}'$ mass parameters have the opposite sign. This results in constraints that are shifted in phase by Arg$(m_{bq}^\prime) \to$ Arg$(m_{bq}^\prime) + \pi$.

We observe that both $m_{bd}^\prime$ and $m_{bs}^\prime$ are strongly constrained by $B_d$ and $B_s$ mixing for large $\tan\beta$ and for heavy Higgs bosons below $\sim 1$~TeV. The fact that these mass parameters have to be much smaller than the generic prediction of our flavor textures, $m_{bd}^\prime \sim m_d$ and $m_{bs}^\prime \sim m_s$ might call for an underlying flavor model.

Similarly to $B$ meson mixing, also the Kaon mixing amplitude obtains additional contributions. The new physics amplitude is
\begin{eqnarray}
 M_{12}^{\text{NP}} &=& m_K^3 \frac{f_K^2}{v^2}\frac{1}{s_\beta^2 c_\beta^2}\Bigg[\frac{1}{4}B_4^K \eta_4^K \Bigg( \frac{c_{\beta-\alpha}^2}{m_h^2} +\frac{s_{\beta-\alpha}^2}{m_H^2}+\frac{1}{m_A^2}\Bigg)\frac{m_{sd}'^\ast m_{ds}'}{m_s^2}\nonumber \\
 &-&  \Bigg( \frac{5}{48} B_2^K\eta_2^K-\frac{1}{48} B_3^K\eta_3^K \Bigg)\Bigg( \frac{c_{\beta-\alpha}^2}{m_h^2} +\frac{s_{\beta-\alpha}^2}{m_H^2}-\frac{1}{m_A^2}\Bigg) \frac{(m_{sd}'^\ast)^2+ (m_{ds}')^2}{m_s^2} \Bigg] ~,
 \label{KaonMixing}
\end{eqnarray}
with the Kaon decay constant $f_K \simeq 155.4$~MeV~\cite{Dowdall:2013rya}. The bag parameters $B_2^K \simeq 0.46$, $B_3^K \simeq 0.79$, $B_4^K \simeq 0.78$ are taken from~\cite{Carrasco:2015pra} (see also~\cite{Jang:2015sla,Garron:2016mva}). The parameters $\eta_2^K \simeq 0.68$, $\eta_3^K \simeq -0.03$, and $\eta_4^K = 1$ (see~\cite{Altmannshofer:2017uvs}) encode one loop renormalization group effects.

The relevant observables in Kaon mixing are the mass difference $\Delta M_K$ and the CP violating parameter $\epsilon_K$. They can be calculated via 
\begin{equation}
 \Delta M_K = \Delta M_K^{\text{SM}}+2 \text{Re}(M_{12}^{\text{NP}}) ~,~~~ \epsilon_K = \epsilon_K^{\text{SM}}+\kappa_\epsilon \frac{\text{Im}(M_{12}^{\text{NP}})}{\sqrt{2}\Delta M_K} ~.
\end{equation}
with $\kappa_\epsilon = 0.94$~\cite{Buras:2010pza}.
In Eqs.~(\ref{downMassParameters1}) and~(\ref{downMassParameters1b}) we saw that the $m'$ parameters that are responsible for Kaon mixing are not independent parameters but given in terms of the parameters that govern $B_s$ and $B_d$ mixing. Given the constraints from $B_s$ and $B_d$ mixing, we find that new physics effects in Kaon mixing are generically below the current bounds.
In particular, we find that new physics effects in $\Delta M_K$ are at most at the permille level, while effects in $\epsilon_K$ are $\lesssim 10\%$.

Analogously to Kaon mixing, the new physics contributions to neutral $D$ meson mixing are given by
\begin{eqnarray}
 M_{12}^{D} &=& m_D^3 \frac{f_D^2}{v^2}\frac{1}{s_{\beta}^2 c_{\beta}^2}\Bigg[ \frac{1}{4} B_4^D \eta_4^D  \Bigg(\frac{c_{\beta-\alpha}^2}{m_h^2} +\frac{s_{\beta-\alpha}^2}{m_H^2}+\frac{1}{m_A^2} \Bigg)\frac{m_{cu}'^\ast m_{uc}'}{m_c^2}\nonumber \\
 &-&\Bigg( \frac{5}{48}B_2^D \eta_2^D -\frac{1}{48}B_3^D \eta_3^D \Bigg)\Bigg(\frac{c_{\beta-\alpha}^2}{m_h^2} +\frac{s_{\beta-\alpha}^2}{m_H^2}-\frac{1}{m_A^2} \Bigg)\frac{(m_{cu}'^\ast)^2+(m_{uc}')^2}{m_c^2} \Bigg] ~. 
\end{eqnarray}
According to Eqs.~(\ref{upMassParametersb}) and~(\ref{upMassParametersc}), the $m_{cu}'$ and $m_{uc}'$ parameters are strongly suppressed, generically of the order of $m_u m_c / m_t$. We find that the resulting new physics contributions to the mixing amplitude are many orders of magnitude below the current sensitivities~\cite{Amhis:2016xyh} in all the models we consider.

\subsection{The rare \texorpdfstring{$B_s \to \mu^+\mu^-$}{Bs --> mu+ mu-} decay}

The rare FCNC decay $B_s \to \mu^+\mu^-$ is known to be a highly sensitive probe of new physics (see e.g.~\cite{Altmannshofer:2017wqy}).
The decay has been observed at the LHC~\cite{CMS:2014xfa} and the latest experimental result for the time integrated branching ratio from LHCb~\cite{Aaij:2017vad}
\begin{equation}
 \text{BR}(B_s \to \mu^+\mu^-)_\text{LHCb} = (3.0 \pm 0.6^{+0.3}_{-0.2}) \times 10^{-9} ~,
\end{equation}
agrees well with the SM prediction~\cite{Bobeth:2013uxa}
\begin{equation}
 \text{BR}(B_s \to \mu^+\mu^-)_\text{SM} = (3.65 \pm 0.23) \times 10^{-9} ~.
\end{equation}
A generic expression for the branching ratio in presence of NP reads~\cite{DeBruyn:2012wk,Altmannshofer:2012az}
\begin{equation} \label{eq:Bsmumu}
 \frac{\text{BR}(B_s \to \mu^+\mu^-)}{\text{BR}(B_s \to \mu^+\mu^-)_\text{SM}} = \left(|S_{\mu\mu}|^2 + |P_{\mu\mu}|^2 \right) \left( \frac{1}{1+y_s} + \frac{y_s}{1+y_s} \frac{\text{Re}(P_{\mu\mu}^2)-\text{Re}(S_{\mu\mu}^2)}{|S_{\mu\mu}|^2 + |P_{\mu\mu}|^2} \right) ~,
\end{equation}
where $y_s$ is the life-time difference of the $B_s$ mesons, $y_s = (6.1 \pm 0.7)\% $~\cite{Aaij:2014zsa}.
In the above expression we do not consider corrections due to a possible non-standard $B_s$ mixing phase $\phi_s$~\cite{Buras:2013uqa}. Given the existing constraint on $\phi_s$~\cite{Amhis:2016xyh}, such corrections to the branching ratio are negligible.

In the SM, the coefficients $P_{\mu\mu}^\text{SM} = 1$ and $S_{\mu\mu}^\text{SM} = 0$.
Corrections due to tree level exchange of the neutral Higgs bosons are collected in the appendix~\ref{appendix}.
As $B_s$ meson mixing puts strong constraints on $m_{bs}'$ we will set it to zero in the following discussion.
In the alignment limit and for $m_H = m_A$, as well as neglecting the life time difference, the expression for BR$(B_s \to \mu^+\mu^-)$ simplifies to
\begin{equation}
 \frac{\text{BR}(B_s \to \mu^+\mu^-)}{\text{BR}(B_s \to \mu^+\mu^-)_\text{SM}} = \left| 1 \pm \frac{1}{C_{10}^\text{SM}} \left(\frac{4\pi^2}{e^2} \right) \frac{m_{B_s}^2}{m_A^2} t_\beta^2 \frac{m_{\mu\mu}'^*}{m_\mu} \right|^2 + \left| \frac{1}{C_{10}^\text{SM}} \left(\frac{4\pi^2}{e^2} \right) \frac{m_{B_s}^2}{m_A^2} t_\beta^2 \frac{m_{\mu\mu}'^*}{m_\mu} \right|^2 ~,
\end{equation}
with the SM Wilson coefficient $C_{10}^\text{SM} \simeq - 4.1$.
The plus (minus) sign in the first term holds in the type~1B and the lepton-specific~B models (type~2B and flipped~B models). Note that the $m_{\mu\mu}'$ parameter is approximately given by $m_\mu$ in the type~1B and flipped~B models. In the type~2B and lepton-specific~B models, $m_{\mu\mu}'$ is a free parameter of $O(m_\mu^2/m_\tau)$.
Consequently, we expect much more stringent constraints in the type~1B and flipped~B models as compared to the type~2B and lepton-specific~B models.

\begin{figure}
\begin{center} 
  \includegraphics[width=0.47\textwidth]{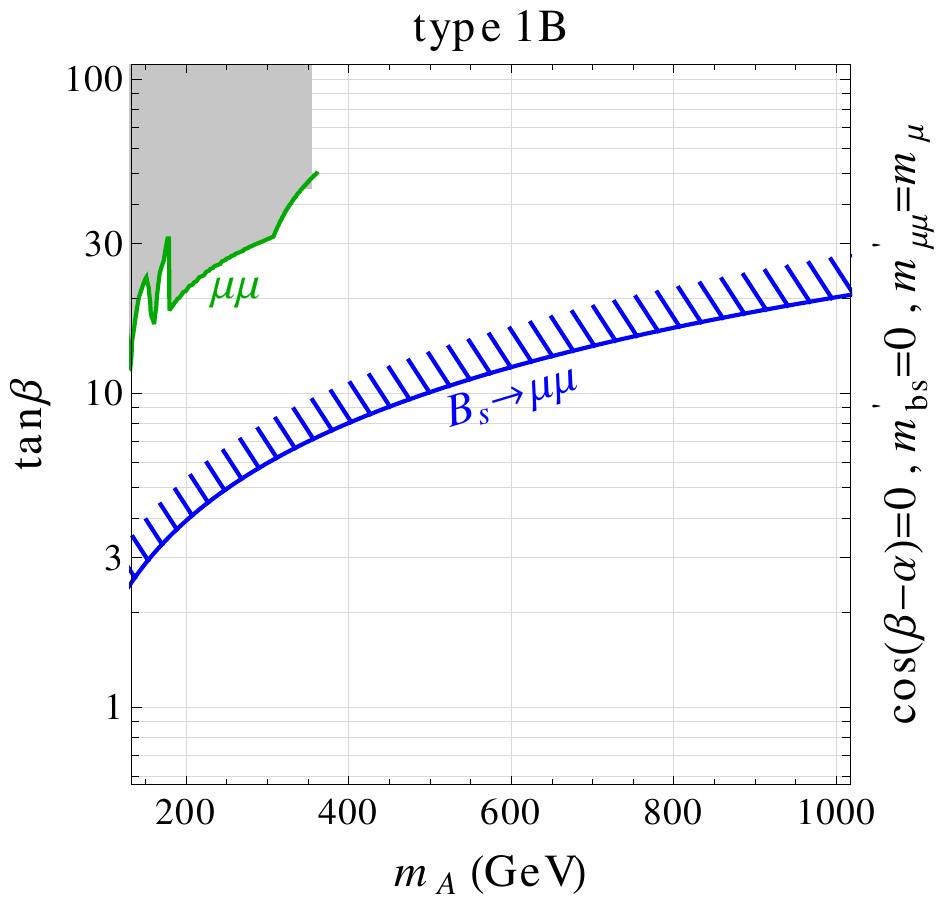} ~~~~~
  \includegraphics[width=0.47\textwidth]{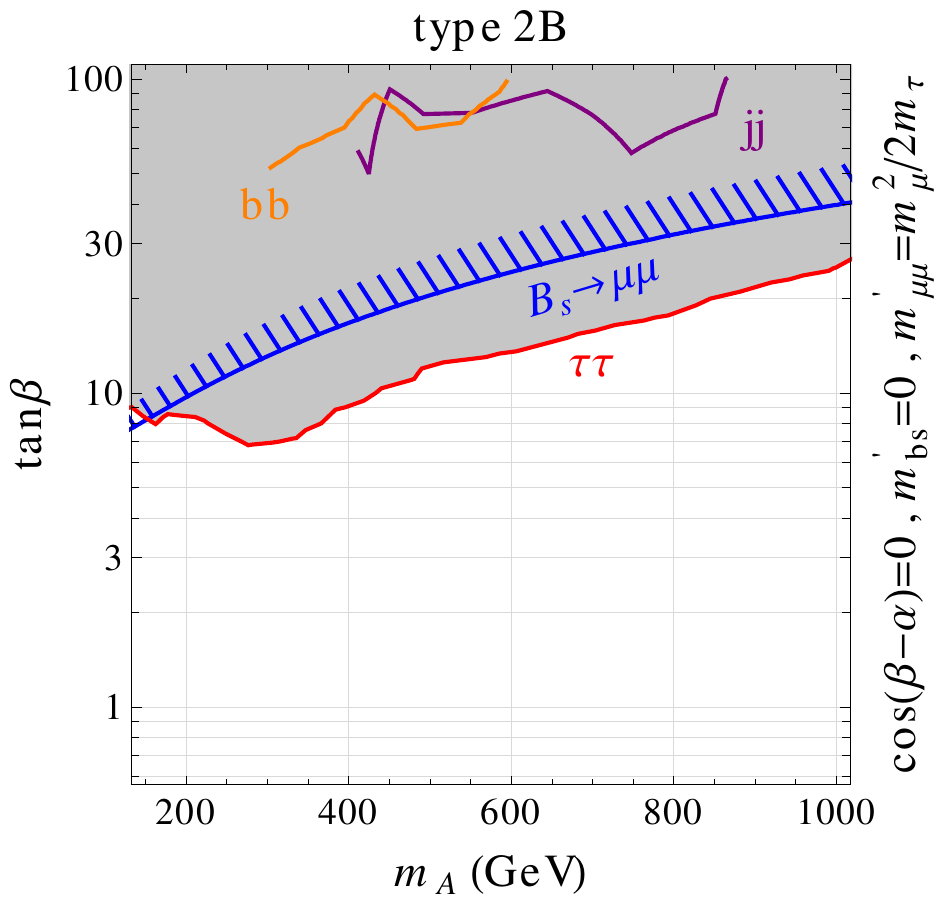}\\[16pt]
  \includegraphics[width=0.47\textwidth]{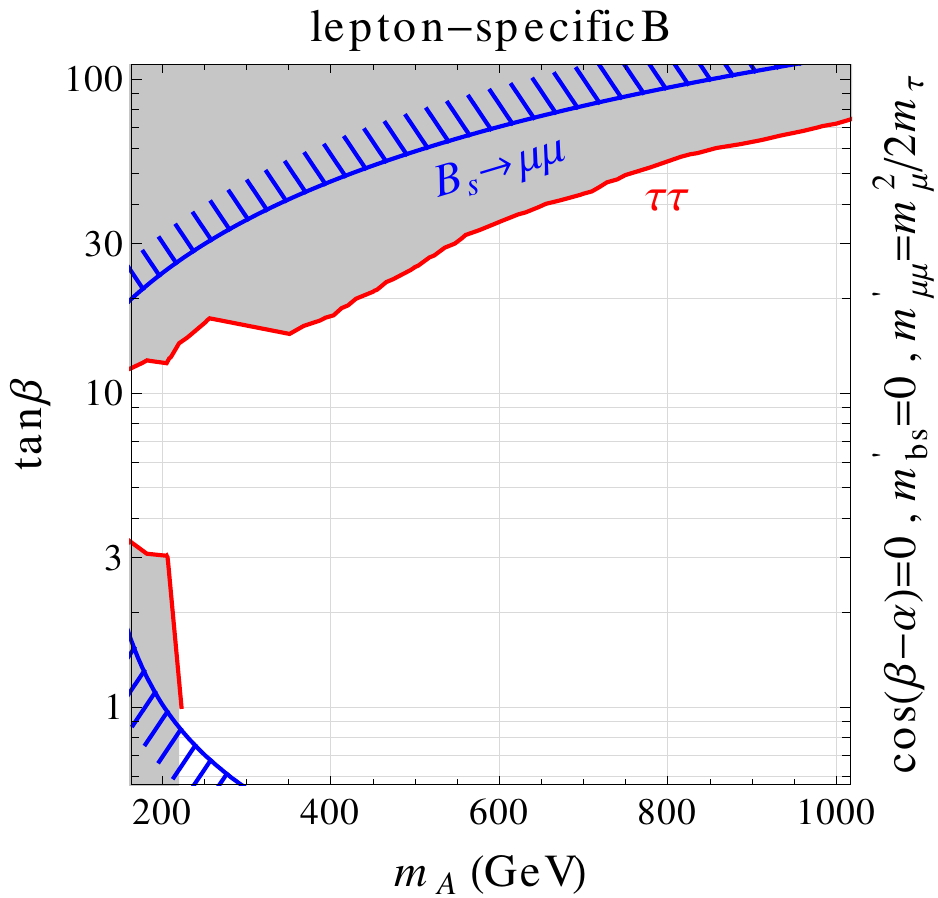} ~~~~~
  \includegraphics[width=0.47\textwidth]{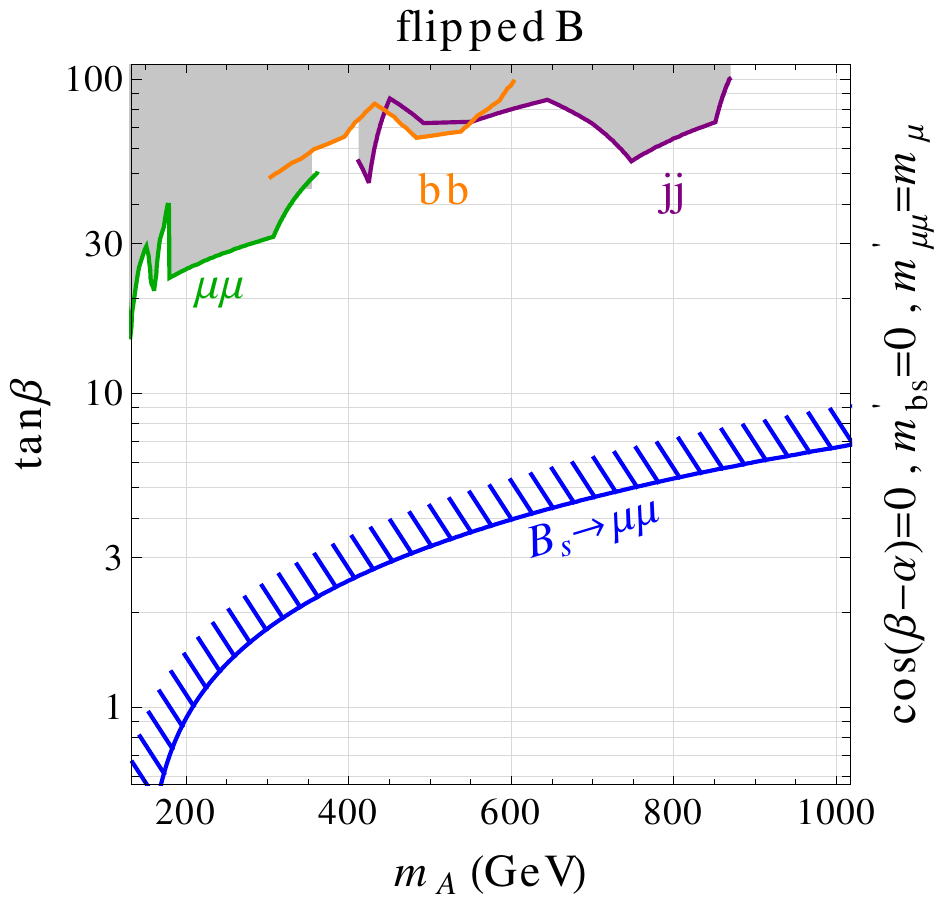} 
\caption{Constraints in the $m_H = m_A$ vs. $\tan\beta$ plane from $B_s \to\mu^+\mu^-$ for benchmark scenarios in the four flavorful models. The regions above the blue hatched contour are excluded by $B_s \to\mu^+\mu^-$ at the $2\sigma$ level. For comparison the region excluded by direct searches for the heavy neutral Higgs bosons is shaded in gray. We show searches in the $\tau^+\tau^-$ channel (red), $\mu^+\mu^-$ channel (green), $b \bar b$ channel (orange), and di-jet channel (purple).}
\label{BstomumuConst}
\end{center}
\end{figure}

In Fig.~\ref{BstomumuConst} we show constraints in the plane of heavy Higgs mass $m_H = m_A$ vs. $\tan\beta$ from $B_s \to\mu^+\mu^-$ in the four flavorful models. In all four models we set $\cos(\beta-\alpha) = 0$ and $m_{bs}' = 0$. In the type~1B and flipped~B models we set the (small) higher order corrections to $m_{\mu\mu}'$ to zero, i.e. $m_{\mu\mu}' = m_\mu$. In the type~2B and lepton-specific~B models we set $m_{\mu\mu}' = + m_\mu^2/2m_\tau$.

The constraints in the type~2B and lepton-specific~B models depend strongly on the choice of $m_{\mu\mu}'$. If $m_{\mu\mu}'$ accidentally vanishes, the $B_s \to\mu^+\mu^-$ constraint can be even completely avoided in these models.
The bounds in the type~1B and flipped~B models, however, are robust. The higher order corrections to $m_{\mu\mu}'$ modify them typically by $10\%$ or less. 
In these models, the shown bounds from $B_s \to\mu^+\mu^-$ can only be avoided by postulating that the CKM matrix is generated in the up-sector.

In comparison to the constraints from direct searches we observe that $B_s \to\mu^+\mu^-$ gives stronger bounds in the type~1B and flipped~B models. In the type~2B and lepton-specific models, the direct searches in the $\tau^+\tau^-$ channel tend to be more constraining, instead. 
  
\subsection{Lepton flavor violating B meson decays}

In the SM, the lepton flavor violating decays based on the $b \to s \tau \mu$ transition are suppressed by the tiny neutrino masses and are far below any imaginable experimental sensitivities. Observation of these decays would be clear sign of new physics. In our setup, tree level exchange of neutral Higgs bosons can induce these decays at levels that might become experimentally accessible.

Similarly to the lepton flavor conserving decay $B_s \to \mu^+\mu^-$ we express the branching ratio of the two body decay $B_s \to \tau^+\mu^-$ as
\begin{eqnarray} \label{eq:Bstaumu}
 \frac{\text{BR}(B_s \to \tau^+\mu^-)}{\text{BR}(B_s \to \mu^+\mu^-)_\text{SM}} &=& \left( 1 - \frac{m^2_\tau}{m_{B_s}^2} \right)^2 \left(|S_{\tau\mu}|^2 + |P_{\tau\mu}|^2 \right) \nonumber \\
 && \qquad \times \left( \frac{1}{1+y_s} + \frac{y_s}{1+y_s} \frac{\text{Re}(P_{\tau\mu}^2)-\text{Re}(S_{\tau\mu}^2)}{|S_{\tau\mu}|^2 + |P_{\tau\mu}|^2} \right) ~,
\end{eqnarray}
where the last line takes into account the effect of a non-zero life time difference in the $B_s$ system.
An analogous expression holds for the decay $B_s \to \mu^+\tau^-$. We will use the notation $B_s \to \tau\mu = B_s \to \tau^+\mu^- + B_s \to \mu^+\tau^-$. 
The expressions for the coefficients $P_{\tau\mu}$ and $S_{\tau\mu}$ are collected in the appendix~\ref{appendix}.

As in our discussion of the $B_s \to \mu^+\mu^-$ decay, we set $m_{bs}' = 0$, $\cos(\beta-\alpha) = 0$, $m_H = m_A$, and neglect the life time difference. In this case we find
\begin{equation}
 \frac{\text{BR}(B_s \to \tau\mu)}{\text{BR}(B_s \to \mu\mu)_\text{SM}} = \left( 1 - \frac{m^2_\tau}{m_{B_s}^2} \right)^2 \frac{1}{|C_{10}^\text{SM}|^2} \left( \frac{4\pi^2}{e^2} \right)^2 \frac{m_{B_s}^4}{m_A^4} t^4_\beta \left( \frac{|m_{\mu\tau}^\prime|^2}{m_\mu^2} + \frac{|m_{\tau\mu}^\prime|^2}{m_\mu^2}  \right) ~.
\end{equation}
This expression holds in all four flavorful 2HDMs.
For all types we have $|m_{\tau\mu}^\prime| \sim |m_{\mu\tau}^\prime| \sim m_\mu$. 
In the type~1B and the flipped~B models, the possible values for BR$(B_s \to \tau\mu)$ are bounded by the measured BR$(B_s \to \mu^+\mu^-)$. Considering $|m_{\tau\mu}^\prime|, |m_{\mu\tau}^\prime| < 3 m_\mu$ and $250$~GeV $< m_A = m_H < 1$~TeV, we find the following upper bounds
\begin{equation} \label{eq:bounds}
 \text{BR}(B_s \to \tau\mu) \lesssim 
 \begin{cases}
 1.5 \times 10^{-7} ~~~~~~~~~~~ \text{type~1B} \\
 4.0 \times 10^{-9} ~~~~~~~~~~~ \text{flipped~B} \\
 \end{cases}
\end{equation}
Note that the given upper limits depend on the ranges of the $m^\prime$ parameters that we have chosen and that we believe to be a representative example of the Yukawa structures that we consider in this work. For example, allowing $|m^\prime_{\mu \tau}|$ and $|m^\prime_{\tau \mu}|$ to be as large as $5 m_\mu$ would result in branching ratios that are larger by almost a factor of 3 compared to the bounds quoted in Eq. (\ref{eq:bounds}).

In the type~2B and lepton-specific~B models, the constraint from $B_s \to \tau\mu$ is much weaker. In those models the strongest constraint comes from direct searches for the heavy Higgs bosons in the $\tau^+\tau^-$ channel (see Fig.~\ref{BstomumuConst}). Values of BR$(B_s \to \tau\mu) \sim \text{few} \times 10^{-6}$ are possible in those models. 

Lepton flavor changing decays involving electrons on the other hand are tiny. Generically we expect in all models
\begin{eqnarray}
 \text{BR}(B_s \to \tau e) &\sim&  \frac{m_e^2}{m_\mu^2} \times \text{BR}(B_s \to \tau \mu) \sim 2 \times 10^{-5} \times \text{BR}(B_s \to \tau \mu) ~, \\
 \text{BR}(B_s \to \mu e) &\sim&  \frac{m_e^2}{m_\tau^2} \times \text{BR}(B_s \to \tau \mu) \sim 8 \times 10^{-8} \times \text{BR}(B_s \to \tau \mu)~.
\end{eqnarray}

In addition to the $B_s \to \tau \mu$ decay, tree level exchange of flavor violating Higgs bosons also leads to three body semi-leptonic $B$ meson decays like $B \to K \tau \mu$, $B \to K^* \tau \mu$, and $B_s \to \phi \tau \mu$.

We find that the $B \to K^* \tau \mu$ and $B_s \to \phi \tau \mu$ branching ratios are directly correlated to the $B_s \to \tau\mu$ branching ratio. Ignoring the life-time difference in the $B_s$ system and using the results from~\cite{Becirevic:2016oho} (see also~\cite{Crivellin:2015era} for a related study) we obtain for the differential branching ratio
\begin{eqnarray}
 \frac{d\text{BR}}{dq^2}(B \to K^* \tau^+\mu^-) &=& \frac{1}{16\pi^2} \lambda^{3/2}\left( 1 , \frac{q^2}{m_B^2} , \frac{m_{K^*}^2}{m_B^2} \right) \left( 1 - \frac{m_\tau^2}{q^2}\right)^2 \left( 1 - \frac{m^2_\tau}{m_{B_s}^2} \right)^{-2} \nonumber \\
 && \times \frac{q^2 A^2_0(q^2)}{m_B^2 f_{B_s}^2} \frac{\tau_{B} m_B^5}{\tau_{B_s} m_{B_s}^5} \times \text{BR}(B_s \to \tau\mu)~,
\end{eqnarray}
where $\lambda(a,b,c) = a^2 + b^2 + c^2 - 2(ab + ac + bc)$.
An analogous expression holds for $B_s \to \phi \tau \mu$.
For the $B_s$ meson decay constant we use $f_{B_s} \simeq 224$~MeV~\cite{Aoki:2016frl}. The $B \to K^*$ and $B_s \to \phi$ form factors $A_0$ are taken from~\cite{Straub:2015ica}. 
Integrating over $q^2$, we find
\begin{eqnarray}
 \text{BR}(B \to K^* \tau\mu) \simeq 2.9 \times 10^{-2} \times \text{BR}(B_s \to \tau\mu) ~, \\
 \text{BR}(B_s \to \phi \tau\mu) \simeq 3.3 \times 10^{-2} \times \text{BR}(B_s \to \tau\mu) ~.
 \label{BRkstartaumu}
\end{eqnarray}
Using the bounds and generic expectations for $B_s \to \tau\mu$ in the different flavorful models discussed above, we find that $\text{BR}(B \to K^* \tau \mu)$ and $\text{BR}(B_s \to \phi \tau\mu)$ can be at most few $\times 10^{-9}$ in the type~1B model and $\sim 10^{-10}$ in the flipped~B model, respectively. In the type~2B and lepton-specific~B models, however, these branching ratios can be as large as $\sim 10^{-7}$. 

We find similar results also for the $B \to K \tau \mu$ decay. The fact that $B \to K$ is a pseudoscalar to pseudoscalar transition, while $B \to K^*$ and $B_s \to \phi$ are pseudoscalar to vector transitions has little impact numerically. We find that $\text{BR}(B \to K \tau \mu)$ can be as large as few $\times 10^{-9}$ in the type~1B model, $\sim 10^{-10}$ in the flipped~B model, and $\sim 10^{-7}$ in the type~2B and lepton-specific~B models\footnote{Also baryonic decays $\Lambda_b \to \Lambda \tau \mu$ can arise. While a detailed discussion of baryonic decays is beyond the scope of this work, we generically expect similar results.}.

\section{Conclusions} \label{sec:conclusions}

Little is known experimentally about the tiny couplings of the Higgs boson to the light flavors of quarks and leptons. It is thus interesting to study possible alternative origins of mass for the light flavors beyond the 125~GeV Higgs boson.
As an example, we analyzed a particular class of 2HDMs with non-trivial flavor structure.  
In analogy to the four, well studied 2HDMs with natural flavor conservation (NFC), we identified four models that preserve an approximate $U(2)^5$ flavor symmetry acting on the first two generations. We refer to them as type~1B, type~2B, lepton-specific~B, and flipped~B. In these flavorful 2HDMs, interesting flavor violating phenomena involving the third generation of fermions can be expected, while the $U(2)^5$ flavor symmetry still protects flavor violating transitions between the first and second generations. 

We studied the production and decay modes of the neutral and charged Higgs bosons of the models, as well as various low energy flavor violating observables, and identified the signatures of the flavorful models that are qualitatively different from the models with NFC. 

With regards to the collider phenomenology we find:
\begin{itemize}
 \item Measurements of Higgs signal strengths give important constraints on the mixing between the two CP-even Higgs bosons, $h$ and $H$. In the type~2, lepton-specific, and flipped models, the constraints are very similar for the models with NFC and our flavorful models. In the type~1 models the constraints are markedly different due to large modifications of the charm and muon couplings in the type~1B model.
 \item The main heavy Higgs production and decay modes in the type~2B and flipped~B models are similar to those in their counterparts with NFC. The highest sensitivity to the type~2B model is achieved in searches for high mass $\tau^+\tau^-$ resonances. The flipped~B model is largely unconstrained at hadron colliders. The most promising search channels are $\mu^+ \mu^-$, $b \bar b$, and di-jet resonances depending on the mass range.
 \item In the lepton-specific~B model, the production of the heavy neutral and charged Higgs bosons at large $\tan\beta$ is much larger than in the corresponding model with NFC. This opens up the possibility to directly probe the large $\tan\beta$ regime of the lepton-specific~B model at hadron colliders in the $\tau^+\tau^-$ channel.
 \item In all flavorful models, the neutral Higgses can have sizable flavor violating branching ratios. In particular, we find that at large $\tan\beta$ typically BR$(H\to tc) \sim 10\%$. Furthermore, BR$(H\to \tau\mu) \sim 0.1\% - 1\%$. These flavor violating branching ratios depend on unknown model parameters and can vary by a factor of few.
\end{itemize}
The most interesting features in the flavor phenomenology are:
\begin{itemize}
 \item In all four flavorful models we find strong constraints from $B_s$ and $B_d$ meson mixing. We find that in the large $\tan\beta$ regime the relevant entries in the down quark mass matrices $m_{bs}^\prime$, and $m_{bd}^\prime$ have to be considerably smaller than their nominal values $m_{bs}^\prime \sim m_s$ and $m_{bd}^\prime \sim m_d$. This might call for an underlying flavor model.
 \item Under the assumption that the CKM matrix is generated in the down sector, the measured value of BR$(B_s \to \mu^+\mu^-)$ gives strong constraints in the $m_A$ vs. $\tan\beta$ parameter space of the type~1B and flipped~B models. In the type~2B and lepton-specific~B models, this constraint is much weaker and can be completely avoided.  
 \item Lepton flavor violating rare $B$ meson decays might be at an experimentally accessible level. In particular, in the type~2B and lepton-specific~B models, BR$(B_s \to \tau\mu)$ could be as large as few$\times 10^{-6}$ while BR$(B \to K^{(*)}\tau\mu)$ and BR$(B_s \to \phi\tau\mu)$ could be as large as $10^{-7}$, potentially in reach of LHCb. Lepton flavor violating decay modes with electrons are predicted to be orders of magnitude smaller. 
\end{itemize}

\section*{Acknowledgements}

We thank Joshua Eby, Stefania Gori, and Douglas Tuckler for useful discussions. We thank Daniil Volkov for pointing out a typo in equation~\eqref{eq:raretop} in previous versions of this paper.
The research of WA is supported by the National Science Foundation under Grant No. PHY-1720252.
WA thanks KITP for hospitality during final stages of this work and acknowledges support by the National Science Foundation under Grant No. NSF PHY11-25915.

\begin{appendix}
\section{New physics contributions to rare meson decays} \label{appendix}

The parameters $S_{\ell\ell'}$ and $P_{\ell\ell'}$ that enter the expressions for the rare $B$ meson branching ratios in Eqs.~(\ref{eq:Bsmumu}) and~(\ref{eq:Bstaumu}) get in general contributions from tree level $h$, $H$ and $A$ exchange
\begin{eqnarray}
 S_{\ell\ell'} &=& \frac{1}{C_{10}^\text{SM}} \left( \frac{4\pi^2}{e^2} \right) \left( S_{\ell\ell'}^h + S_{\ell\ell'}^H + S_{\ell\ell'}^A \right) ~, \\
P_{\ell\ell'} &=& 1 + \frac{1}{C_{10}^\text{SM}} \left( \frac{4\pi^2}{e^2} \right) \left( P_{\ell\ell'}^h + P_{\ell\ell'}^H + P_{\ell\ell'}^A \right)~,
\end{eqnarray}
with the SM Wilson coefficient $C_{10}^\text{SM} \simeq - 4.1$.
In the flavor conserving case $\ell\ell' = \mu\mu$ we find
\begin{eqnarray}
 S^h_{\mu\mu} &=& - \frac{m_{B_s}^2}{m_h^2} \left( \frac{c_\alpha}{s_\beta} - \frac{\text{Re}(m_{\mu\mu}^\prime)}{m_\mu} \frac{c_{\beta-\alpha}}{s_\beta c_\beta} \right) \frac{c_{\beta-\alpha}}{s_\beta c_\beta} \left( \frac{m_{sb}^\prime - m_{bs}^\prime {}^*}{m_b V_{tb} V_{ts}^*} \right) ~, \\
 S^H_{\mu\mu} &=& \frac{m_{B_s}^2}{m_H^2} \left( \frac{s_\alpha}{s_\beta} + \frac{\text{Re}(m_{\mu\mu}^\prime)}{m_\mu} \frac{s_{\beta-\alpha}}{s_\beta c_\beta} \right) \frac{s_{\beta-\alpha}}{s_\beta c_\beta}\left( \frac{m_{sb}^\prime - m_{bs}^\prime {}^*}{m_b V_{tb} V_{ts}^*} \right) ~, \\
 S^A_{\mu\mu} &=& - \frac{m_{B_s}^2}{m_A^2}  \frac{1}{s_\beta^2 c_\beta^2} \frac{i \text{Im}(m_{\mu\mu}^\prime)}{m_\mu} \left( \frac{m_{sb}^\prime + m_{bs}^\prime {}^*}{m_b V_{tb} V_{ts}^*} \right) ~,  \\
 P^h_{\mu\mu} &=& \frac{m_{B_s}^2}{m_h^2}  \frac{c_{\beta-\alpha}^2}{s_\beta^2 c_\beta^2} \frac{i \text{Im}(m_{\mu\mu}^\prime)}{m_\mu} \left( \frac{m_{sb}^\prime - m_{bs}^\prime {}^*}{m_b V_{tb} V_{ts}^*} \right) ~, \\
 P^H_{\mu\mu} &=& \frac{m_{B_s}^2}{m_H^2}  \frac{s_{\beta-\alpha}^2}{s_\beta^2 c_\beta^2} \frac{i \text{Im}(m_{\mu\mu}^\prime)}{m_\mu} \left( \frac{m_{sb}^\prime - m_{bs}^\prime {}^*}{m_b V_{tb} V_{ts}^*} \right) ~, \\
 P^A_{\mu\mu} &=& \frac{m_{B_s}^2}{m_A^2} \left( \frac{1}{t_\beta} - \frac{\text{Re}(m_{\mu\mu}^\prime)}{m_\mu} \frac{1}{s_\beta c_\beta} \right) \frac{1}{s_\beta c_\beta}\left( \frac{m_{sb}^\prime + m_{bs}^\prime {}^*}{m_b V_{tb} V_{ts}^*} \right) ~.
\end{eqnarray}

\noindent
In the flavor violating cases $\ell\ell' = \mu\tau, \tau\mu$ we find instead
\begin{eqnarray}
 S_{\ell\ell'}^h &=& \frac{m_{B_s}^2}{2 m_h^2} \frac{c_{\beta-\alpha}^2}{c_\beta^2 s_\beta^2} \left( \frac{m_{bs}^\prime {}^* - m_{sb}^\prime}{m_b V_{tb} V_{ts}^*} \right) \left( \frac{m_{\ell\ell'}^\prime {}^* + m_{\ell'\ell}^\prime}{m_\mu}\right) ~, \\
S_{\ell\ell'}^H &=& \frac{m_{B_s}^2}{2 m_H^2} \frac{s_{\beta-\alpha}^2}{c_\beta^2 s_\beta^2} \left( \frac{m_{bs}^\prime {}^* - m_{sb}^\prime}{m_b V_{tb} V_{ts}^*} \right) \left( \frac{m_{\ell\ell'}^\prime {}^* + m_{\ell'\ell}^\prime}{m_\mu}\right) ~, \\
S_{\ell\ell'}^A &=& \frac{m_{B_s}^2}{2 m_A^2} \frac{1}{c_\beta^2 s_\beta^2} \left( \frac{m_{bs}^\prime {}^* + m_{sb}^\prime}{m_b V_{tb} V_{ts}^*} \right) \left( \frac{m_{\ell'\ell}^\prime - m_{\ell\ell'}^\prime {}^*}{m_\mu}\right) ~, \\
 P_{\ell\ell'}^h &=& \frac{m_{B_s}^2}{2 m_h^2} \frac{c_{\beta-\alpha}^2}{c_\beta^2 s_\beta^2} \left( \frac{m_{bs}^\prime {}^* - m_{sb}^\prime}{m_b V_{tb} V_{ts}^*} \right) \left( \frac{m_{\ell'\ell}^\prime - m_{\ell\ell'}^\prime {}^* }{m_\mu}\right) ~, \\
P_{\ell\ell'}^H &=& \frac{m_{B_s}^2}{2 m_H^2} \frac{s_{\beta-\alpha}^2}{c_\beta^2 s_\beta^2} \left( \frac{m_{bs}^\prime {}^* - m_{sb}^\prime}{m_b V_{tb} V_{ts}^*} \right) \left( \frac{m_{\ell'\ell}^\prime - m_{\ell\ell'}^\prime {}^* }{m_\mu}\right) ~, \\
P_{\ell\ell'}^A &=& \frac{m_{B_s}^2}{2 m_A^2} \frac{1}{c_\beta^2 s_\beta^2} \left( \frac{m_{bs}^\prime {}^* + m_{sb}^\prime}{m_b V_{tb} V_{ts}^*} \right) \left( \frac{m_{\ell'\ell}^\prime + m_{\ell\ell'}^\prime {}^*}{m_\mu}\right) ~.
\end{eqnarray}

\end{appendix}


\end{document}